\documentclass[longauth,fleqn,usenatbib,oneside,openany]{aa}
\usepackage{float}
\usepackage{graphicx}
\usepackage{amsmath}
\usepackage{amssymb}

\usepackage{txfonts}
\usepackage{longtable}
\usepackage{graphicx}
\usepackage{supertabular,booktabs}
\usepackage[section]{placeins}

\makeatletter
\renewcommand*\aa@pageof{, page \thepage{} of \pageref*{LastPage}}
\makeatother
\DeclareRobustCommand{\VAN}[3]{#2}
\let\VANthebibliography\thebibliography
\def\thebibliography{\DeclareRobustCommand{\VAN}[3]{##3}\VANthebibliography}

\usepackage[unicode=true,psdextra]{hyperref}

\begin{document} 
   \title{Long-term evolution of the SN 2009ip-like transient SN 2016cvk}
   \author{K.~K. Matilainen
          \inst{1,2}\fnmsep\thanks{E-mail: katja.matilainen@utu.fi}
          \and
            E. Kankare\inst{1} 
           \and
          S. Mattila\inst{1,3}
           \and
           A. Reguitti\inst{4,5}
           \and
           G. Pignata\inst{6}
           \and
           J. Brimacombe\inst{7}
           \and
           A. Pastorello\inst{5}
           \and
           M. Fraser\inst{8}
           \and
           S.~J. Brennan\inst{9}
           \and
           J.~P. Anderson\inst{10,11}
           \and
           B. Ayala-Inostroza\inst{12}
           \and
           R. Cartier\inst{13}
           \and
           P. Charalampopoulos\inst{1}
           \and
           T.-W. Chen\inst{14} 
           \and
           M. Gromadzki\inst{15} 
           \and
           C.~P. Guti\'errez\inst{16,17} 
           \and
           C. Inserra\inst{18} 
           \and
           T.~E. Müller-Bravo\inst{19,20}
           \and
           M. Nicholl\inst{21}
           \and
           J.~L. Prieto\inst{22,11}
           \and
           F. Ragosta\inst{23,24}         
           \and
           T.~M. Reynolds\inst{1,25,26} 
           \and           
           I. Salmaso\inst{5,24} 
           \and
           D.~R. Young\inst{21} 
          }
   \institute{Department of Physics and Astronomy, University of Turku, FI-20014 University of Turku, Finland
         \and
            Nordic Optical Telescope, Aarhus Universitet, Rambla Jos\'e Ana Fern\`{a}ndez P\'erez 7, local 5, 
            E-38711 San Antonio, Bre\~{n}a Baja 
            Santa Cruz de Tenerife, Spain 
        \and
            School of Sciences, European University Cyprus, Diogenes street, Engomi, 1516 Nicosia, Cyprus
         \and
            INAF - Osservatorio Astronomico di Brera, Via E. Bianchi 46, I-23807 Merate (LC), Italy
         \and
             INAF - Osservatorio Astronomico di Padova, Vicolo dell'Osservatorio 5, I-35122 Padova, Italy        
        \and
            Instituto de Alta Investigaci\'on, Universidad de Tarapac\'a, Casilla 7D, Arica, Chile
        \and
            James Cook University, Cairns, Queensland, Australia
        \and
            School of Physics, University College Dublin, LMI Main Building, Beech Hill Road, Dublin 4, D04 P7W1, Ireland
        \and
            Max-Planck-Institut f\"ur extraterrestrische Physik, Giessenbachstrasse 1, 85748 Garching, Germany
        \and
            European Southern Observatory, Alonso de C\'{o}rdova 3107, Casilla 19, Santiago, Chile
        \and
            Millennium Institute of Astrophysics MAS, Nuncio Monse\~nor Sotero Sanz 100, Off. 104, Providencia, Santiago, Chile
        \and
            Instituto de Astrofísica, Departamento de F\'isica, Facultad de Ciencias Exactas, Universidad Andrés Bello, Fern\'andez Concha 700, Las Condes, Santiago RM, Chile
        \and
            Centro de Astronom\'ia (CITEVA), Universidad de Antofagasta, Avenida Angamos 601, Antofagasta, Chile
        \and
            Graduate Institute of Astronomy, National Central University, 300 Jhongda Road, 32001 Jhongli, Taiwan
        \and
            Astronomical Observatory, University of Warsaw, Al. Ujazdowskie 4,
00-478 Warszawa, Poland
        \and
            Institut d'Estudis Espacials de Catalunya (IEEC), Edifici RDIT, Campus UPC, 08860 Castelldefels (Barcelona), Spain
        \and
            Institute of Space Sciences (ICE, CSIC), Campus UAB, Carrer de Can Magrans, s/n, E-08193 Barcelona, Spain
        \and
            Cardiff Hub for Astrophysics Research and Technology, School of Physics \& Astronomy, Cardiff University, Queens Buildings, The Parade, Cardiff, CF24 3AA, UK
        \and
             School of Physics, Trinity College Dublin, The University of Dublin, Dublin 2, Ireland
        \and
            Instituto de Ciencias Exactas y Naturales (ICEN), Universidad Arturo Prat, Chile
        \and
            Astrophysics Research Centre, School of Mathematics and Physics, Queens University Belfast, Belfast BT7 1NN, UK
        \and
            Instituto de Estudios Astrof\'isicos, Facultad de Ingenier\'ia y Ciencias, Universidad Diego Portales, Avenida Ejercito Libertador 441, Santiago, Chile
        \and
            Dipartimento di Fisica “Ettore Pancini”, Università di Napoli Federico II, Via Cinthia 9, 80126 Naples, Italy
        \and
            INAF - Osservatorio Astronomico di Capodimonte, Via Moiariello 16, I-80131 Naples, Italy
        \and
            Cosmic Dawn Center (DAWN)
        \and
            Niels Bohr Institute, University of Copenhagen, Jagtvej 128, 2200 København N, Denmark
        }
   \date{Received 11.6.2025; accepted 29.08.2025}
   
  \abstract
   {}
   {The interacting transient SN 2016cvk (ASASSN-16jt) is a member of the peculiar SN 2009ip-like events. We present our follow-up data and aim to draw conclusions about the physical nature of the progenitor system.}
   {Our spectrophotometric data set of SN 2016cvk covers the ultraviolet, optical, and near-infrared wavelength region extending to $+1681\,$d from the light curve peak; the data is analysed and compared to other SN 2009ip-like transients. Archival data reveals pre-outbursts of the progenitor with the first detection at $-$1219~d.}
   {The light curve evolution of SN 2016cvk consists of two consecutive luminous events A and B with peak magnitudes of $M_V < -15.6$ and $M_r = -18.3$~mag, respectively. The spectra are dominated by Balmer emission lines that have a complex, multi-component evolution similar to other SN 2009ip-like targets. SN 2016cvk is among the first detected SN 2009ip-like events that show early `flash ionisation' features of C~{\sc iii}, N~{\sc iii}, and He~{\sc ii}, lasting for $16 \pm 5$~d. Our late-time +405~d spectrum shows forbidden [Ca~{\sc ii}], [Fe~{\sc ii}], and [O~{\sc i}] features with the latter detected particularly clearly for a SN 2009ip-like event.}
   {The evolution of SN 2016cvk is similar to other SN 2009ip-like transients, with some uncommon traits. The lack of a double-peaked structure in the Balmer lines is likely caused by differences in the circumstellar medium structure or viewing angle. The flash features in the early spectra propose abundances consistent with a red, yellow, or blue supergiant progenitor rather than for example a luminous blue variable. The detection of [O~{\sc i}] in the +405~d spectrum suggests possible evidence of nucleosynthesised material generated in a SN explosion.}
   \maketitle

\section{Introduction}

A supernova (SN) is a stellar explosion that terminally destroys the star. The most massive (zero-age main sequence mass, $M_\text{ZAMS} > 8M_\odot$) stars in the Universe end their life cycles as core-collapse supernova (CCSN) explosions, which terminally destroy the progenitor star \citep{Smartt2009}. SNe that show narrow ($\lesssim$1000~km~s$^{-1}$) Balmer lines are classified as Type IIn. These lines are produced when the expanding ejecta collides and interacts with the H-rich circumstellar medium (CSM). 
The presence of the dense CSM indicates that the Type IIn SN progenitors had high mass-loss rates shortly before their terminal explosion. For example, a progenitor star has  been observed in possible quiescence before its explosion as the Type IIn SN 2005gl \citep{Gal-Yam2009}. The precursor was identified as a luminous blue variable (LBV) star in this case. However, according to classical models of stellar evolution, LBVs are not expected to explode as SNe at that stage of their lives. Instead, they are expected to first lose most of their outer hydrogen envelope and evolve into highly luminous Wolf-Rayet stars, and later explode as H-poor Type Ib/c SNe \citep[e.g. ][]{Groh2013}. 
On the other hand, large progenitor mass-loss rates inferred from follow up observations of some SNe IIn are only compatible with LBVs \citep[e.g.][]{Fransson2014}. However, the LBV origin of Type IIn SNe has been contested by \cite{Dwarkadas2011}, who pointed out that a high CSM density need not necessarily imply a high wind mass-loss rate; high densities may also arise from wind clumps or a previous LBV phase before the SN explodes as a Wolf-Rayet star. 

One of the most well known and thoroughly studied interacting transients is SN 2009ip. It was initially classified as a Type IIn SN \citep{Maza2009}; however, after its discovery, SN 2009ip underwent several episodes of fading away and re-brightening within a period of three years. During this stage the spectrum of SN 2009ip was hot and dominated by narrow lines, similar to some eruptions of LBV stars. This was followed by a major brightening event in 2012, resulting in a debate on the nature of the event and possible origin between terminal CCSN explosion and various supernova impostor scenarios \citep[e.g.][]{Pastorello2013, Prieto2013, Mauerhan2013, Fraser2013, Smith2013, Levesque2014, Margutti2014, Smith2014, Graham2014, Fraser2015, Graham2017, Smith2022, Pessi2023b}. 

In recent years a growing number of transients have been discovered that share similarities with SN 2009ip, such as SN 2010mc \citep{Ofek2013}, SN 2011fh \citep{Pessi2022,Reguitti2024}, LSQ13zm \citep{Tartaglia2016}, SN 2015bh \citep{EliasRosa2016, Thone2017, Boian2018}, SN 2016bdu \citep{Pastorello2018}, and SN 2016jbu \citep{Kilpatrick2018, Brennan2022a, Brennan2022b}. 
In addition, very recent studies of SN 2022mop \citep{Brennan2025}, SN 2023vbg \citep{Goto2025}, and SN 2023ldh \citep{Pastorello2025} have further highlighted the diversity of SN 2009ip-like events. 

SN 2009ip-like transients have a preceding brightening `event A' shortly before the main `event B'. The peak brightness of event B is almost identical for these transients, accompanied by a quite similar spectral evolution 
\citep[e.g. ][]{Pastorello2013}. This high degree of similarity between events is rather surprising, as the varying temporal evolution of polarization implies that SN 2009ip was a complex transient with a highly structured CSM \citep{Mauerhan2014}. Monte Carlo simulations based on the spectropolarimetric observations of SN 2009ip have suggested a disc-like CSM structure, and a bi-polar main event explosion where the fast-moving ejecta expand in a perpendicular direction compared to the orientation of the CSM disc \citep{Reilly2017}. From this point of view, an observationally diverse set of transients could be expected, due to differences in mass loss history, CSM geometry, and our viewing angle of the events. 

There has been no clear evidence of nucleosynthesised material, such as prominent features of the forbidden lines of [O~{\sc i}] at 6300 and 6364 Å or [Fe~{\sc ii}] at 7155 Å, in any of the SN 2009ip-like events even in the very late-stage observations, and thus their terminal SN nature has been debated \citep{Graham2014, Fraser2015, Brennan2022a, Smith2022}. However, for example, weak features of [O~{\sc i}] have been detected in the late-time spectra of  SN 2009ip \citep{Fraser2015}, SN 2016bdu \citep{Pastorello2018}, and SN 2016jbu \citep{Brennan2022b}. If not genuine SNe, the luminosities of SN 2009ip-like events would likely arise from an explosive mass-loss episode of the star, during which the ejected material strongly interacts with ambient gas around the progenitor. However, the late-time Hubble Space Telescope observations of SN 2009ip, SN 2015bh, and SN 2016jbu have shown that these events have faded below the brightness level of their progenitors identified in archival data, strongly supporting a terminal explosion scenario \citep{Brennan2022c, Jencson2022, Smith2022}.

A new addition to the expanding list of SN 2009ip-like events is the transient SN 2016cvk presented in this work. The data of SN 2016cvk are summarised in Sect. \ref{sec:observations}. The light curve and spectroscopic evolution are analysed in Sect. \ref{sec:lightcurves} and \ref{sec:spectra}, respectively. Discussion and final conclusions are presented in Sect. \ref{sec:discussion} and \ref{sec:conclusions}, respectively.

\section{Data}\label{sec:observations}

\begin{figure}
\includegraphics[trim={4.5cm 8cm 5cm 8cm},clip,width=\linewidth]{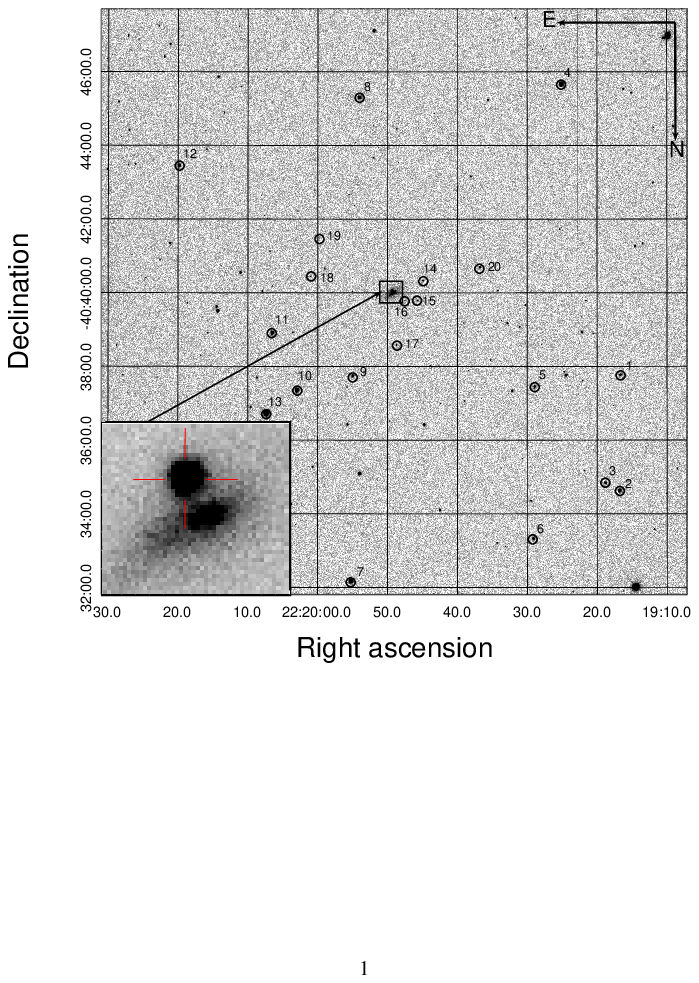}
\caption{Image of the field of SN 2016cvk and its host galaxy ESO 344 - G 021 in $r$-band observed with a Las Cumbres Observatory 1m\#5 telescope in CTIO, Chile, on 2016 September 10. Location of the transient ($\alpha = 22^{\mathrm{h}}19^{\mathrm{m}}49\fs39$ and $\delta = -40\degr40\arcmin03\farcs2$) is marked in the subpanel image with a red crosshair. The field stars used to calibrate the photometry are circled in the image.}
\label{fig:FC}
\end{figure}

SN 2016cvk was discovered by \cite{Parker2016} in ESO 344 - G 021 at an unfiltered magnitude of 17.6~mag using a 35~cm Celestron telescope on 2016 June 12.64 UT (JD = 2457552.14) preceded by the last non-detection of $>$18~mag on 2016 June 3.59 UT (JD = 2457543.09). The All Sky Automated Survey for SuperNovae (ASAS-SN) reported the discovery of the transient ASASSN-16jt at a $V$-band magnitude of 16.2~mag on 2016 August 31.09 UT (2460918.59) with a non-detection of $>$17.8~mag on 2016 August 26.24 UT (JD = 2460913.74), which they associated with SN 2016cvk. \cite{Brown2016} reported additional ASAS-SN $V$-band non-detections of the event and spectra obtained with the du Pont 2.5 m telescope on 2016 June 18 and 2016 September 1, noting the similarity of the former spectrum with that of SN 2009ip before the luminous peak, and the characteristics of the latter spectrum including a blue continuum, Balmer lines with broad and narrow emission components, and He~{\sc i} and He~{\sc ii} 4686 Å features. A Type IIn-pec classification spectrum for the Transient Name Server was reported by \cite{Bersier2016} obtained with the New Technology Telescope (NTT) of the European Southern Observatory (ESO) on 2016 September 6.0 UT (JD = 2460924.5). Radio non-detection of the event was reported by \cite{Ryder2016} using the Australia Telescope Compact Array resulting in 3$\sigma$ upper limits of 68 and 75 microJy/beam at 5.5 and 9.0 GHz, respectively, on 2016 September 4.6 UT (JD = 2460923.1). \cite{Higgins2019} reported based on optical observations no detection of intrinsic polarization with an upper limit of $P \leq 1.90$~\% in SN 2016cvk during the event A on 2016 June 20.3 (JD 2457559.8, i.e. $-$84~d from the event B peak). SN 2016cvk is located at the edge of a small, weakly barred host ESO 344 - G 021, which is morphologically classified as a Sab: galaxy in the NASA Extragalactic Database (NED)\footnote{\url{https://ned.ipac.caltech.edu}}, see Fig. \ref{fig:FC}. The metallicity $12 + \log(\text{O/H})_{\text{N}2} = 8.529 \pm 0.028$ of the local environment of the transient (measured within the 0.3 kpc radius around the SN position) is very close to the solar value ($\sim$8.7), and the star formation rate surface density of the host galaxy is around $10^{-4.0}$~$M_\odot$~$\text{yr}^{-1}$~$\text{kpc}^{-2}$ \citep{Moriya2023}.

\subsection{Distance and extinction}\label{sec:redshift}

\begin{table*}
\caption{Coordinates, redshifts, luminosity distances, distance moduli, line-of-sight extinction values, and $r$-band peak magnitudes and epochs for SN 2009ip-like events.}
\centering
\begin{tabular}{llllllllc}
\hline 
\multicolumn{1}{c}{Object} & \multicolumn{1}{c}{RA} & \multicolumn{1}{c}{Dec} & \multicolumn{1}{c}{$z$} & \multicolumn{1}{c}{$D_L$} & \multicolumn{1}{c}{$\mu$} & \multicolumn{1}{c}{$A_v$} & \multicolumn{1}{c}{$M_{r,\text{peak}}$} & \multicolumn{1}{c}{$\text{JD}_\text{peak}$} \\
& \multicolumn{1}{c}{} & \multicolumn{1}{c}{} &  & \multicolumn{1}{c}{(Mpc)} & \multicolumn{1}{c}{(mag)} & \multicolumn{1}{c}{(mag)} & \multicolumn{1}{c}{(mag)} & \multicolumn{1}{c}{($-2400000$ d)} \\
\hline
SN 2016cvk & $22^{\mathrm{h}}19^{\mathrm{m}}49\fs39$ & $-40\degr40\arcmin03\farcs2$ & 0.0108$^b$   & 43.3 & 33.18 & 0.035$^f$ & $-18.4$ & 57643.4 \\
SN 2009ip  &$22^{\mathrm{h}}23^{\mathrm{m}}08\fs26$ & $-28\degr56\arcmin52 \farcs40 $ & 0.00590$^b$   & 24.0 & 31.90 & 0.055$^f$ &$-18.0$ & 56209.7 \\ 
SN 2016bdu   &$13^{\mathrm{h}}10^{\mathrm{m}}13\fs95$ & $+32\degr31\arcmin14 \farcs07 $ & 0.0173$^c$     & 76.1$^c$ & 34.37$^{c}$ & 0.041$^{c}$ & $-17.9$ & 57542.7 \\
SN 2016jbu   &$07^{\mathrm{h}}36^{\mathrm{m}}25\fs96$ & $-69\degr32\arcmin55 \farcs25 $ & 0.00489$^b$   & 17.3 & 31.60 & 0.555$^f$ & $-18.0$ & 57783.9 \\
SN 2015bh    &$09^{\mathrm{h}}09^{\mathrm{m}}35\fs06$ & $+33\degr07\arcmin22 \farcs12 $ & 0.00649$^d$   & 29.4 & 32.35 & $0.713^{f,g}$ & $-17.8$ & 57168.5 \\
LSQ13zm     &$10^{\mathrm{h}}26^{\mathrm{m}}54\fs59$ & $+19\degr52\arcmin54 \farcs91 $ & 0.029$^a$      & 122 & 35.43 & 0.052$^f$ & $-18.4$ & 56409.7 \\
SN 2010mc    &$17^{\mathrm{h}}21^{\mathrm{m}}30\fs68$ & $+48\degr07\arcmin47 \farcs39 $ & 0.035$^e$   & 152  & 35.91 & 0.046$^f$ & $-18.4$ & 55447.0 \\
\hline
\end{tabular}

$^a$ \protect\cite{Tartaglia2016}, $^b$ \protect\cite{Wong2006}, $^c$ \protect\cite{Pastorello2018}, $^d$ \protect\cite{Haynes1997}, $^e$ \protect\cite{Ofek2013b}, $^f$\protect\cite{Schlafly2011}, $^g$\protect\cite{Thone2017}

\label{tab:rs_table}
\end{table*}

\begin{figure*}
\centering
\begin{minipage}{0.5\linewidth}
\includegraphics[trim={0cm 0cm 0cm 0cm},clip,width=\linewidth]{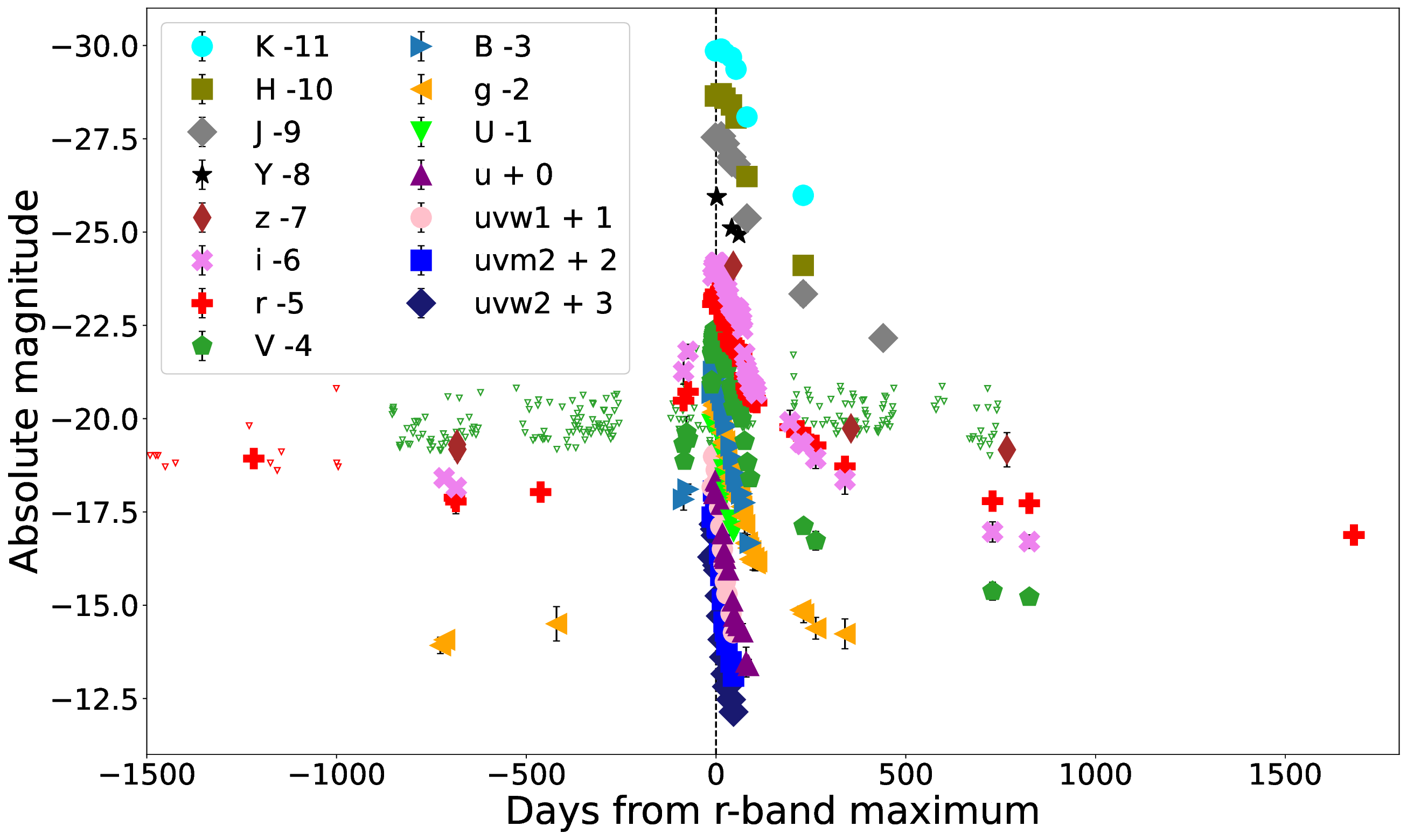}
\end{minipage}\begin{minipage}{0.5\linewidth}

\includegraphics[trim={0cm 0cm 0cm 0cm},clip,width=\linewidth]{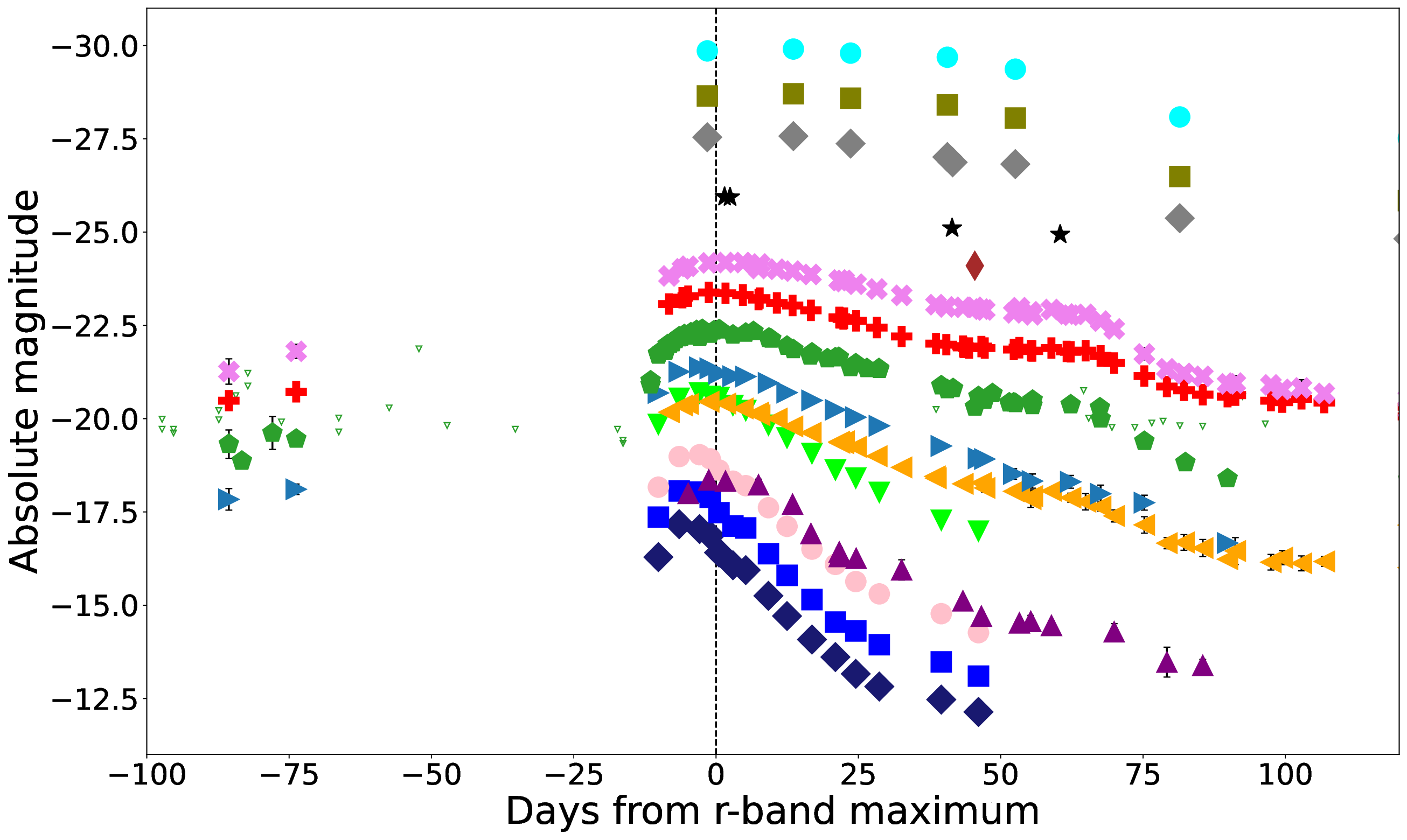}
\end{minipage}
\caption{Extinction-corrected absolute $u, g, r, i, z, Y$ (AB magnitudes) and $UVW2, UVM2, UVW1, U, B, V, J, H, K$ (Vega magnitudes) light curves for SN 2016cvk, shifted for clarity as indicated in the figure legend. The dashed vertical line at $t=0$ signifies the event B peak. Downward-pointing triangles indicate upper limits.}
\label{fig:abs2016cvk}
\end{figure*}

\begin{figure*}
\centering
\begin{minipage}{0.5\linewidth}
\includegraphics[trim={0cm 0cm 0cm 0cm},clip,width=\linewidth]{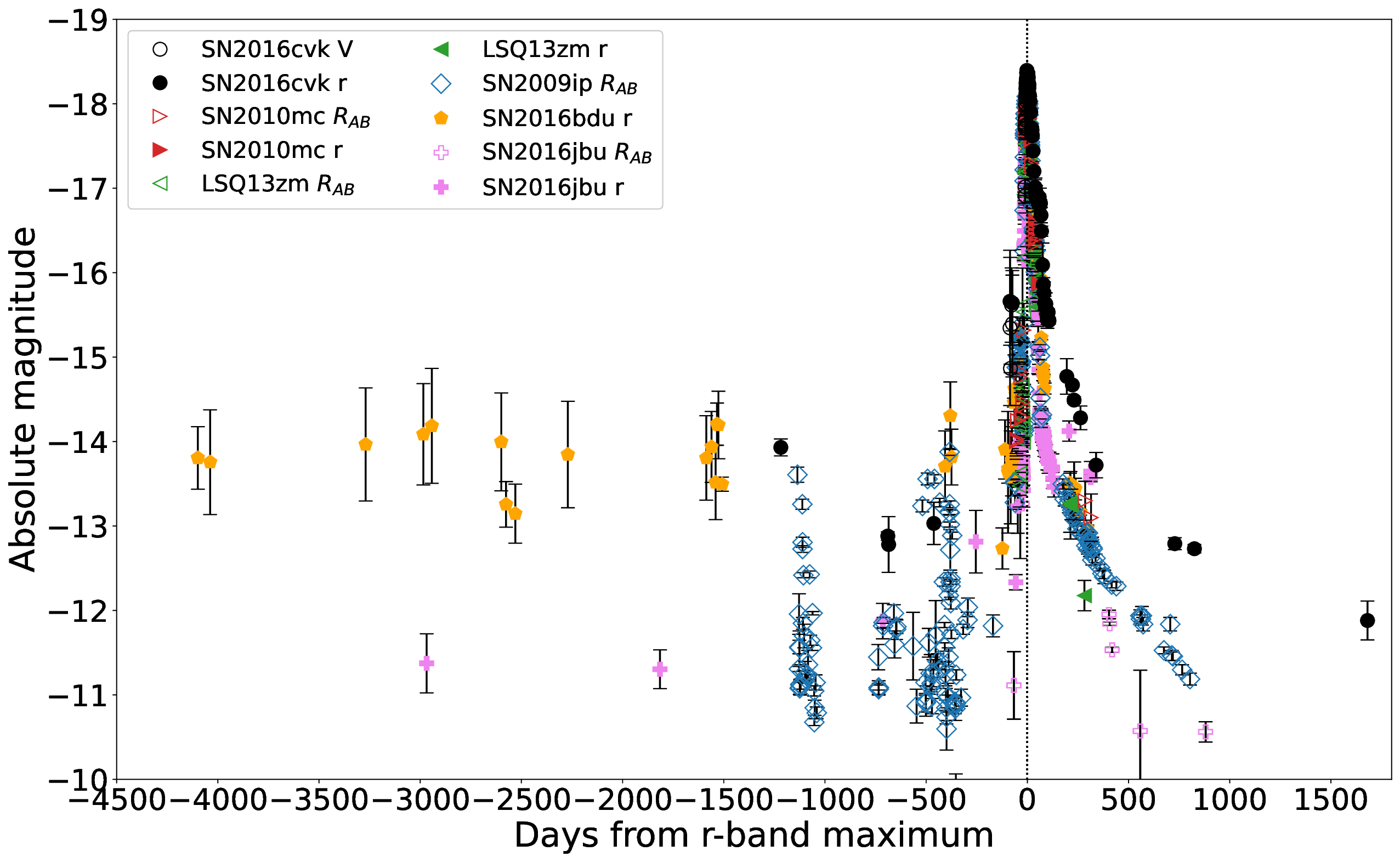}
\end{minipage}\begin{minipage}{0.5\linewidth}
\includegraphics[trim={0cm 0cm 0cm 0cm},clip,width=\linewidth]{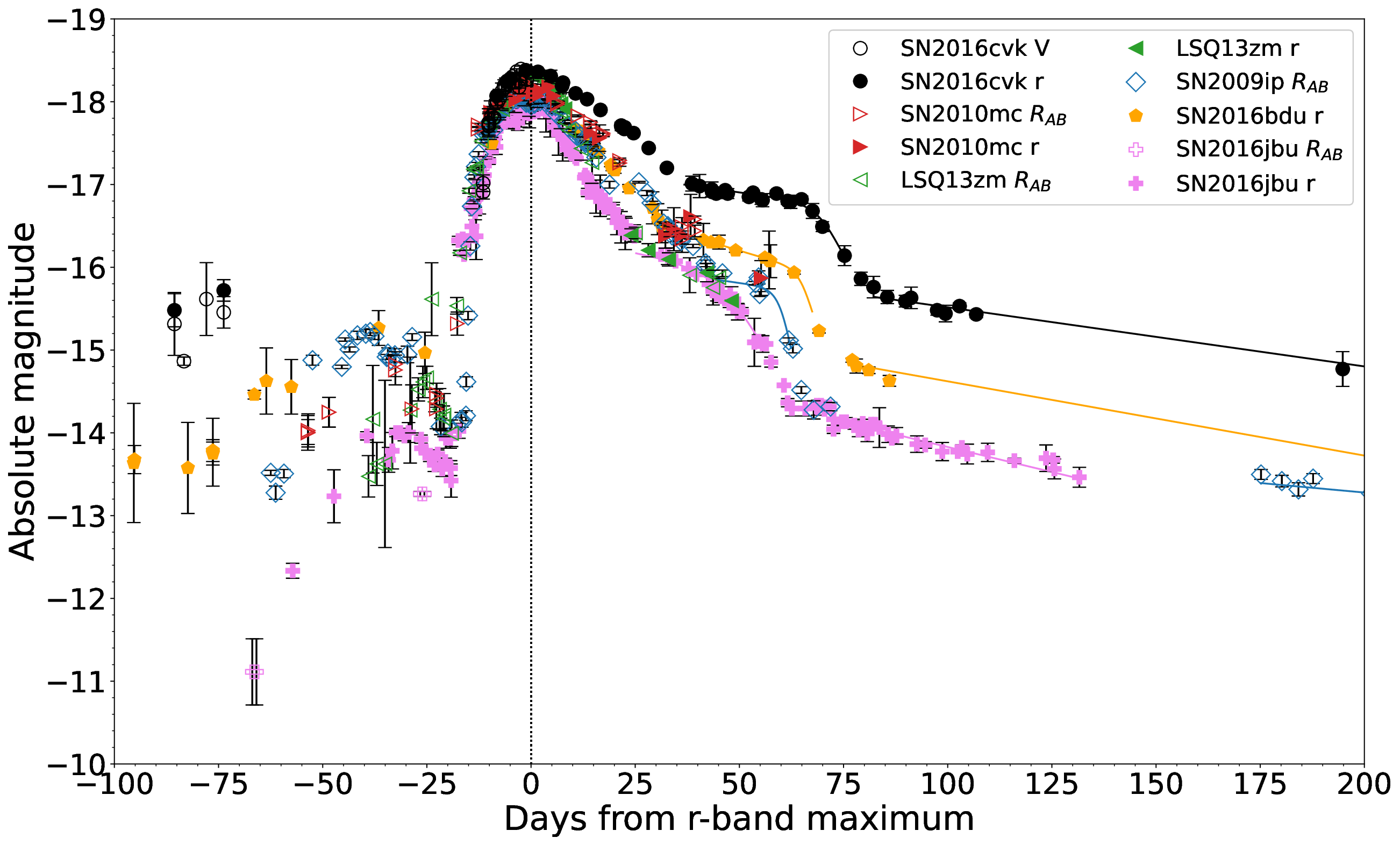}
\end{minipage}
\caption{Extinction-corrected absolute $r$ (or $R_{\text{AB}}$) band light curves for a selection of SN 2009ip-like transients. Early $V$-band data is also included for SN 2016cvk. In the right figure the fits for the plateau and the tail phases are shown with solid curves. The dashed vertical line at $t=0$ signifies the event B peak.}
\label{fig:LCcomp}
\end{figure*}

In Table \ref{tab:rs_table} the coordinates (RA, Dec), redshifts ($z$), luminosity distances ($D_L$), distance moduli ($\mu$), and line-of-sight $V$-band extinctions ($A_V$) for SN 2016cvk and a sample of SN 2009ip-like events are summarized, along with the peak $r$-band magnitudes and their corresponding JD epochs. The data collected from literature and public databases has been homogenised, assuming $H_0 = 73 \, \text{km s}^{-1} \text{Mpc}^{-1}$, $\Omega_M = 0.27$, and $\Omega_\Lambda = 0.73$. Virgo infall corrected host luminosity distances were adopted, if readily available via NED. The luminosity distance of SN 2016bdu is adopted from \protect\cite{Pastorello2018}. The adopted distances are generally consistent with the values used in the previous studies of the comparison sample events. For targets with redshifts based observations of the transient, the luminosity distances were derived using the Ned Wright's Cosmology Calculator \citep{Wright2006}. The Galactic dust map based $A_V$ values \citep{Schlafly2011} were retrieved from NED using the coordinates of the transients. For reddening, a standard extinction law by \cite{Cardelli1989} was adopted with $R_V=3.1$. For SN 2016cvk and the other selected SN 2009ip-like targets in the literature a negligible host galaxy extinction has been assumed, typically based on the lack of Na~{\sc i}~D absorption lines, with the exception of SN 2015bh, which has an associated host galaxy extinction of $A_V = 0.651$~mag \citep{Thone2017}. The epochs of our sample transients are provided in the observer frame; the redshift time-dilation effect for the nearby events is quite small.

\subsection{Photometric observations}\label{sec:16cvkphotometry}

Optical imaging of the field of SN 2016cvk was taken between 2014 September 16 and 2021 April 9 in $ugVrizY$ bands through several observing programmes and in near-infrared $JHK$ bands between 2016 September 9 and 2017 November 25, see Appendix for details on the observations and data reduction.
For both the optical and near-infrared ground-based images, we carried out point spread function (PSF) photometry with the QUBA pipeline \citep{Valenti2011}. For this step, the $ugri$ magnitudes of the field stars were calibrated using observations of photometric standard stars from the Las Cumbres Observatory facilities. Furthermore, the $g$ and $r$-band field star magnitudes were converted into $B$ and $V$-band magnitudes via the transformations by \citet{Jester2005}. A selection of $z$ and $Y$-band field star magnitudes were adopted from the Dark Energy Survey Data Release 2 \citep{Abbott2021} for the analysis. In $JHK$ bands the Two Micron All Sky Survey \citep[2MASS; ][]{Skrutskie2006} magnitudes of the field stars were adopted to calibrate the field.  Archival data was also obtained for the analysis from the Neil Gehrels Swift Observatory in $UVW2$, $UVM2$, $UVW1$, $U$, $B$, and $V$ bands. A least-squares 3rd order polynomial fit was carried out to the $r$-band light curve near maximum; the resulting JD$_\text{peak}$ 2457643.4 was adopted as the reference epoch ($t = 0$ d). A complete table of photometric observations of SN 2016cvk is given in Table \ref{tab:photom} for the ground-based optical range ($u$, $B$, $g$, $V$, $r$, $i$, $z$ and $Y$ bands), in Table \ref{tab:photomJHK} for the near-infrared area ($J$, $H$ and $K$ bands) and in Table \ref{tab:xray} for the Swift observations ($UVW2$, $UVM2$, $UVW1$, $U$, $B$ and $V$ bands). The absolute light curves of SN 2016cvk in different bands are shown in Fig. \ref{fig:abs2016cvk}. Light curve comparisons with other SN 2009ip-like events are found in Fig. \ref{fig:LCcomp}. In the comparison plots, both the Sloan $r$ and Bessel $R$ bands are shown for some of the events for a better coverage of the different light curve phases. Magnitudes in the $u$, $g$, $r$, $i$, $z$, and $Y$ bands are given in the AB system and in the other bands in the Vega system unless otherwise noted.

\subsection{Spectroscopic observations}

Most of the spectroscopic monitoring of SN 2016cvk was carried out with the NTT using the ESO Faint Object Spectrograph and Camera 2 (EFOSC2) and the Son of ISAAC (SOFI) instruments via the extended Public ESO Spectroscopic Survey of Transient Objects (ePESSTO) programme \citep{Smartt2015}. A single late phase spectrum of the target at around +405~d from peak was obtained using the Very Large Telescope (VLT) with the FOcal Reducer and low dispersion Spectrograph 2 (FORS2) instrument at the Cerro Paranal observatory, Chile. Details on the data reduction are provided in the Appendix. We could also include in our analysis the early spectra reported by Brown et al. (2016), which were obtained with the du Pont 2.5 m telescope using the Wide-Field CCD (WFCCD) instrument at the Las Campanas Observatory, Chile. In Table \ref{tab:specinstr} the full details of all the spectroscopic observations of SN 2016cvk are listed.

\section{Light curve evolution}\label{sec:lightcurves}

\begin{figure}
\centering
\includegraphics[trim={0cm 0cm 0cm 0cm},clip,width=\linewidth]{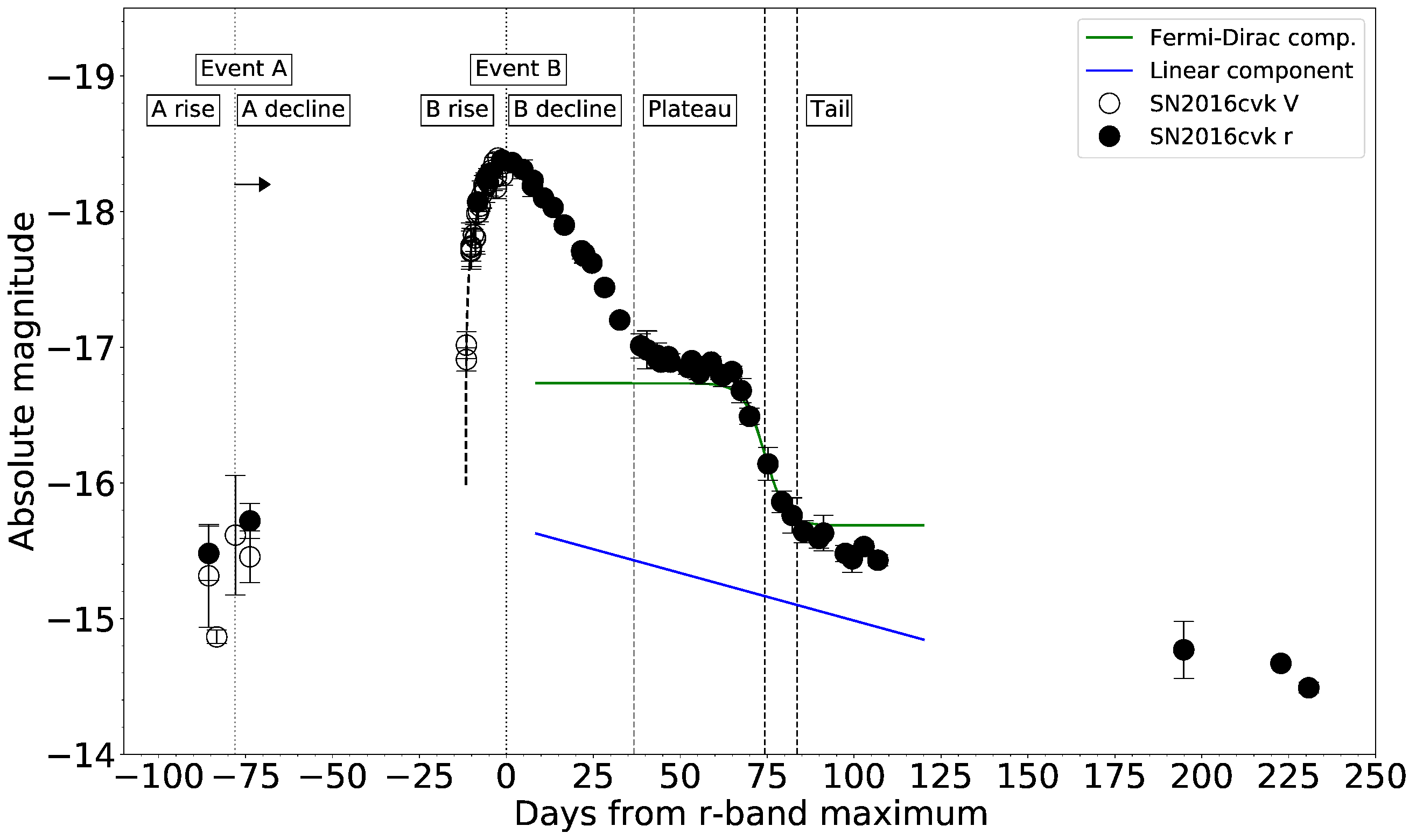}
\caption{Light curve phases of SN 2016cvk. Fits for the event B rise phase (black dashed curve) and the plateau phase (black solid curve) of the transient are included.}
\label{fig:LCphases16cvk}
\end{figure}

\begin{table*}
\caption{Epochs, durations, and magnitudes of $r$-band light curve phases for the sample of SN 2009ip-like transients with typical errors of 0.1~mag for the magnitudes and 1~d for the epochs, unless otherwise stated.}
\centering
\begin{tabular}{lcccccccc}
\hline
\hline
Transient & $t_\text{peak, A}$ & $M_{\text{peak A}}$ & $M_{\text{peak B}}$ & $t_\text{mid, plateau}$  & $\Delta t_\text{plateau}$ & $M_{\text{plateau}}$ &$t_\text{start, tail}$ & $\gamma_\text{tail}$\\
 & (d) & (mag) & (mag) & (d) & (d) & (mag) & (d) & (mag / 100 d) \\
\hline
SN 2016cvk & $>-78$$^a$ & $< -15.6$$^a$ & $-18.3 $ & $+54 $ & $38 $ & $-16.9 $ & $+82 $&$0.7 \pm 0.1$ \\
SN 2009ip & $-38 \pm 1$ & $-15.1 $ & $-18.0 $ & $+53 $ & $18 $ & $-15.8 $ &$+67 $&$0.5 \pm 0.1$\\
SN 2016bdu & $-39 \pm 5$ & $-15.2 $ & $-17.9 $ & $+49 $ & $36 $ & $-16.2 $ &$+73 $&$0.9 \pm 0.1$\\
SN 2016jbu & $-34 \pm 1$ & $-13.9 $ & $-18.0 $ & $ +40 $ & $30$ & $-15.9 $ &$+73 $& $1.3 \pm 0.1$\\
LSQ13zm & $-28 \pm 1$ & $-14.6 $ & $-18.2 $ & $>+37$ & $>16$ & $>-15.9$ &$<+211$&$-$\\
SN 2010mc & $-35 \pm 2$ & $-14.7 $ &  $-18.1 $ & $>+35$ & $>8$ & $>-16.5$ &$<+285$&$-$\\
\hline
\end{tabular}

$^a$ \protect{in $V$-band}
\label{tab:LCphases}
\end{table*}

SN 2009ip-like targets have a very similar evolution in their light curves, with clearly distinguishable light curve phases for which we adopt the nomenclature by \citet{Graham2014}. There are two clear major events in the evolution of SN 2009ip-like transients in our sample: a less bright initial event A, and the main event B. After the initial sharp decline from the event B peak, all the SN 2009ip-like events in our sample flatten to a more moderately decreasing or almost horizontal plateau phase. The plateau later turns into another sharp decline in the light curve, called the "knee/ankle-stage" in previous literature, followed by a late linear tail phase with a slowly evolving decline. 

\subsection{Light curve phases}\label{sec:lcphases}

To define the phases of SN 2016cvk and to carry out comparisons between SN 2009ip-like events, selected functions were fitted to these light curve sections for our sample of targets. SN 2015bh was not included in this part of the analysis, since the reported photometric observations did not cover the epochs during the short plateau phase shown by SN 2009ip-like events. Other potentially SN 2009ip-like SNe with even more limited spectrophotometric coverage, such as SN 2011fh, were left out of the sample entirely. 
For phases A and B, the fits were carried out with second or third degree polynomials via minimizing the $\chi^2$ value. The uncertainties for these parameters were determined via the error limits of the polynomial fit coefficients.

We fitted the plateau and the knee-ankle phase with a combination of a Fermi-Dirac function and a linear part, similar to the approach used by \cite{Anderson2014} and \cite{Olivares2010} for the plateau of Type IIP SNe via equation 
\begin{equation}\label{eq:FD}
f_{FD}(t) = M_0 - p_1 (t-t_{PT}) - \frac{a_0}{e^{\frac{1}{\omega_0}(t-t_{PT})}+1}
\end{equation}
A visual representation of the parameters is found in Fig. \ref{fig:LCphases16cvk} for SN 2016cvk. $M_0$ is the magnitude level of the fit, which the tail of the Fermi-Dirac fit approaches. Coefficient $p_1$ is the slope of the linear component. The parameter $t_{PT}$ is the midpoint in time during the drop from plateau to tail phase, $a_0$ is the change in magnitude during the Fermi-Dirac fit, and $\omega_0$ describes the width of the transition between plateau and tail phases, so that at epoch $t=t_{PT}-3 \omega_0$ the magnitude is $M_r \approx M_0 - 0.953 a_0$. The epoch $t = t_{PT}$ was chosen as the end of the plateau phase to make definitions clear and easily comparable between the different events. Hence, the soft knee transition from plateau into decline is also included in the plateau definition.

The starting epoch of the plateau phase was estimated as the crossing point of a second degree polynomial fit of the B phase and the Fermi-Dirac fit of the plateau phase. The uncertainties of the crossing times were determined based on the error margins of the fit parameters. The plateau phase magnitude was set as the value of the Fermi-Dirac fit at the midpoint of the plateau. The final tail phase was treated as a simple first-order polynomial fit. The beginning of the tail phase was defined as $t = t_{PT}+3 \omega_0$. 
Our derived epochs from the $r$-band peak of event B, durations and absolute $r$-band magnitudes of the different light curve phases are listed in Table \ref{tab:LCphases}. We carried out some checks for correlations between key epochs or magnitudes of the different light curve phases of our sample, see Appendix; however, no robust trends were identified.

\subsubsection{Event A}
Event A of SN 2009ip-like transients is a preceding phase to the brighter main event B. The event A absolute peak magnitudes in our sample range from $-13.9 \pm 0.1$ to $<-15.6$~mag (see Table \ref{tab:LCphases}). The rise to the event A peak is slow for SN 2016bdu \citep{Pastorello2018,Brennan2022a}. For SN 2009ip, SN 2016jbu, and LSQ13zm \citep{Pastorello2013,Brennan2022a,Tartaglia2016}, the A phase is well defined with a clear, moderately steep rise and decline phase. For SN 2016cvk the A phase coverage is unfortunately scarce. Detections in $V$-band as early as $-90$ and $-78$~d from event B peak suggest that SN 2016cvk had a more moderate rise to the event A peak, similar to SN 2016bdu. For SN 2016cvk, the brightest $V$-band data point in the A phase is used as a lower limit for the peak magnitude of event A ($M_{V,\text{A}} < -15.6$~mag). The timing of the A phase is similarly taken from this detection and given as a lower limit of $t_\text{peak,A} > -78$~d, which does not provide a tight constraint. 

\subsubsection{Event B and plateau phase}
Event B is the main event in the light curve evolution of SN 2009ip-like transients. The rise to the event B peak is much shorter and steeper than the rise to the A phase peak. The event B peak within a 0.4~mag $r$-band range between $-17.9 \pm 0.1$ and $-18.3 \pm 0.1$~mag. This is a notably narrower distribution than the wide $>$1.7~mag range of event A peak magnitudes. The peak magnitude for SN 2016cvk is $M_{r,\text{B}} = -18.3 \pm 0.1$~mag, which makes it the brightest transient in the sample.

The initial steep decline from the main event peak evolves into either a less rapid linear descent or a horizontal plateau. There is a distribution of plateau properties for the objects in our sample; however, SN 2016cvk stands out as having the longest and most luminous plateau, as well as the latest phase for the plateau midpoint. The plateau phase of SN 2016cvk is also somewhat brighter than that of the rest of the events in the sample, as its plateau midpoint absolute magnitude is $-16.9 \pm 0.1$~mag. For the remainder of the sample the measured values are very close to each other, between $-15.8 \pm 0.1$ and $-16.2 \pm 0.1$~mag.

\subsubsection{Tail phase}

We have sufficient data to fit both the Fermi-Dirac function and the tail phase for SN 2016cvk, as well as several other SN 2009ip-like events. SN 2016cvk is the most luminous of these objects at the beginning of the tail phase, and exhibits a slow decline rate 0.7~mag per 100~d. Using equation 2 from \cite{Hamuy2003}, we estimate $< 0.07 M_\odot$ as a conservative upper limit for ejected $^{56}$Ni mass during event B of SN 2016cvk, using the bolometric tail luminosity $L_t$ estimated from the $+82$~d photometry (see Sect. \ref{sec:bb}). The final value for the $^{56}$Ni mass is an upper limit due to the ongoing CSM interaction in the tail phase.

\subsection{Blackbody evolution}
\label{sec:bb}

\begin{figure}
\includegraphics[trim={0.5cm 0.5cm 0cm 0.5cm},clip,width=\linewidth]
{{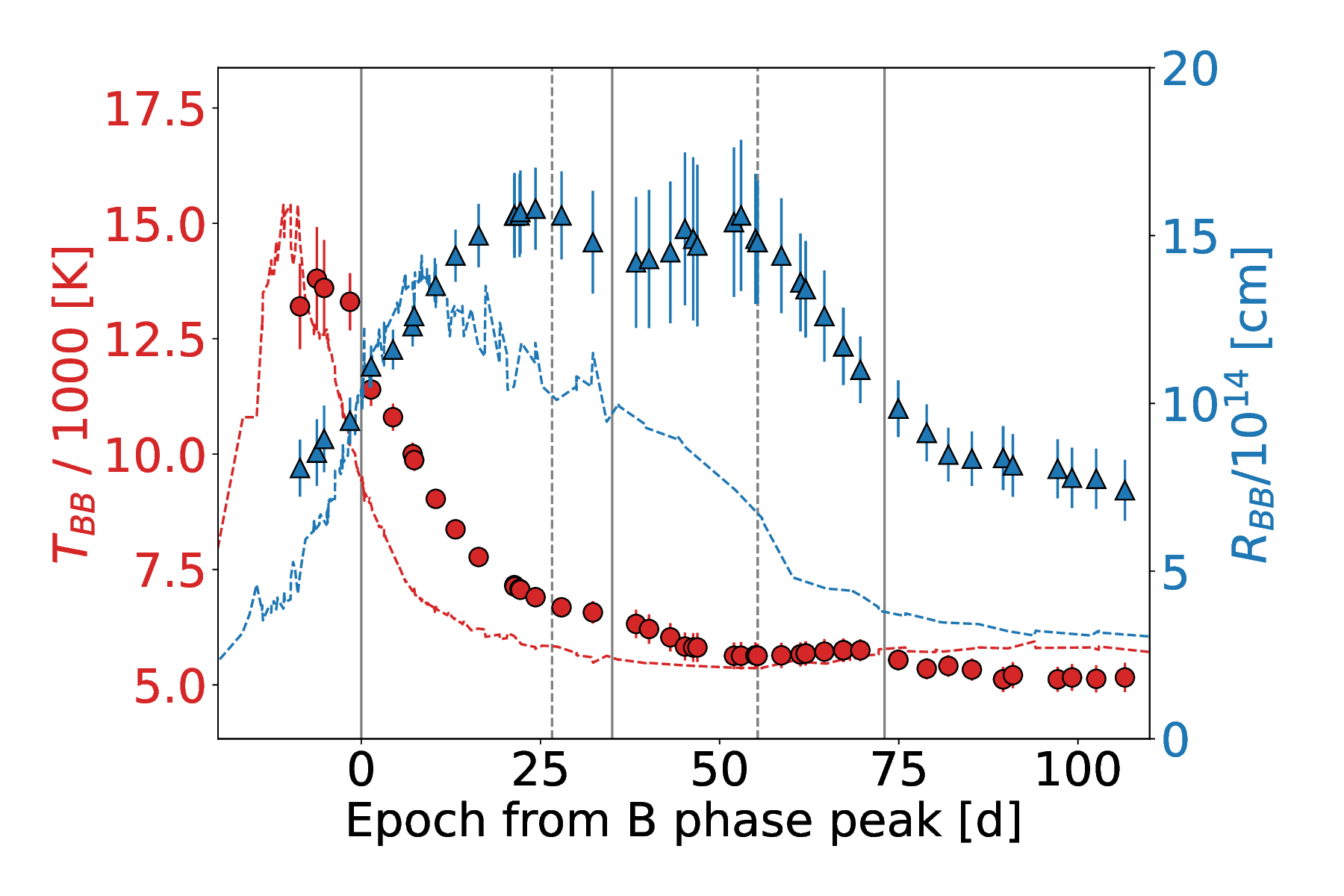}}
\caption{Evolution of the blackbody temperature $T_\text{BB}$ (red circles) and radius $R_\text{BB}$ (blue triangles) of SN 2016cvk. Values reported for SN 2016jbu \citep{Brennan2022b} are shown for comparison with dashed curves.}
\label{fig:Tbb}
\end{figure}

To roughly characterize the temperature and radial evolution of SN 2016cvk, blackbody fits were carried out to the available photometry between $-10.5$ and $+106.7$ d using the SuperBol package \citep{Nicholl2018}. 
The package generates an interpolated light curve for the bands used in the fits within the given time range, and creates a blackbody fit for each interpolated epoch. The photometry used in the fit includes ultraviolet data between $-10.4$ and $+45.8$ d (Table \ref{tab:xray}), optical data between $-10.5$ and $+106.7$ d (Table \ref{tab:photom}), and near-infrared data between $-1.7$ and $+81.2$ d (Table \ref{tab:photomJHK}).

The evolution of the blackbody parameters of SN 2016cvk are similar to those of other SN 2009ip-like events. For all transients in our sample, the blackbody temperature is highest between $-14$ and $-6$ d before the phase B peak (see Table \ref{tab:maxt}), and decreases rapidly before the beginning of the plateau phase followed by a more moderate temperature decline \citep{Brennan2022b, Fraser2013, Ofek2013, Tartaglia2016, Thone2017}. The results for the blackbody temperature, $T_\text{BB}$, and radius, $R_\text{BB}$, of SN 2016cvk are found in Table \ref{tab:bb} and in Fig. \ref{fig:Tbb}.

For SN 2016cvk the peak temperature is close to 14000~K around $-6$ d, and decreases quickly to roughly 6300~K at $+38$ d from peak, which corresponds to an epoch shortly after the start of the plateau phase. Due to the relatively large errors of the early $T_\text{BB}$ values, it cannot be excluded that the blackbody temperature of SN 2016cvk would have peaked somewhat earlier. For comparison, the blackbody temperature evolution of SN 2016jbu has a similar fast drop from 12000 to 6000~K during the first $\sim$30~d from the phase B peak \citep{Brennan2022a}, until the start of the plateau phase at +31~d (Table~\ref{tab:LCphases}). A similar trend during the first $\sim$30 or 40~d after the event B peak has also been found for SN 2009ip, SN 2010mc, LSQ13zm, SN 2016bdu and SN 2015bh \citep{Fraser2013, Ofek2013, Tartaglia2016, Pastorello2018, Thone2017}. During the plateau phase the decrease in temperature is more moderate for all SN 2009ip-like transients in our sample. At the end of the plateau phase the temperature has reached approximately $5000-6000$~K in all the events where this epoch is sufficiently documented. The blackbody temperature of SN 2016cvk flattens to values around 5700~K during the plateau phase. 

The blackbody radius of SN 2016cvk has a similar evolution as observed in other SN 2009ip-like transients. The radius expands rapidly to $1.6 \times 10^{15} \text{cm}$ at $+24$~d, approximately a week before the plateau phase begins. After this epoch the radius drops slightly and remains fairly constant throughout the plateau phase. Towards the end of the plateau phase, the blackbody radius starts to recede rapidly and reaches $\sim$8.5$\times 10^{14}$~cm at +82~d in the beginning of the tail phase. In the case of SN 2016jbu, the blackbody radius continues to increase also after the phase B peak, and reaches a maximum value of $1.2 \times 10^{15}$~cm at +7~d, about 20~d before the beginning of the plateau phase. This radius remains roughly constant around $10^{15}$~cm throughout the plateau, and begins to rapidly decrease after the knee phase in the light curve \citep{Brennan2022a}. A blackbody fit is unlikely to describe reliably the late-time observations and the epochs beyond +107~d were not included in the analysis. 

We also carried out a blackbody fit to the $giz$-band photometric data of the historical precursor detection between $-727$ and $-682$~d, estimating a blackbody temperature of $7600 \pm 1300$~K and a radius of $1.2 \pm 0.3 \times 10^{14}$~cm. The luminosity of the historical detections is consistent with a relatively large radius, which suggests that the star was observed during an outburst. The $r$-band data showed excess compared to the rest of the photometric spectral energy distribution, likely arising from strong H$_\alpha$ emission and the band was not included in the fit.  

\section{Spectral evolution}\label{sec:spectra}

\begin{figure*}
\includegraphics[trim={2cm 4cm 0.5cm 3cm},clip,width=\linewidth]{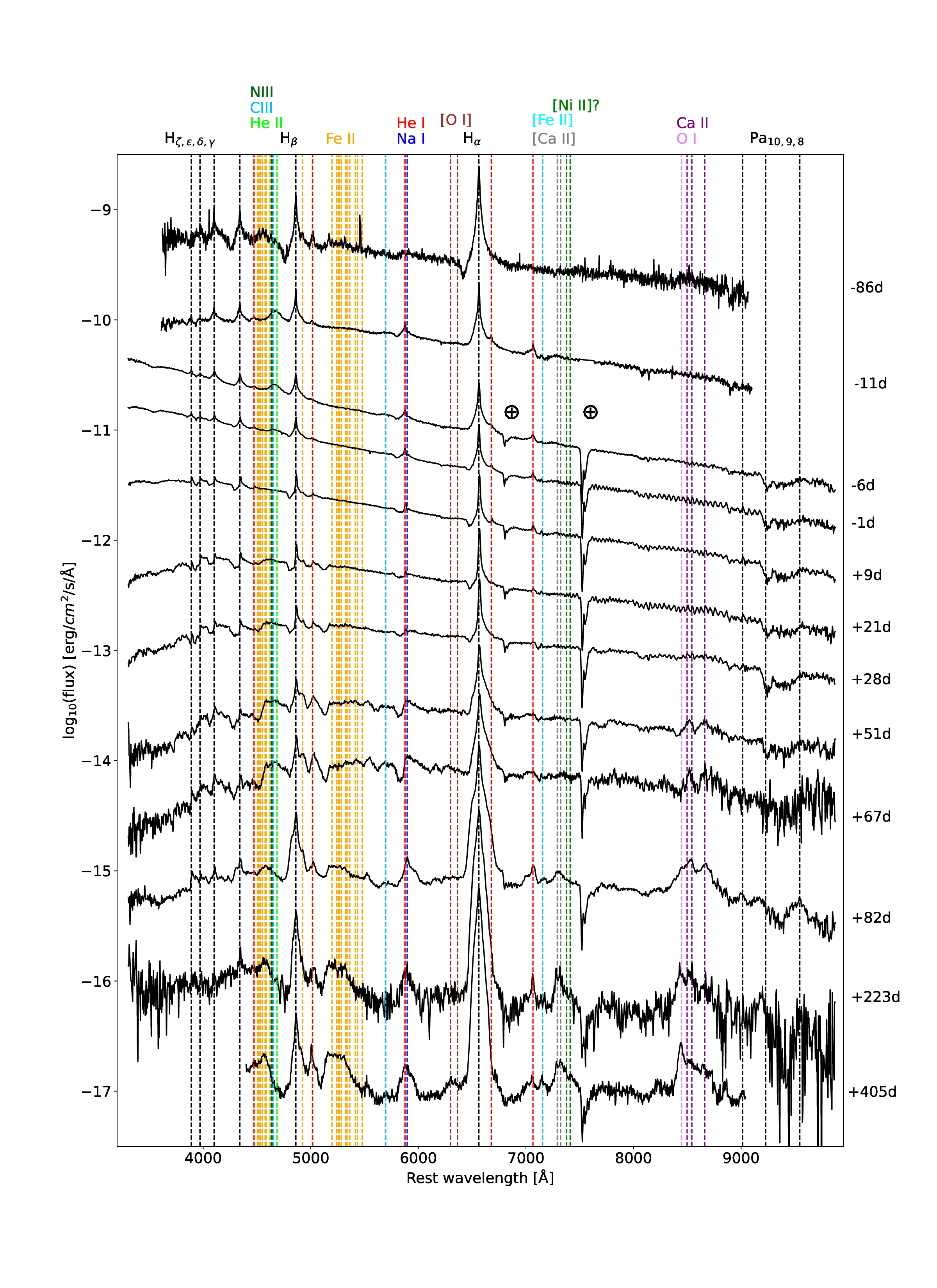}

\caption{Spectral time series of SN 2016cvk with the epochs relative to the $r$-band maximum. The spectra have been dereddened and corrected to the rest frame wavelengths. The wavelengths of the most prominent spectral lines are indicated with vertical dashed lines and the telluric features with a $\oplus$ symbol. Logarithmic scale is used for flux and the spectra have been vertically shifted for clarity.}
\label{fig:logspectra2016cvk}
\end{figure*} 

Similar to other SN 2009ip-like events, SN 2016cvk shares spectroscopic similarity to Type IIn SNe and undergoes a complex evolution with Balmer lines as the most prominent features, with the H$_\alpha$ and H$_\beta$ multi-component development detailed in Sect. \ref{sec:balmer}. The H$_\alpha$/H$_\beta$ Balmer decrement value evolves over time and has its minimum value near the main event peak (see Sect. \ref{sec:balmerdec}). In the early epochs of SN 2016cvk a so called `flash ionisation feature' is also visible at around $\sim$4700~Å, and is discussed in Sect. \ref{sec:flashfeature}. Other notable features detected during the evolution of SN 2016cvk, such as He~{\sc i}, Na~{\sc i}~D, Ca~{\sc ii}, and Fe~{\sc ii} lines are discussed in Sect. \ref{sec:otherfeatures}, and in Sect. \ref{sec:nir} near-infrared features are identified. The late-time spectrum obtained at +405~d and the detected plausible emission feature of nucleosynthesised [O~{\sc i}] are discussed in Sect. \ref{sec:latephase}. Optical time-series of spectra of SN 2016cvk are presented in Fig. \ref{fig:logspectra2016cvk} and comparisons with a selection of other SN 2009ip-like transients are presented in Fig. \ref{fig:all_spectra_allSN}.

\subsection{Evolution of the \protect{H$_\alpha$} and \protect{H$_\beta$} lines}\label{sec:balmer}

\begin{figure*}
\begin{minipage}{0.5\linewidth}
    \includegraphics[trim={0cm 1cm 1cm 0.5cm},clip,width=\linewidth]{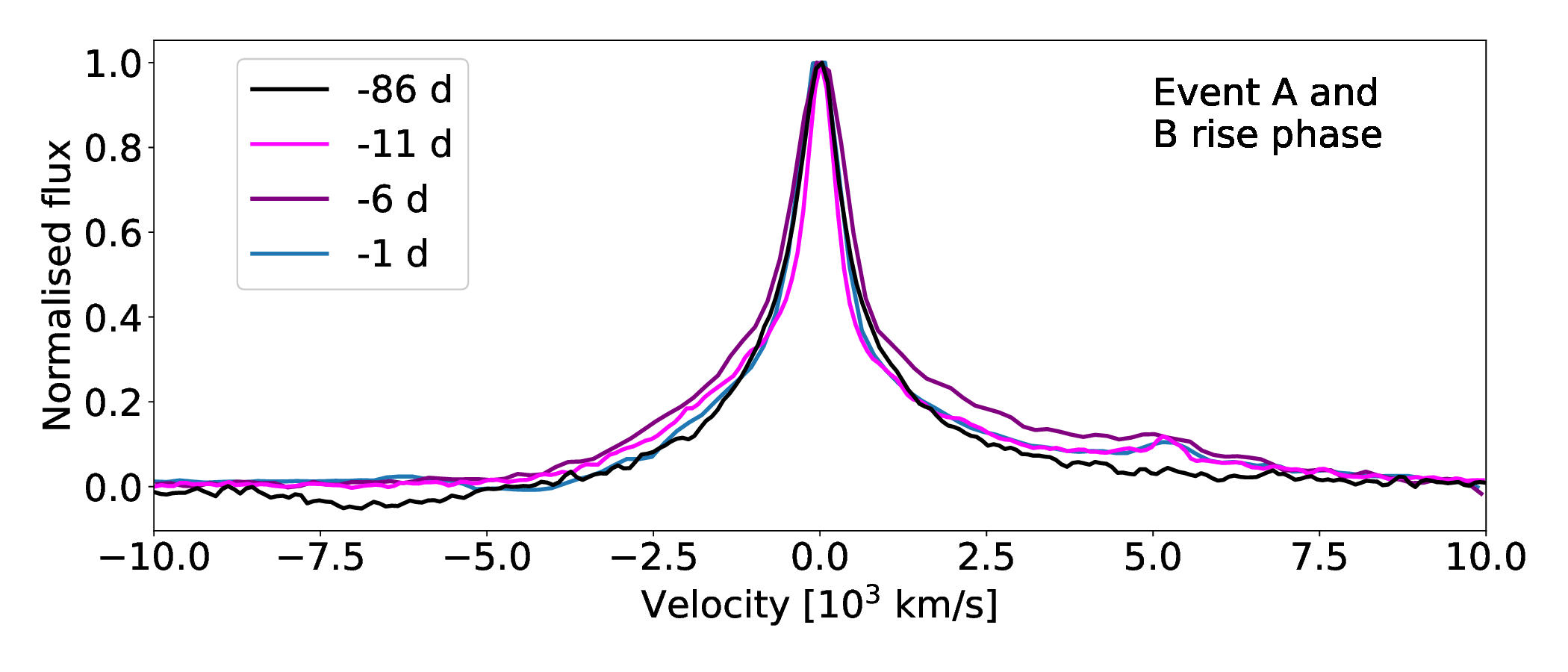}
    
    \includegraphics[trim={0cm 1cm 1cm 0.5cm},clip,width=\linewidth]{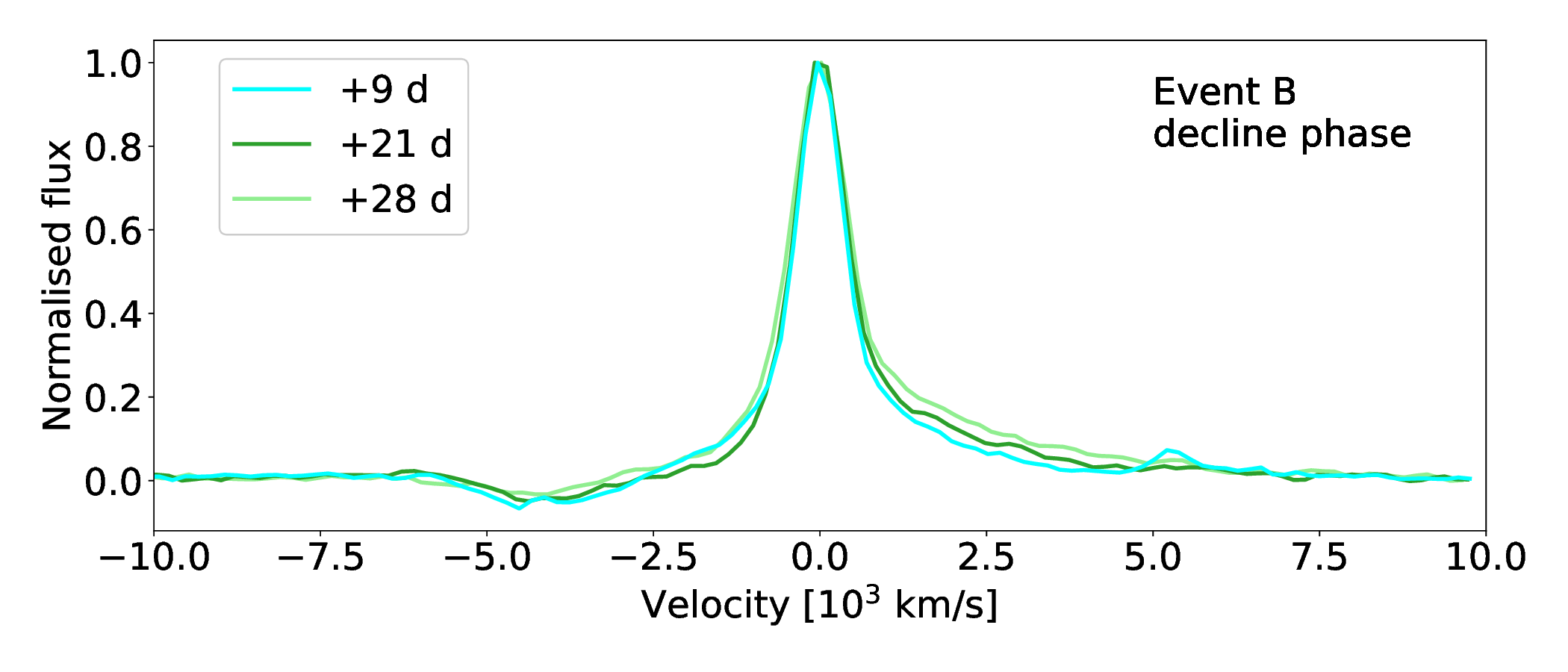}

    \includegraphics[trim={0cm 1cm 1cm 0.5cm},clip,width=\linewidth]{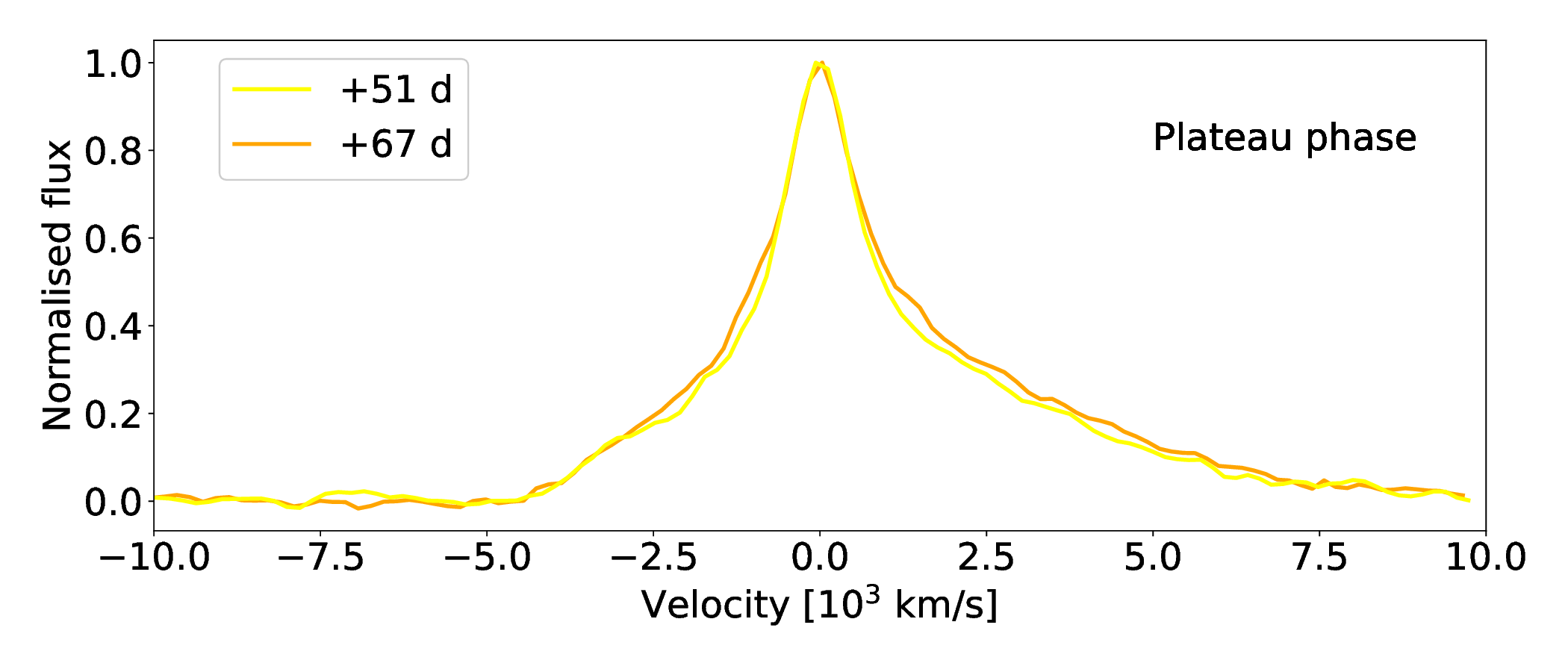}

    \includegraphics[trim={0cm 1cm 1cm 0.5cm},clip,width=\linewidth]{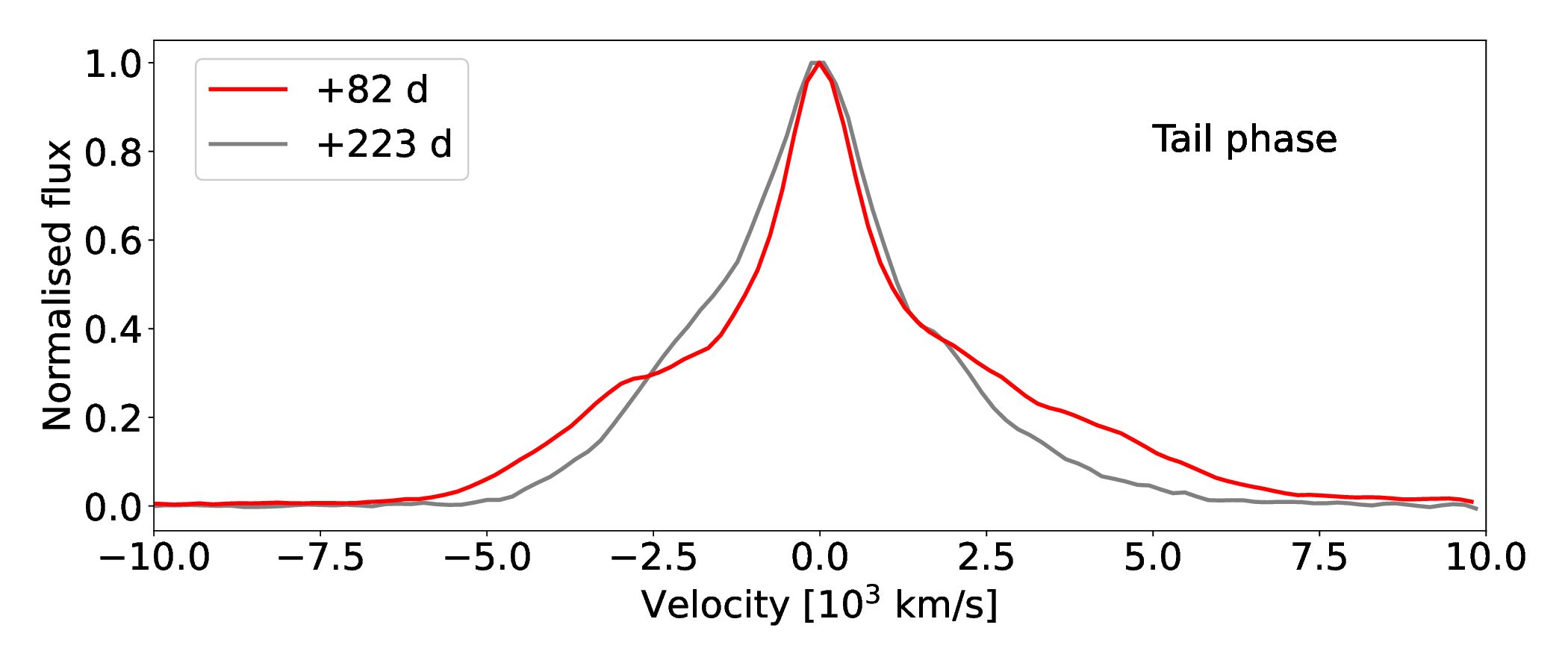}
\end{minipage}\begin{minipage}{0.5\linewidth}
    \includegraphics[trim={0cm 1cm 1cm 0.5cm},clip,width=\linewidth]{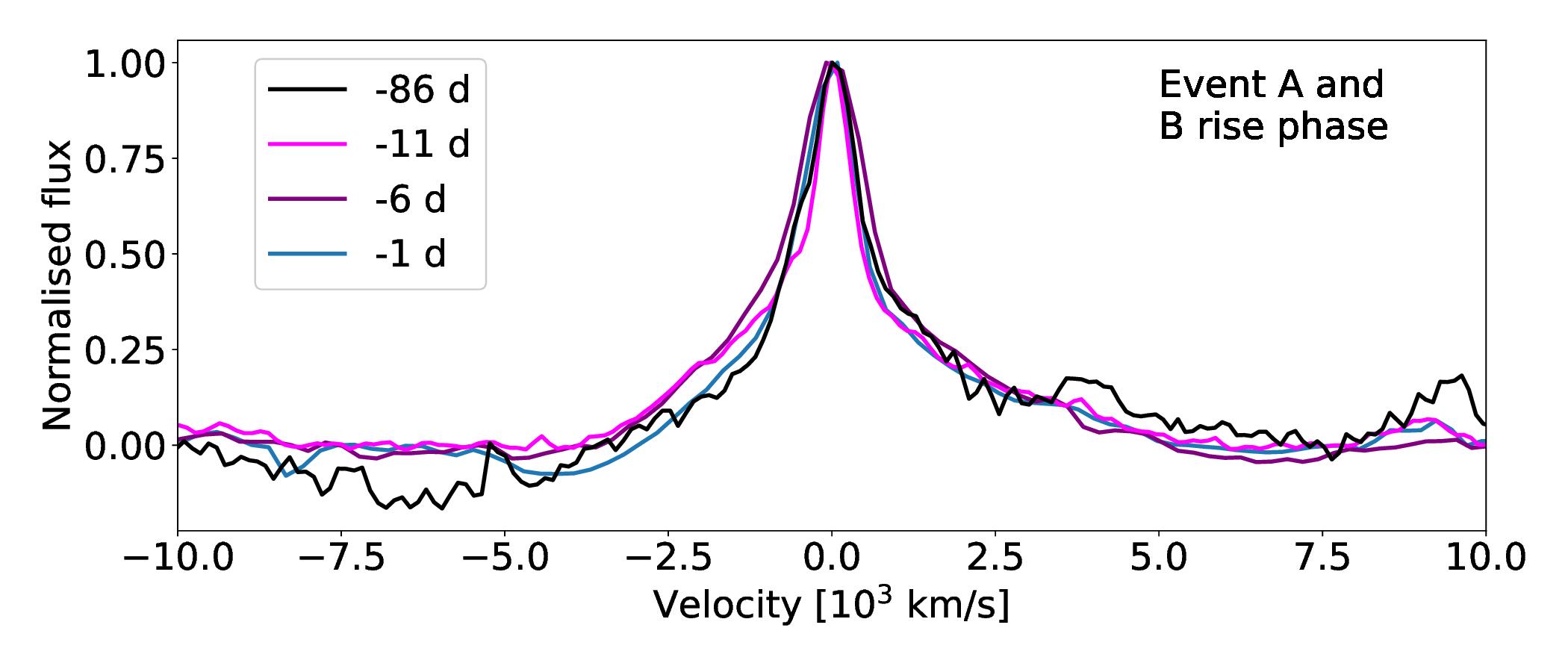}

    \includegraphics[trim={0cm 1cm 1cm 0.5cm},clip,width=\linewidth]{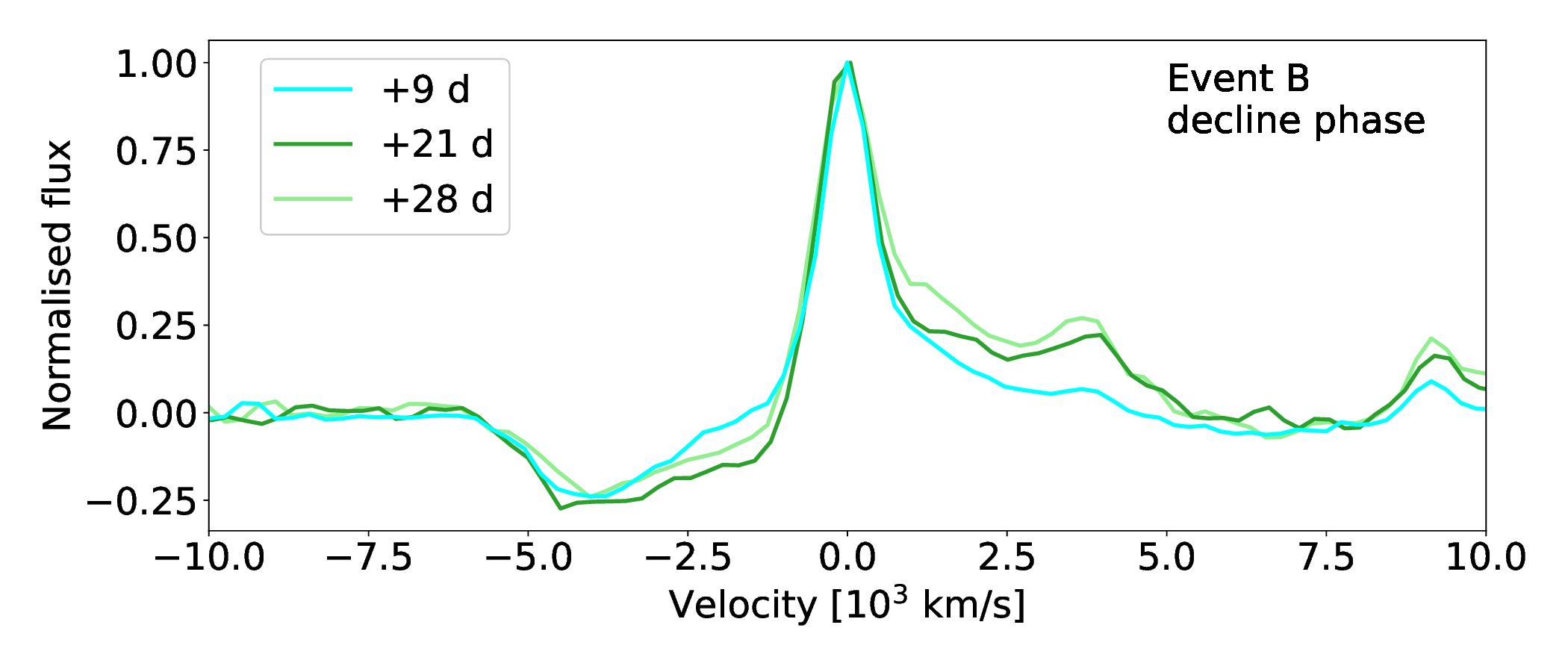}

    \includegraphics[trim={0cm 1cm 1cm 0.5cm},clip,width=\linewidth]{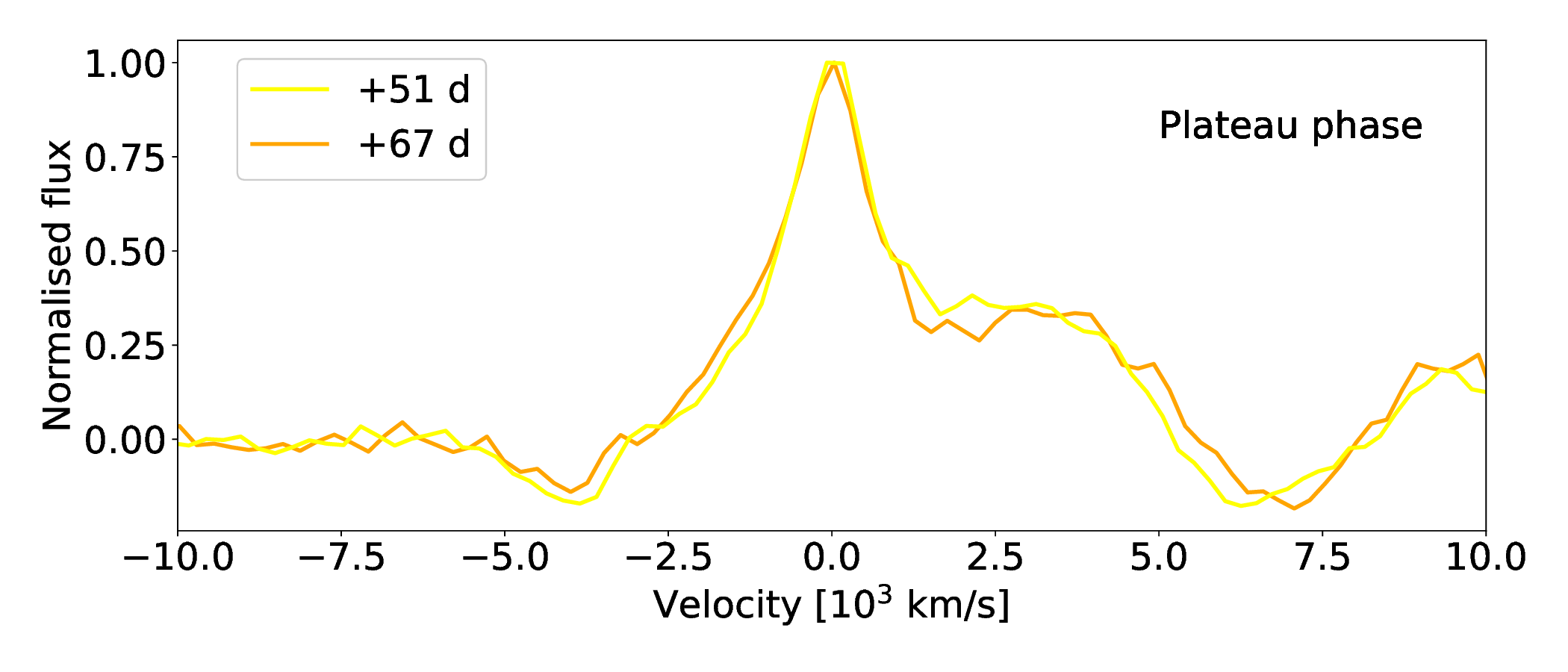}

    \includegraphics[trim={0cm 1cm 1cm 0.5cm},clip,width=\linewidth]{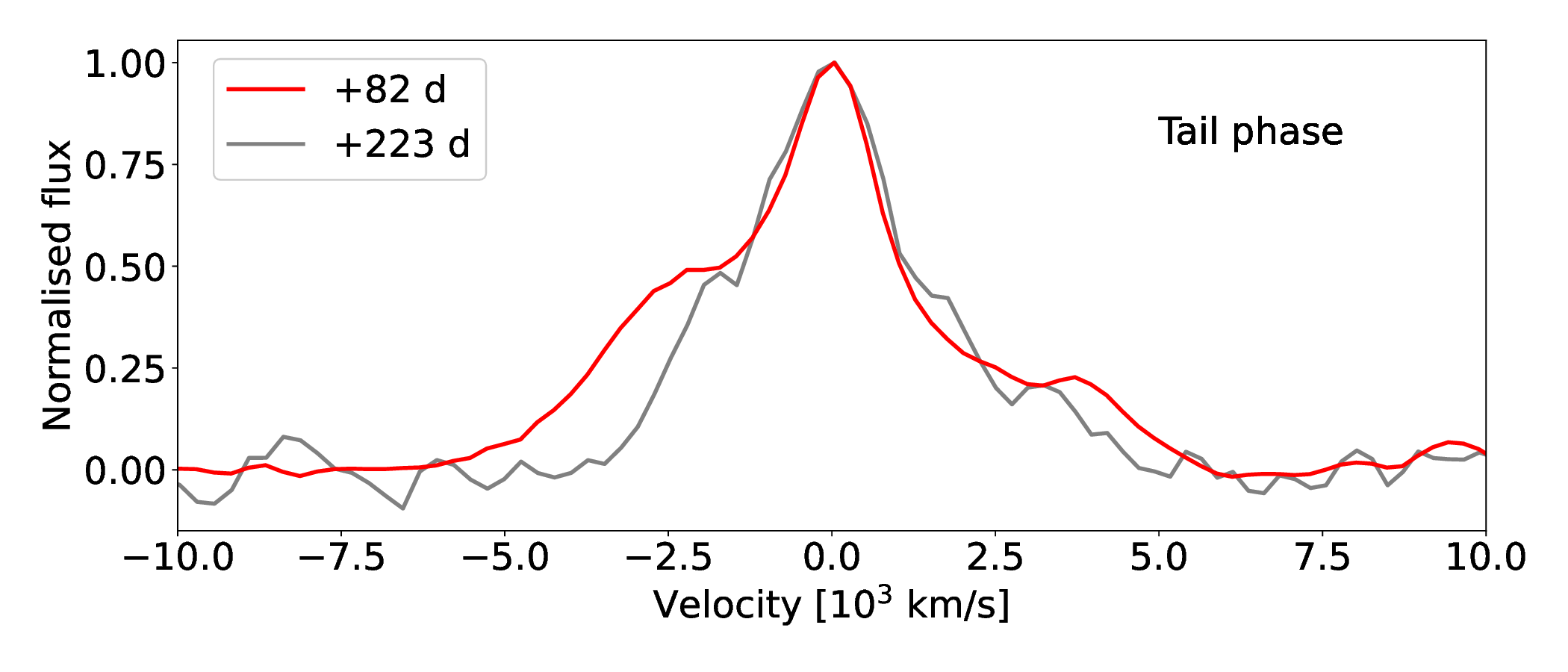}
\end{minipage}
\caption{Evolution of the continuum-subtracted and peak-normalised H$_\alpha$ (left) and H$_\beta$ (right) lines of SN 2016cvk.}
\label{fig:halpha_hbeta}
\end{figure*}

\begin{figure}[hbt!]
\includegraphics[trim={2cm 0cm 3cm 3cm},clip,width=\linewidth]{{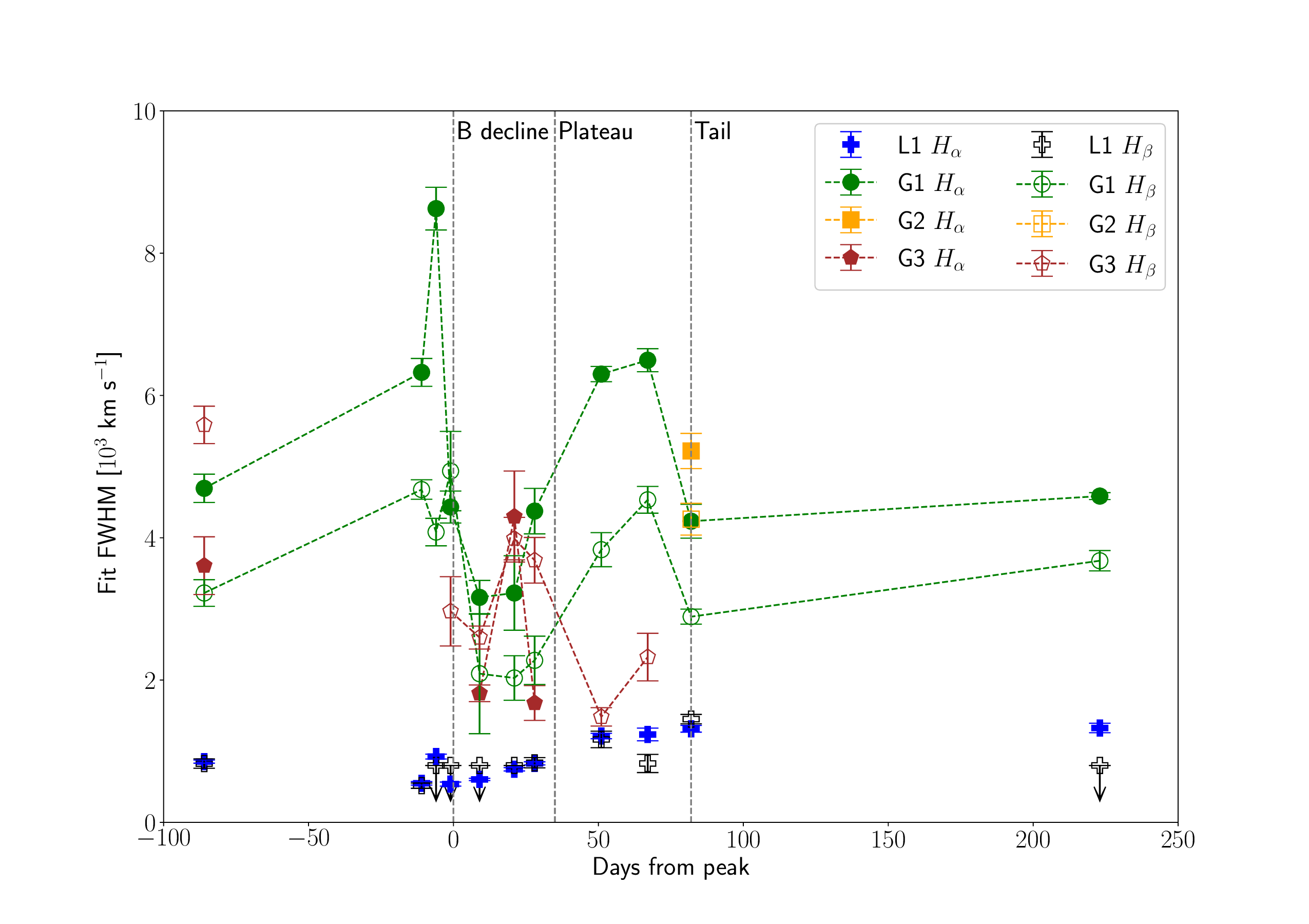}}

\includegraphics[trim={2cm 0cm 3cm 3cm},clip,width=\linewidth]{{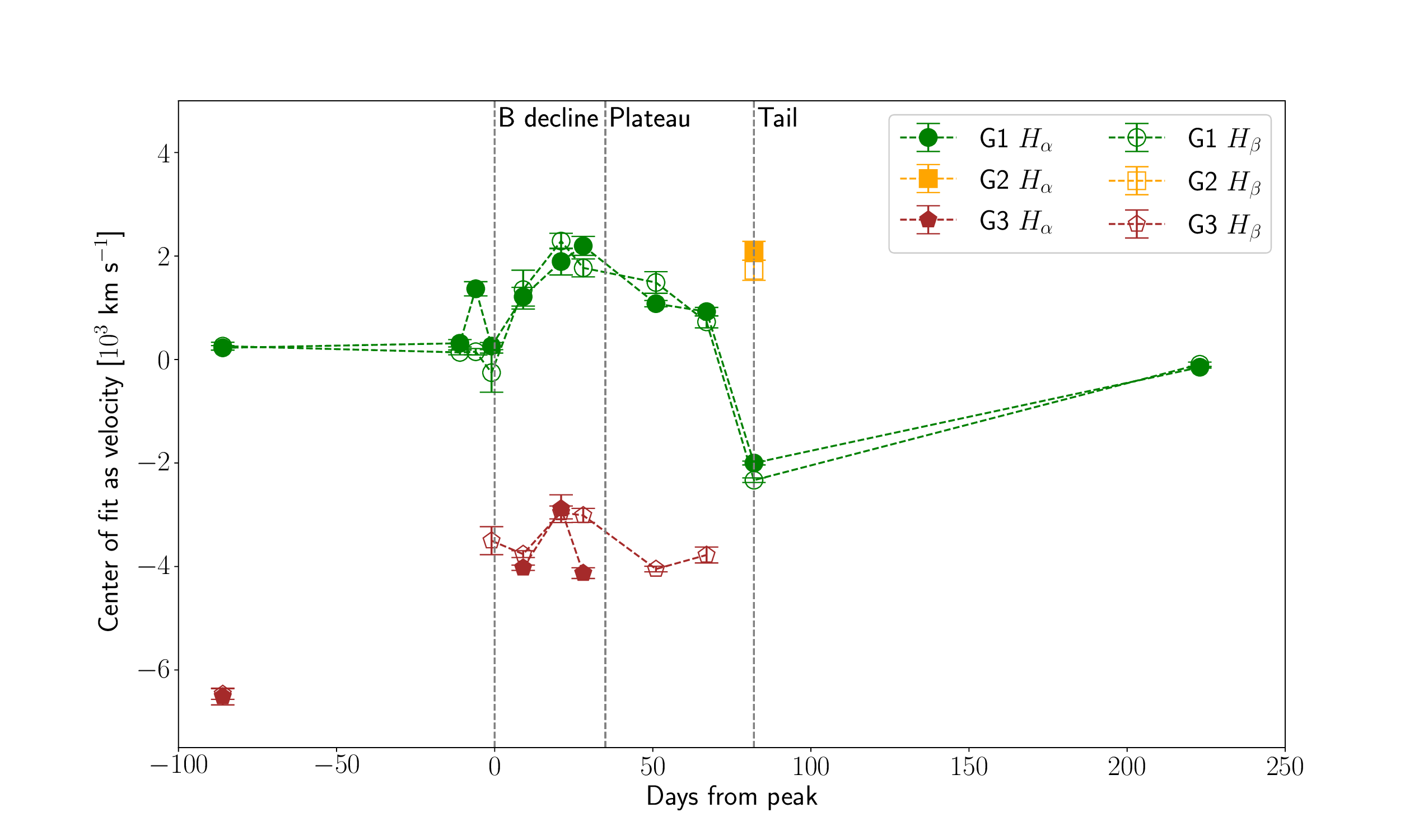}}

\includegraphics[trim={0cm 0cm 0cm 0cm},clip,width=\linewidth]{{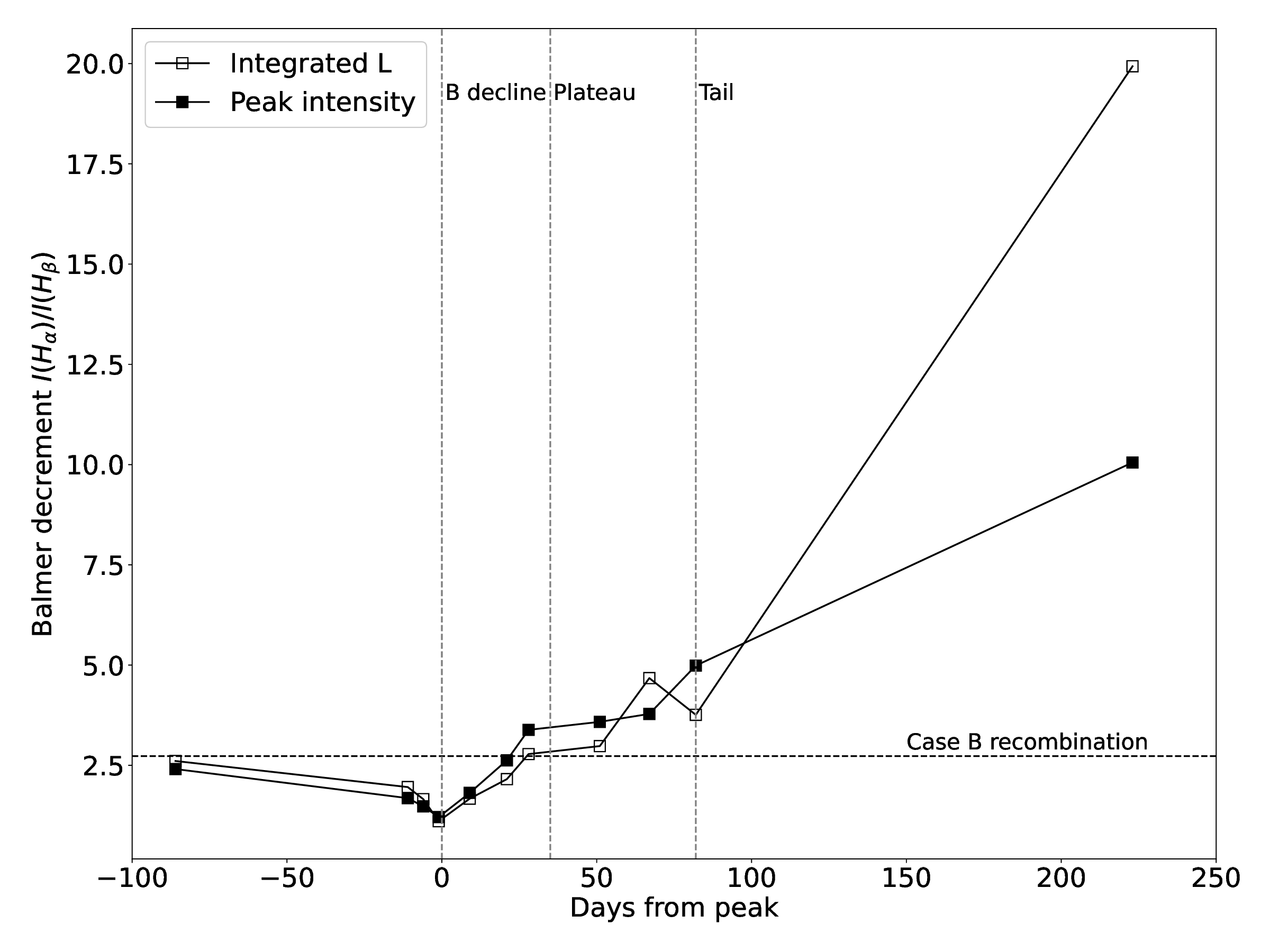}}

\caption{Time evolution of FWHM (top) and centre (middle) of Gaussian and Lorentzian components yielded by the fit to the H$_\alpha$ and H$_\beta$ lines, along with the evolution of the $I(H_\alpha) / I(H_\beta)$ Balmer decrement (bottom) for SN 2016cvk.}
\label{fig:fwhmcomp}
\end{figure}

The H$_\alpha$ line undergoes a complex evolution over time (see Fig. \ref{fig:halpha_hbeta}) and can be fitted with a combination of multiple Gaussian and Lorentzian components. For comparison, similar multi-component fits were made for the H$_\beta$ line to examine if both of these lines show a similar evolution. 
Using a similar approach as \cite{Brennan2022a}, we created a custom Python script that iteratively runs a multi-component fit for the H$_\alpha$ and H$_\beta$ lines, making use of the lmfit package \citep{Newville2014}. The number of used Gaussian and Lorentzian components was fixed manually by testing different combinations for each epoch with the aim to minimize the number of used components while sufficiently fitting the data. 
We corrected the spectra for redshift and extinction with IRAF, and subtracted the local continuum around the H$_\alpha$ and H$_\beta$ lines separately from the extracted 1D spectra. 
The wavelengths of both the H$_\alpha$ and H$_\beta$ spectral line peaks were fixed at $v = 0$~km~s$^{-1}$ for each epoch, and the continuum-subtracted line flux was normalised between 0 and 1. The fits are shown in Fig. \ref{fig:MCMC}. The evolution of the instrumental-resolution corrected full width at half maximum (FWHM) values and the centre of the components of the H$_\alpha$ and H$_\beta$ fits are presented in Fig. \ref{fig:fwhmcomp}, and the results are listed in Tables \ref{tab:MCMCHalpha} and \ref{tab:MCMCHbeta}.

Both the H$_\alpha$ and the H$_\beta$ line have a similar evolution during the different light curve phases (see Fig. \ref{fig:halpha_hbeta}). During event A the line profiles consist of a narrow Lorentzian emission combined with a broad Gaussian emission component. We also see a blueshifted very broad Gaussian absorption component at this time. During the event B rise phase the absorption component disappears, while the broad emission component becomes broader. 
During the event B decline the absorption component briefly re-appears, being more narrow and slightly less blueshifted. At the same time the broad emission component moves redward and becomes more narrow. During the plateau there is no notable evolution in the H$_\alpha$ and the H$_\beta$ lines. At the start of the tail phase the blue absorption component disappears again, and the broad Gaussian emission component splits into two. In the later tail phase there is no clear multi-component structure visible, and the spectral line has become fairly broad. A more detailed description is provided in the following subsections.

\subsubsection{Event A}
During the earliest observed epoch at $-86$~d the emission line profile of H$_\alpha$ is best described using a narrow Lorentzian component (FWHM$_{\text{L1}}=850$~km~s$^{-1}$) combined with a wider Gaussian (FWHM$_{\text{G1}}=4700$~km~s$^{-1}$) emission feature and a similarly wide Gaussian absorption feature (FWHM$_{\text{G3}}=3600$~km~s$^{-1}$) located blueward from the rest wavelength at around $-6500$~km~s$^{-1}$. The centre of the Lorentzian (L1) was fixed at the peak of the spectral line at $v=0$~km~s$^{-1}$, while the peak intensity and width of the component were allowed to vary during the fit. The H$_\beta$ line has similar components, though the broad emission component G1 remains not as broad through the whole observed evolution of SN~2016cvk, and the absorption component G3 is broader and present more frequently in the profiles. From the blue edge of the G3 component, we estimate a maximum velocity of $-12500$~km~s$^{-1}$ for the absorption component of the H$_\beta$ line at $-86$~d. The G3 component of H$_\alpha$ has a slightly lower maximum velocity of $-10000$~km~s$^{-1}$. 

\subsubsection{Event B rise and peak}
During the event B rise phase at $-11$ and $-6$~d the absorption component G3 has disappeared from both H$_\alpha$ and H$_\beta$ lines, and the wide emission component G1 has become broader (FWHM$_\text{G1}=8600$ and 4100 ~km~s$^{-1}$ for H$_\alpha$ and H$_\beta$, respectively, at $-6$ d). The line broadening could be related to electron scattering effects in the spectral line profile. The width of the narrow L1 component reduces slightly in both lines to around FWHM$_\text{L1} = 500$~km~s$^{-1}$, although the FWHM of the component in H$_\beta$ is not resolved from $-6$ to $+21$~d (i.e. FWHM$_\text{L1} < 800$~km~s$^{-1}$).

Near the main event peak at $-1$~d, the Gaussian component G1 of H$_\alpha$ has become more narrow (FWHM$_\text{G1} = 4400$~km~s$^{-1}$), while the same component of H$_\beta$ remains unchanged within errors between $-6$ and $-1$~d.  Furthermore, the G3 absorption component has re-appeared in the line profile of H$_\beta$ and moved more redward to $-3500$~km~s$^{-1}$. The narrow emission features of Fe~{\sc ii} at 4924 and 5018 \AA, and He~{\sc i} at 6678 \AA\ have appeared near the Balmer lines at $-1$~d.

In common with SN 2016cvk, near the light curve maximum, the Balmer lines of other SN 2009ip-like events have been fit with a narrow emission component and accompanying broad wings or a broad emission component. While the absorption components of Balmer lines might not be clearly detected near light curve peak of some of these events such features appear post maximum. Throughout the observed spectral evolution, SN 2016cvk does not show more than one absorption component associated with the Balmer lines; however, many SN 2009ip-like events show two components, for example SN 2009ip with two minima at $-8000$ and $-12 500$~km~s$^{-1}$ \citep{Fraser2013} or SN 2016jbu with a narrow P Cygni profile and a broad absorption at $-3200$~km~s$^{-1}$ \citep{Brennan2022a}. 

\subsubsection{Event B decline and plateau phase}
At epoch $+9$~d the broad emission component G1 of both H$_\alpha$ and H$_\beta$ has become narrower (FWHM$_\text{G1} = 3200$ and 2100~km~s$^{-1}$, respectively), and the G3 component has re-appeared to H$_\alpha$ as well. The centre of the G1 component of H$_\alpha$ and H$_\beta$ features has moved redwards at $+21$ and $+28$~d, while the G3 component has become broader (FWHM$_\text{G3} = 4300$ and 4000~km~s$^{-1}$, respectively). 
From the blue edge of the G3 component of H$_\alpha$ we estimate $-7500$~km~s$^{-1}$ as the maximum velocity for the absorption component during the event B. 

Similar to SN 2016cvk, the broad Gaussian emission component becomes narrower in other SN 2009ip-like events during the event B decline phase. However, a second broad emission component appeared for example in the H$_\alpha$ profile of SN 2009ip, SN 2015bh, and SN 2016jbu \citep{Fraser2013,EliasRosa2016,Brennan2022a} with the latter two evolving to show a clearly double-peaked line profile, accompanied by the disappearance of the second higher velocity absorption component; this evolution is not seen in SN 2016cvk.

The plateau phase of SN 2016cvk begins at around $+35$~d from main event peak. In the spectra obtained during this phase, at $+51$~d and onwards the absorption feature G3 of H$_\alpha$ is not detected. The FWHM of the narrow Lorentzian and broad Gaussian emission components have increased to 1100 and 6500~km~s$^{-1}$, respectively, with the former also resolved. At $+67$~d no major evolution is seen. In the H$_\beta$ line similar trends are seen, accompanied by a weakening of the absorption component G3.

\subsubsection{Tail phase}
From $+82$~d onward the light curve of SN 2016cvk is in its tail phase and the absorption component G3 has disappeared also from the H$_\beta$ line. The H$_\alpha$ line fit of the Lorentzian component yields FWHM$_\text{L1}=$ 1200~km~s$^{-1}$ at $+82$~d, and the L1 component shows no major tail-phase evolution as the $+223$~d value is consistent with this within errors. At $+82$~d the fit requires a second relatively broad Gaussian emission feature G2 in addition to the G1 component for both H$_\alpha$ (FWHM$_\text{G1}$ = 4200~km~s$^{-1}$ and FWHM$_\text{G2}$ = 5200~km~s$^{-1}$) and H$_\beta$ (FWHM$_\text{G1}$ = 2900~km~s$^{-1}$ and FWHM$_\text{G2}$ = 4300~km~s$^{-1}$) lines. Later in the evolution at $+223$~d the G2 component is not required by the fit and the G1 component of both Balmer lines has broadened to 4500 and 3700~km~s$^{-1}$ for H$_\alpha$ and H$_\beta$, respectively.

The H$_\alpha$ late-time profiles of SN 2016cvk are similar to those of SN 2009ip, LSQ13zm, and SN 2016bdu, which showed a combination of narrow and intermediate-width or broad emission components \citep{Fraser2013,Tartaglia2016,Pastorello2018}. However, the H$_\alpha$ and H$_\beta$ lines of SN 2016cvk had a second Gaussian component only at $+82$~d, and the lines never evolved to a double-peaked profile similar to those of SN 2015bh and SN 2016jbu. Additionally, only the $+82$~d H$_\beta$ line of SN 2016cvk had a clear blue shoulder in emission whereas SN 2009ip and SN 2016bdu persistently had this characteristic in Balmer lines.

\subsection{Evolution of the Balmer decrement}\label{sec:balmerdec}

We use the same approach as \cite{Levesque2014} and \cite{Thone2017}, and define the Balmer decrement as the ratio of maximum intensities of the two spectral lines. This approach is better suited for consistently comparing the low-velocity CSM regions that dominate the line centres, compared to the integrated fluxes that would be strongly affected by the broad, high-velocity components.  
During the rise to the event B peak the Balmer decrement of SN 2009ip-like targets decreases, and is close to unity near maximum brightness. For SN 2009ip the Balmer decrement is between 1.1 and 1.3 near the event B peak \citep{Levesque2014}, and in the case of SN 2015bh the value remains between 1 and 2 \citep{Thone2017}. For comparison, the classical Case B re-combination value is 2.87 for photo-ionized nebulae assuming $T = 10000 K$ \citep{Osterbrock2006}. 
\cite{Levesque2014} interpreted the low Balmer decrement of SN 2009ip as a likely result of high-density CSM concentrated in a disc around the progenitor. 
In the case of SN 2016cvk the Balmer decrement value remained similarly between 1.2 and 1.8 near phase B peak (see Table \ref{tab:bb} and Fig. \ref{fig:fwhmcomp}), consistent with collisional de-excitation process at high electron densities \citep{Drake1980}. 

After the light curve maximum, the Balmer decrement value for SN 2016cvk increases, potentially related to a decrease in electron density, and varied within a range of 3.4 to 3.8 during the plateau phase. The late-time +223~d tail-phase spectrum of SN 2016cvk has a Balmer decrement of 10.1, which is in line with other SN 2009ip-like transients. \cite{Fraser2015} measured a Balmer decrement of $\sim$12 from +200~d onward for SN 2009ip. SN 2016bdu has a similar large late-time $H_\alpha/H_\beta$ line ratio of $\sim$11 at roughly 200 to 300~d, which \cite{Pastorello2018} concluded to be common in Type IIn SNe and often viewed as indicative of collisional excitation.

For comparison, we also calculated the Balmer decrement using the integrated fluxes of the narrow Lorentzian emission components. During event A, event B and plateau phase the evolution is very similar using this method. In the late tail phase at +223~d the Balmer decrement rises to an even higher value of $\sim$20 using the integrated Lorentzian components (see Fig. \ref{fig:fwhmcomp}).

\subsection{Flash feature}
\label{sec:flashfeature}

\begin{figure}
\includegraphics[trim={1cm 0.5cm 3cm 2cm},clip,width=\linewidth]{{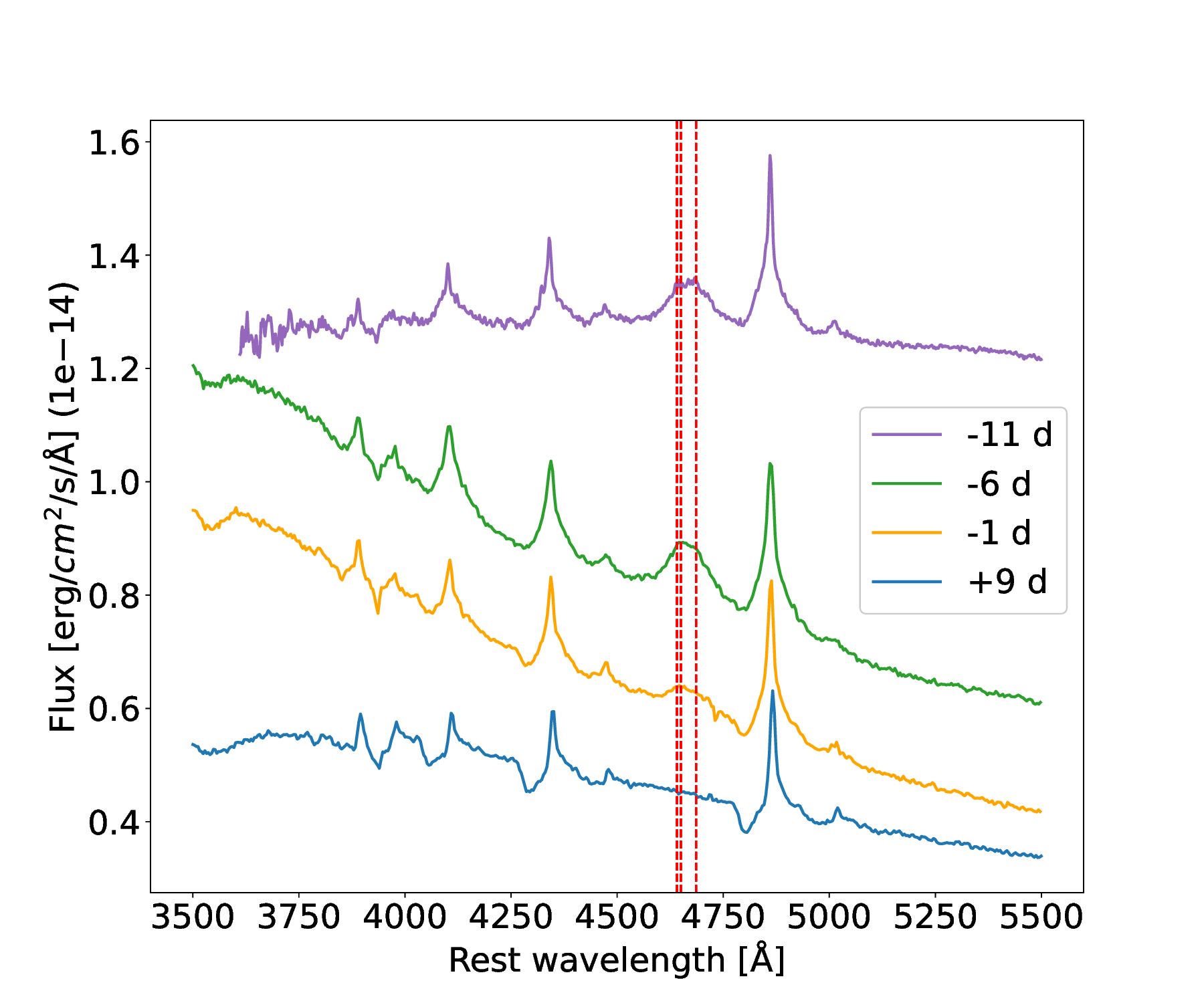}}
\caption{Partial spectra of SN 2016cvk in four earliest B phase epochs. The blended `flash ionisation' feature is shown at $\protect\sim$4700~Å. The locations of the N~{\sc iii} line at 4641 Å, C~{\sc iii} at 4650 Å, and He~{\sc ii} at 4686 Å are marked with red dashed lines.}
\label{fig:flashfeature}
\end{figure}

\begin{figure}
\includegraphics[trim={0cm 0cm 0cm 0cm},clip,width=\linewidth]{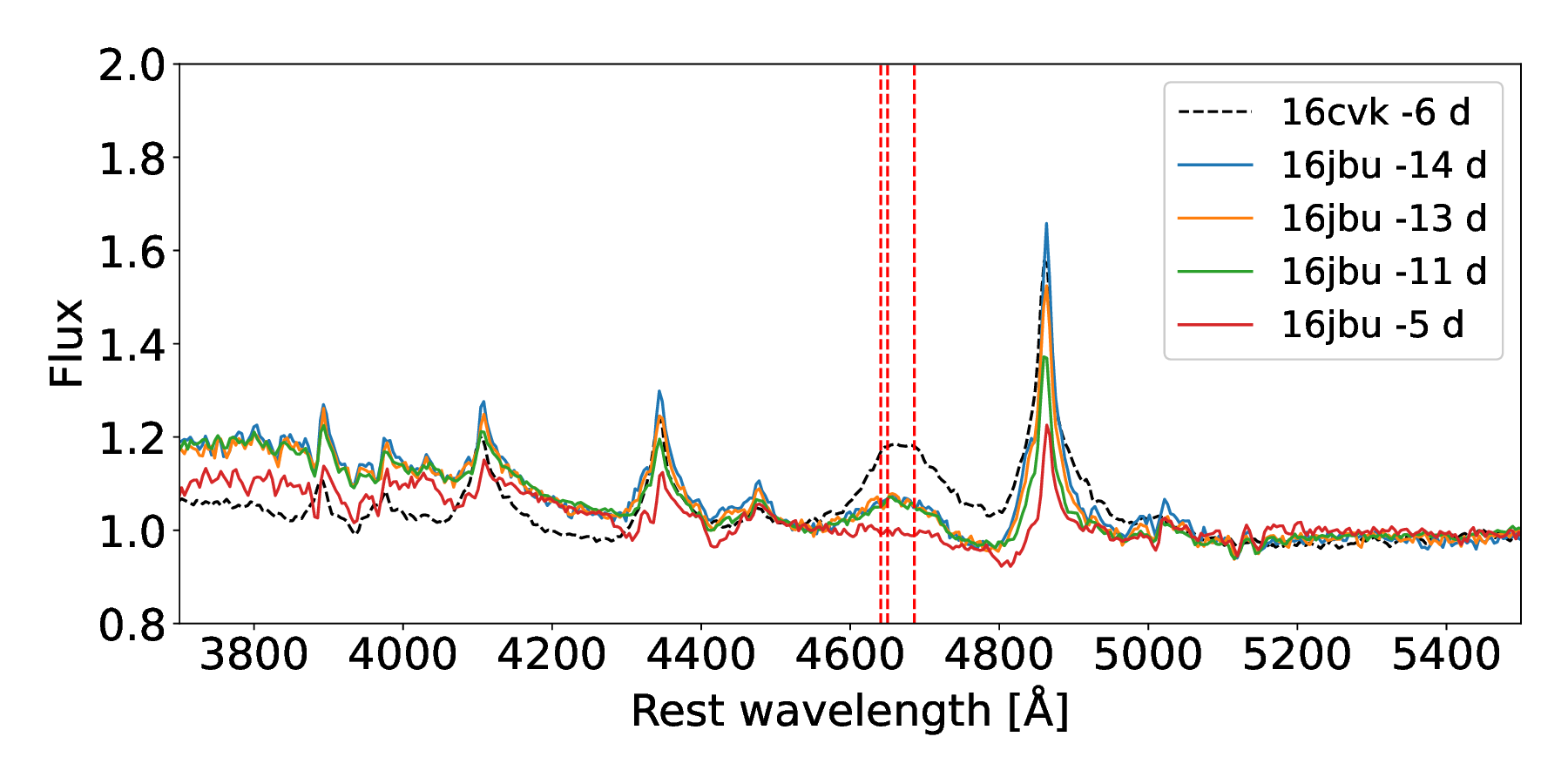}

\includegraphics[trim={0cm 0cm 0cm 0cm},clip,width=\linewidth]{{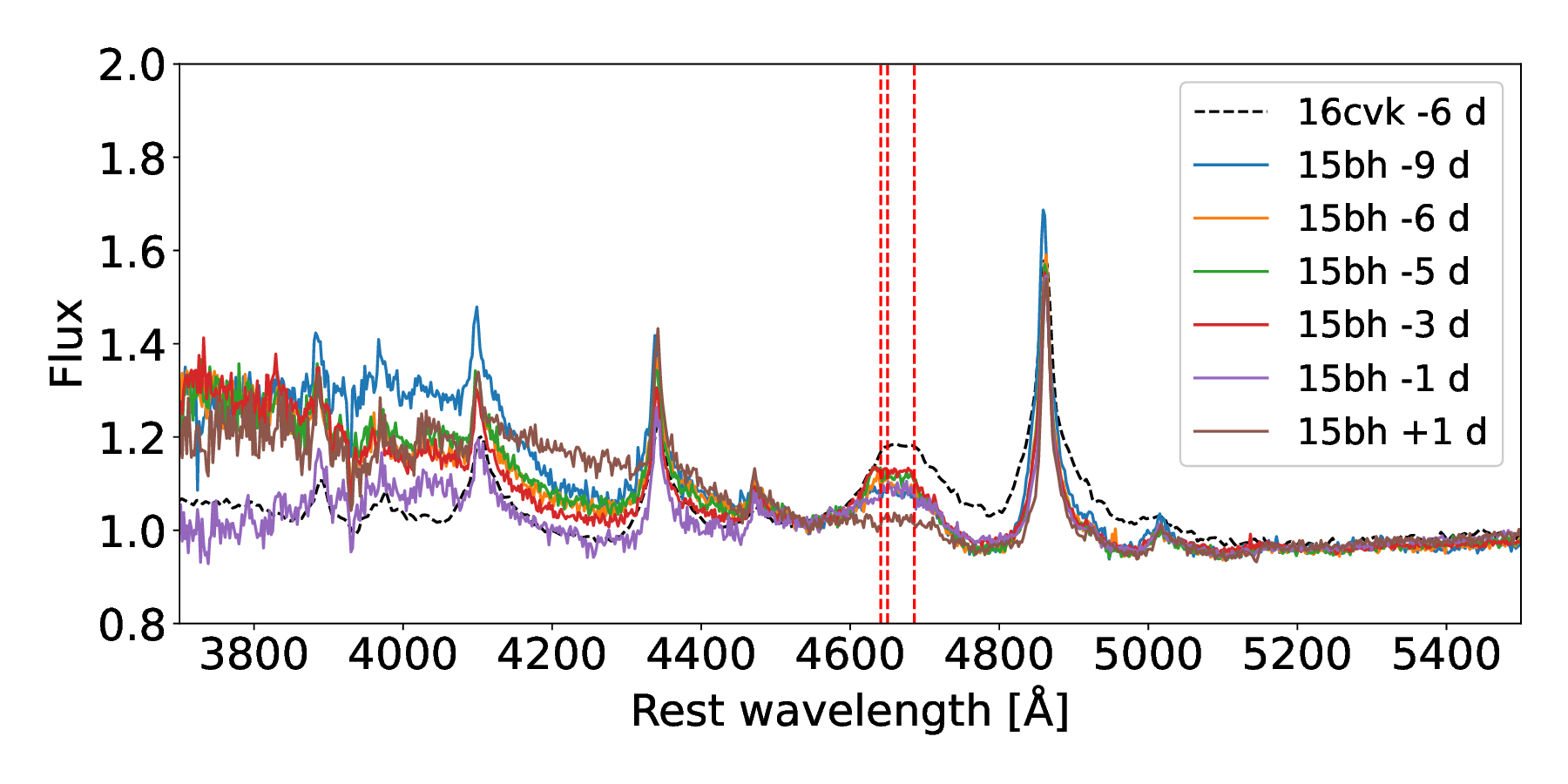}}

\includegraphics[trim={0cm 0cm 0cm 0cm},clip,width=\linewidth]{{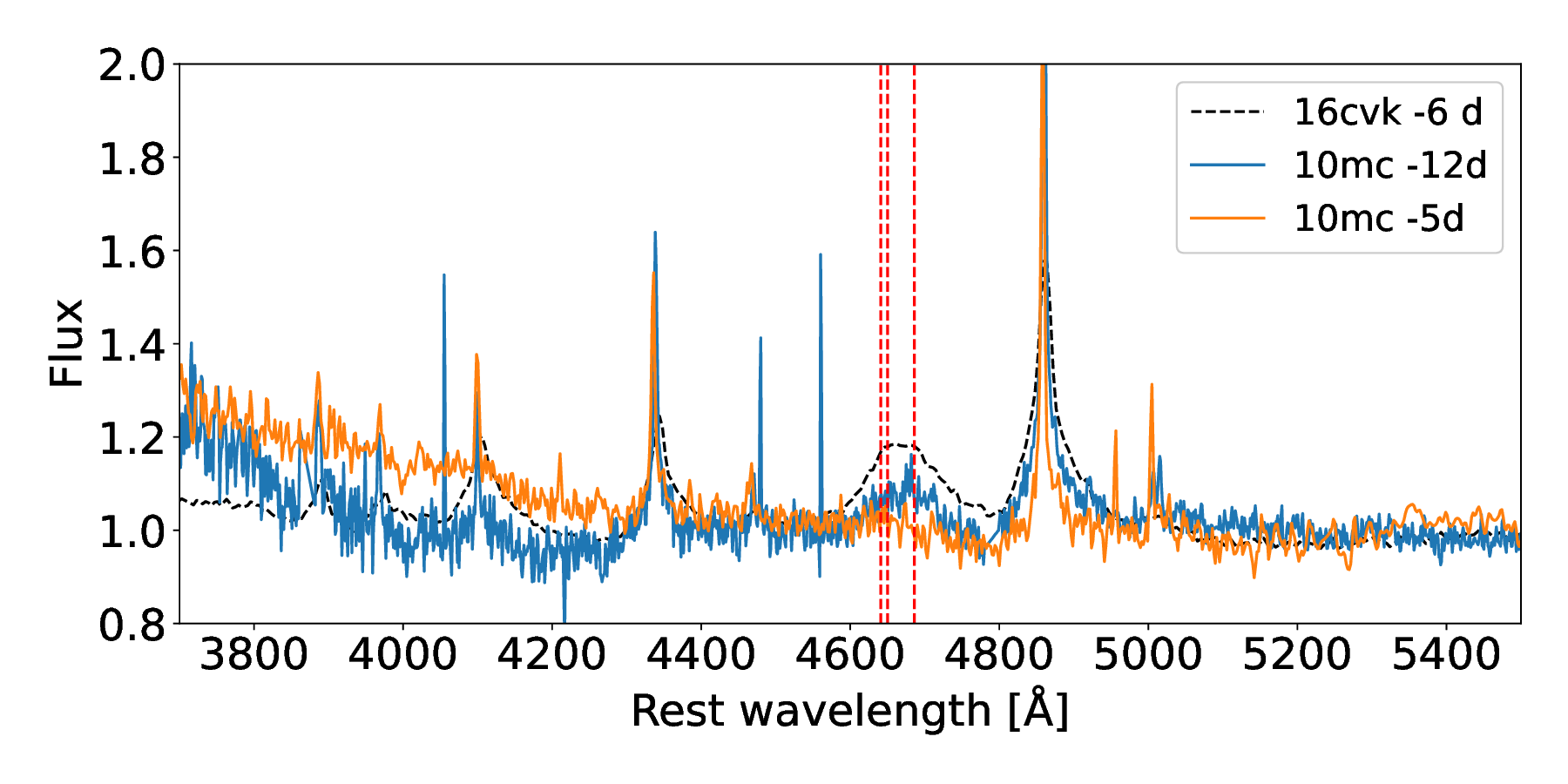}}

\caption{Partial spectra of SN 2016jbu \citep{Brennan2022a}, SN 2015bh \citep{Thone2017}, and SN 2010mc \citep{Ofek2013} during event B rise phase. A flash ionisation feature is seen around $4700$ Å in the $-14$, $-13$, and $-11$~d spectra of SN 2016jbu, in the $-9$, $-6$, $-5$, $-3$, and $-1$~d spectra of SN 2015bh, and in the $-12$ d spectrum of SN 2010mc. The $-6$~d spectrum of SN 2016cvk is included for comparison. The locations of the N~{\sc iii}, C~{\sc iii}, and He~{\sc ii} lines at 4641, 4650, and 4686~Å, respectively, are indicated with red dashed lines. Fluxes have been normalised by dividing the spectra with a continuum estimated based on the line-free regions between 4500 to 9000~Å.}
\label{fig:ff16jbu15bh}
\end{figure}

H-rich SNe with dense CSM often show so called "flash features" in their early spectra ($ \lesssim 10$~d from explosion). These features arise from recombination of the CSM that was ionized by either the shock breakout UV flash of the SN \citep[`flash ionisation', e.g.][]{Gal-Yam2014} or by radiation originating from ejecta-CSM interaction \citep['shock ionisation', e.g.][]{Jacobson2022}; however, to maintain the flash features for several days as in the case of SN 2016cvk, the latter ongoing ejecta-CSM interaction process is required due to the relatively short timescale for an ionized H-rich CSM to recombine. Furthermore, the optically thick CSM has to be sufficiently extended for long-lived flash features \citep{Dessart2023}. The profiles of these features are narrow emission lines with Lorentzian wings \citep{Khazov2016}, and they are present in the first days or weeks in the spectra of at least $30\%$ of Type II SNe \citep{Bruch2023}. Flash features are relatively common in Type IIn SNe, including SN 1998S \citep{Fassia2001}, which is often dubbed as a prototypical SN IIn. The large sample of Type II SNe in \cite{Khazov2016} included seven IIn SNe, two of which showed signs of flash features in their spectra. This ratio is similar to the rest of their sample of Type II SNe. Similarly, \cite{Bruch2023} found one IIn supernova with flash features in a sample of 11 Type IIn SNe. Flash features are associated with highly ionized states of e.g. He, C, N, and O, which return to lower energy levels through emission lines such as the N~{\sc iii} 4641~Å and He~{\sc ii} 4686~Å lines.

In SN 2016cvk such a flash ionisation feature is present at around $\sim$4700~Å in the $-11$, $-6$, and $-1$~d spectra, see Fig. \ref{fig:flashfeature}. The flash feature remains prominent in the first two spectra with a very similar flux ratio between the $\sim$4700~Å feature and the H$_\beta$ line in the $-11$ and $-6$ d spectra. This blended feature likely includes the He~{\sc ii} line at 4686~Å and probably emission from N~{\sc iii} (4634~Å and 4641~Å) and C~{\sc iii} (4648~Å and 4650~Å) similar to that found in early epochs of for example the Type II SN 2023ixf \citep[e.g. ][]{Bostroem2023, Hiramatsu2023, Jacobson2023, Smith2023} or Type II SN 2024ggi \citep[e.g. ][]{Jacobson2024, Pessi2024, Shrestha2024}, which are uniquely well documented due to their close proximity. Higher ionisation state lines of C~{\sc iv} at 5801 and 5812 Å or N~{\sc iv} at 7109 and 7123 Å are not detected in our spectra of SN 2016cvk. Furthermore, the He~{\sc ii} line at 5412 Å is not detected; however, we tentatively identify a possible contribution from a weak C~{\sc iii} feature at 5696 Å in the $-$11 and $-$6~d spectra. The equivalent width (EW) of the blended feature at $\sim$4700~Å is $-$18 Å in the earliest spectrum at $-11$ and $-6$~d, after which it evolves to $-$6 Å at $-1$~d. For comparison, the EW of the  H$_\alpha$ line evolves from $-$130 Å in the $-6$~d spectrum to $-$72 Å in the $-1$~d spectrum, and the EW of H$_\beta$ develops from $-$42 to $-$26 Å at the same time.

Flash features have previously been reported in SN 2009ip-like events only in SN 2015bh \citep{Thone2017}, and in SN 2010mc \citep{Khazov2016}. \cite{Bruch2023} reported that for normal Type II SNe typical flash events last for approximately 5 days from the estimated explosion date, while some of these events have flash features present in their spectra for around 10 days. To get an estimate of the duration of flash features in SN 2016cvk spectra, we fit an exponential function $f(t) = a (t-t_\text{exp})^n$ to the event B rise phase flux \citep{Bruch2023}. This fit is also presented in absolute magnitudes in Fig. \ref{fig:LCphases16cvk}. From this fit we determined that $t_{\text{exp}}$, which we adopt as the onset of event B, occurred roughly 11~d before the event B peak. The 4700 Å emission feature is detected in the $-1$~d spectrum; however, it has disappeared by the $+9$ day epoch. Therefore, the flash feature is  visible for around $16 \pm 5$ d.

For comparison, we checked all the available early ($\leq$ +10 d) event B spectra of SN 2009ip \citep{Fraser2013, Fraser2015, Graham2014, Margutti2014} and SN 2016jbu \citep{Kilpatrick2018, Brennan2022b}. We found no traces of the aforementioned He~{\sc ii}, N~{\sc iii}, C~{\sc iii}, C~{\sc iv}, or N~{\sc iv} emission lines in the 11 spectra of SN 2009ip obtained from $-19$~d onwards. Contrary to this, we identified a blended flash ionisation feature of He~{\sc ii}, C~{\sc iii}, and N~{\sc iii} in the early spectra of SN 2016jbu, similar to the feature seen in SN 2016cvk at $\sim$4700 Å. \cite{Kilpatrick2018} interpreted that high-ionisation lines were entirely absent in the spectra of SN 2016jbu; however, we detected a faint flash feature bump in the $-14$, $-13$, and $-11$~d spectra, which has disappeared by $-5$~d (see Fig. \ref{fig:ff16jbu15bh}). By fitting an exponential function to the event B rise phase of SN 2016jbu, we estimated an onset time of $t_\text{exp} = -19$~d before the phase B peak. Therefore, the flash feature is visible for approximately $12 \pm 3$~d, which is similar in duration to that of SN 2016cvk. 
In SN 2015bh the blended 4700 Å flash feature is visible in the $-9$, $-6$, $-5$, $-3$, and $-1$ d spectra (see Fig. \ref{fig:ff16jbu15bh}). From an exponential fit to the rise phase of the transient we get $-13.7$~d as the onset time; therefore, the flash feature is visible for $14 \pm 1$~d. 
\cite{Khazov2016} also detected a flash ionisation feature of He~{\sc ii} with an EW of $-2.93 \pm 0.27$~Å in the $-12$ d spectrum of SN 2010mc. This feature is no longer visible in the  $-5$ d spectrum (see Fig. \ref{fig:ff16jbu15bh}).

\begin{figure*}
\includegraphics[trim={1cm 0cm 1.5cm 0cm},clip,width=\linewidth]{{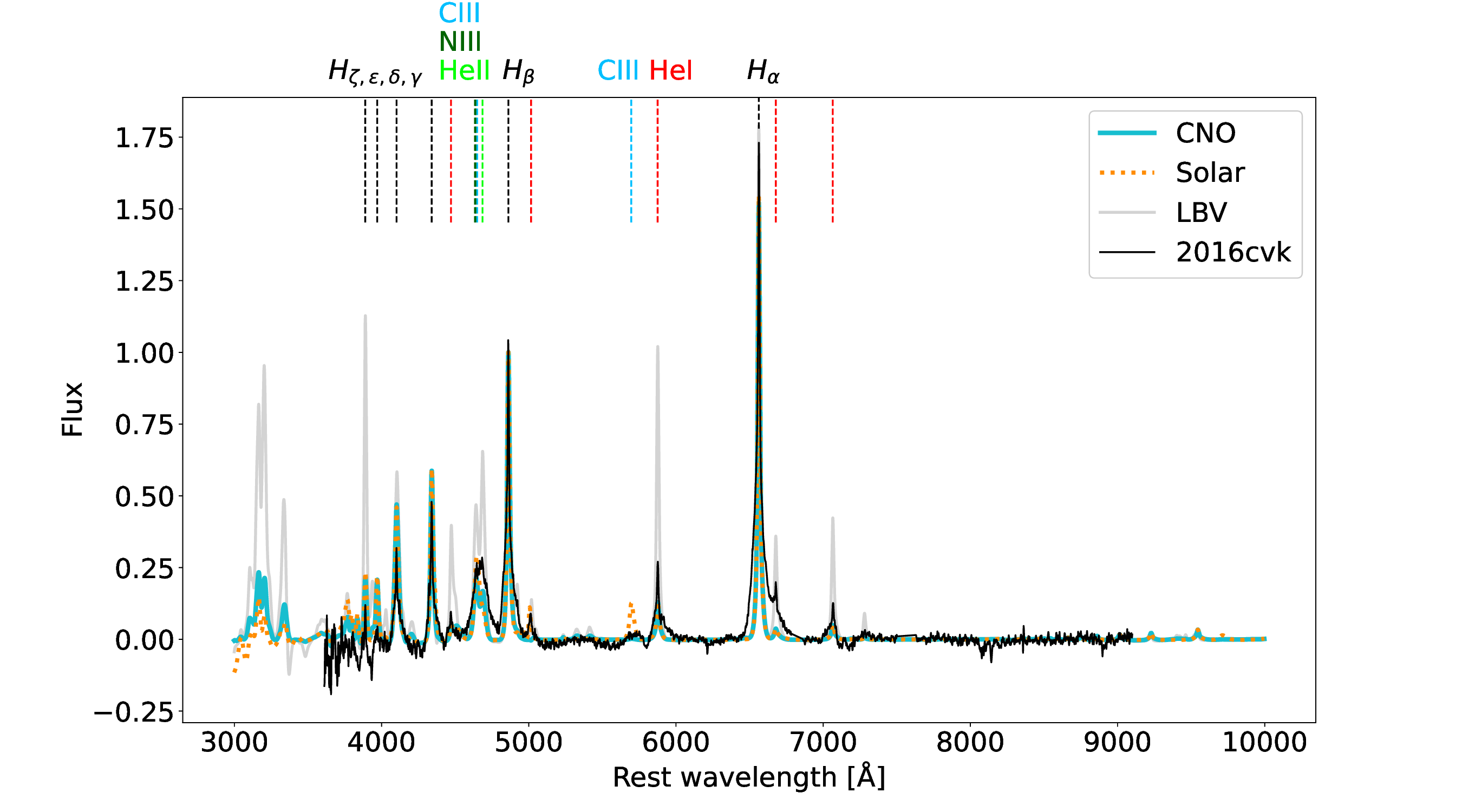}}
\caption{The continuum-subtracted and H$_\beta$-normalised $-11$~d spectrum of SN 2016cvk (black) compared to the early SN models by \cite{Boian2019} of lower-mass RSG with solar abundance (orange), LBV star (light gray) and a CNO-processed high-mass RSG, YSG, or BSG star (light blue), see text for details.}
\label{fig:prog}
\end{figure*}

To constrain the possible progenitor star type of SN 2016cvk, we compared the first $-11$~d spectrum with flash features to the models of \cite{Boian2019} calculated for different types of progenitors. However, we do not aim to match the temperatures and we compared the continuum subtracted spectra with matched spectral resolution, normalised to the H$_\beta$ line maxima to have only a rough indication of the progenitor star abundance. We found the best fit with a low-mass ($8-15 \, M_\odot$) red supergiant (RSG) model with solar metallicity (which also assumed $\dot{M} = 3 \times 10^{-3} M_\odot$ and $L = 0.39 \times 10^9 L_\odot$), see Fig. \ref{fig:prog}. However, another possible model fit to the early spectrum of SN 2016cvk could be a high-mass RSG, or a yellow or blue supergiant (YSG, BSG, respectively) with CNO-processed abundances ($15-30 \, M_\odot$). The model used the same parameters for mass-loss rate and luminosity as above. For comparison, \cite{Brennan2022b} characterized the progenitor of SN 2016jbu to be a $\sim 22$ to $25 \, M_\odot$ yellow hypergiant. Interestingly, helium-rich model consistent with an LBV was not a good fit for SN 2016cvk, whereas an LBV like precursor has been proposed for SN 2009ip based on the progenitor detection \citep{Mauerhan2013}.

\subsection{Evolution of the He, Na, Ca, and Fe features}
\label{sec:otherfeatures}

Emission lines of He~{\sc i} at 5876 and 7065 Å are clearly visible throughout the observed evolution of SN 2016cvk from $-11$ to $+405$~d (see Fig. \ref{fig:logspectra2016cvk}). Around light curve maximum from $-6$ to $+9$~d the He~{\sc i} lines at 4471, 5015, and 6678 Å are also seen. However, the 5015 Å feature blends with Fe~{\sc ii} 5018 Å, and the 6678 Å line is heavily blended with the strong H$_\alpha$ line (see Fig. \ref{fig:MCMC}). The 5876 Å line is potentially blended with the Na~{\sc i}~D feature at 5890 and 5895 Å throughout the spectral series. The characteristics of the He~{\sc i} features are similar to those of other SN 2009ip-like SNe with visible helium emission from the beginning of the phase B to the very late tail phase at +400~d \citep{Fraser2013, Brennan2022a}, which has been interpreted as a sign of ongoing interaction with the CSM at late times. The He~{\sc i} lines at 5876 and 7065 Å are clearly visible in the spectra of LSQ13zm, SN 2010mc, SN 2015bh, SN 2016jbu, SN 2009ip, and SN 2016cvk in Fig. \ref{fig:all_spectra_allSN} in all four comparison epochs. P Cygni absorption features are somewhat visible for all He~{\sc i} emission lines, but most clearly detected for the 5876 Å line. The location of the absorption minimum of the feature remains roughly constant around $-3600$~km~s$^{-1}$ from $-11$ to +67~d. During the tail phase from +82~d onward the absorption component disappears from the He~{\sc i} line, similar to the evolution seen in the H$_\beta$ line at this time. The two last spectra at +223~d and +405~d are quite noisy, and the location of the absorption feature is no longer clearly defined. The P Cygni velocities are similar to those of the Balmer lines.

The Ca~{\sc ii} NIR triplet at 8498, 8542, and 8662 Å is clearly detected during the light curve plateau of SN 2016cvk from +51~d onwards with P Cygni absorptions; however, a weak blended bump associated with the Ca~{\sc ii} NIR triplet could tentatively be present also at +21 and +28~d. This is similar to that of the reported observations of SN 2009ip-like events with the Ca~{\sc ii} NIR triplet that appears weakly in the event A spectra and disappears around the event B peak, only to re-appear around +20~d as seen with SN 2009ip and SN 2016jbu \citep{Fraser2013, Kilpatrick2018, Brennan2022a}. 
Within the Ca~{\sc ii} NIR lines, the P Cygni absorption component of the 8498 Å line is less blended; therefore, we used it to measure the shift of the absorption minima. In the earliest potential detections at +21 and +28~d the absorption component is not visible. At +51~d the absorption minima is shifted to $-5100$~km~s$^{-1}$ and remains constant during the plateau phase. Furthermore, in SN 2009ip-like events the forbidden [Ca~{\sc ii}] lines at 7291 and 7323 Å are visible during event A, after which they disappear and re-appear in the late plateau or tail phase. Similarly, in SN 2016cvk the [Ca~{\sc ii}] feature is not detectable in the early spectra during event B. At the end of the plateau phase at +67~d the wavelength region associated with this feature has a quite low signal-to-noise ratio, and possible detection of the line doublet is tentative at best. However, in the tail phase from +82 to +405~d the blended [Ca~{\sc ii}] feature is clearly identifiable. Therefore, in short, the beginning and the end of the plateau phase of SN 2016cvk coincide with epochs when the Ca~{\sc ii} NIR and [Ca~{\sc ii}] features evolve to be markedly detectable, respectively, which seems to be a common characteristic of SN 2009ip-like events.

Prominent Fe~{\sc ii} lines with P Cygni profiles appear at the +21~d spectrum of SN 2016cvk at 4924, 5018, and 5169 Å and these spectral features remain detected through the observed evolution. +21~d marks also the epoch when the broad emission line regions appear from roughly 4500 to 4700 Å and 5200 to 5400 Å consisting likely of a plethora of Fe~{\sc ii} and other metal lines blended together. In Fig. \ref{fig:all_spectra_allSN} these features are seen in the spectra of LSQ13zm, SN 2016jbu, SN 2009ip, and SN 2016cvk around $+28$~d. For example, \cite{Pastorello2013} and \cite{Fraser2013} associated the most dominant features within these regions of SN 2009ip to the Fe~{\sc ii} multiplets 37, 38, 41, 48, and 49 \citep{Moore1945}. The spectrophotometric evolution of SN 2016cvk at post-peak and plateau phases shares particular similarity to those of the SN 2009ip-like SN 2023ldh \citep{Pastorello2025}; at the plateau phase, their spectra with broad P Cygni metal lines resemble also some Type II SNe during the recombination of the hydrogen envelope, though the Balmer lines are dominated by their emission components. In the +82~d spectrum of SN 2016cvk, a new feature appears, which could be associated with the forbidden [Fe~{\sc ii}] line at 7155 Å and is seen also in the +223 and +405~d spectra, though poorly in the former due to the low signal-to-noise of the region. The [Fe~{\sc ii}] 7155 Å line was identified in the spectra of SN 2009ip by \cite{Fraser2015} and \cite{Graham2017}; however, appearing only around +250~d from the phase B peak. Similarly a late-time detection of the [Fe~{\sc ii}] line in SN 2015bh was indicated by \cite{EliasRosa2016}.

The red end of our spectra of SN 2016cvk have a relatively poor signal-to-noise ratio; however, the Paschen lines at 9546, 9229, and 9015 Å can be identified clearly in the +82~d spectrum. This is quite similar to SN 2009ip, which showed Paschen lines through the early evolution of phase B with the lines becoming notably more prominent towards the end of the plateau phase \citep{Fraser2013, Margutti2014}.

\subsection{Near-infrared spectroscopy}
\label{sec:nir}

\begin{figure}
\includegraphics[trim={1cm 0.5cm 3cm 0cm},clip,width=\linewidth]{{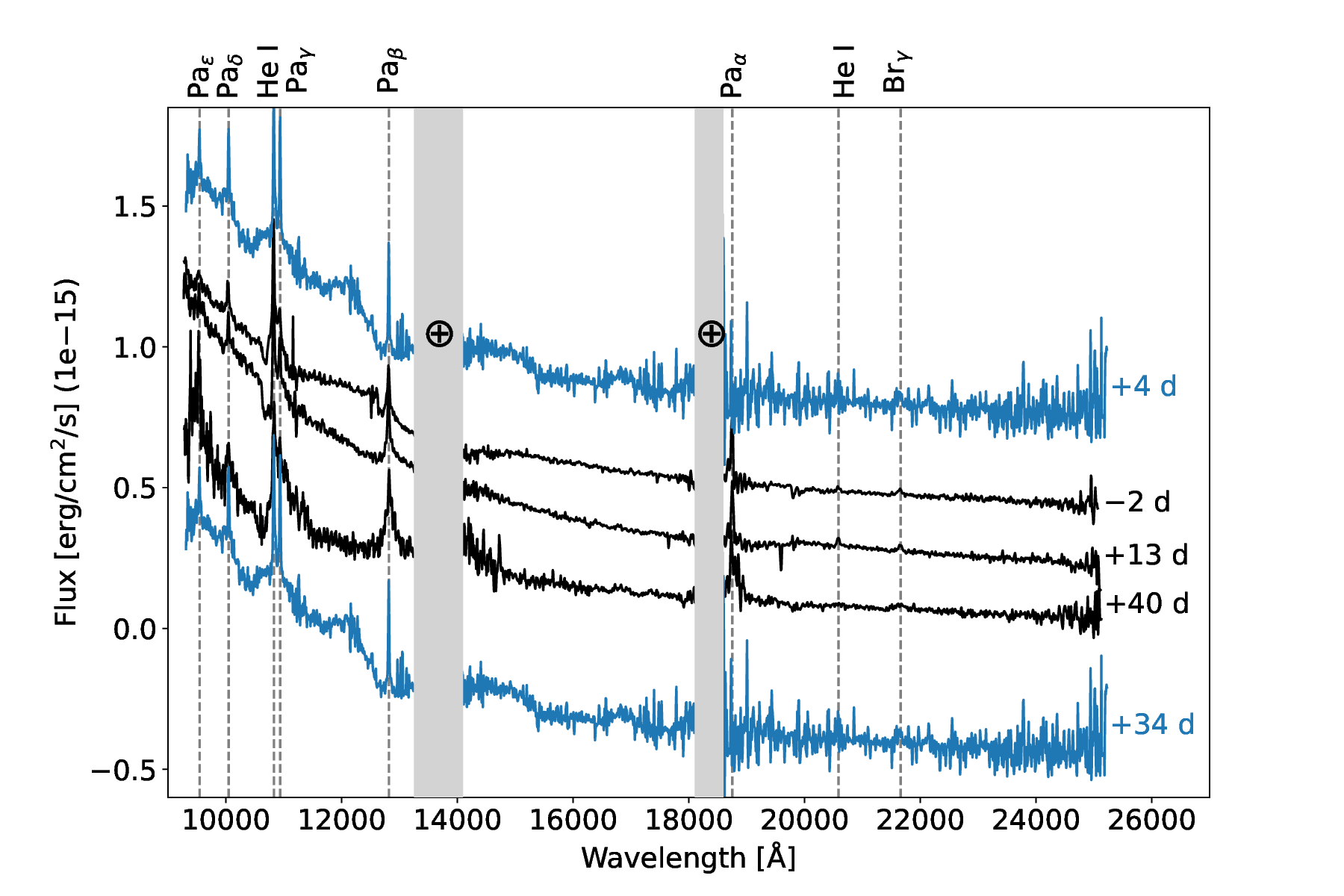}}
\caption{Near-infrared spectra of SN 2016cvk (black) compared to SN 2009ip (blue). The wavelengths of the most prominent telluric regions ($13200 - 14100$ Å and $18100 - 18600$ Å) are covered with gray bands and indicated with the $\oplus$ symbol. The spectra have been vertically shifted for clarity.}
\label{fig:nirspec}
\end{figure}

In addition to optical spectroscopy, we obtained three epochs of near-infrared spectra at $-2$, $+13$, and $+41$~d from peak. Therefore, SN 2016cvk is the third SN 2009ip-like transient that has been monitored with near-infrared spectroscopy, along with SN 2016jbu \citep{Brennan2022a} and SN 2009ip \citep{Fraser2013, Fraser2015, Pastorello2013, Smith2014, Levesque2014, Margutti2014}. 
The most notable characteristics in the near-infrared spectra are hydrogen features, including several Paschen lines at 9546, 10938, 10049, 12818, and 18751 Å, and the Brackett $\gamma$ line at 21661 Å. Furthermore, the He~{\sc i} 10830 Å feature is blended with Pa$\gamma$, and a weak He~{\sc i} line at 20587 Å is also visible in the $-2$ and +13~d spectra. The near-infrared spectral features and their evolution is very similar to those of SN 2009ip and SN 2016jbu, providing evidence for a common evolution at these wavelengths among SN 2009ip-like transients, see Fig. \ref{fig:nirspec}.

\subsection{Late-time spectroscopy}\label{sec:latephase}

\begin{figure}
\includegraphics[trim={0cm 0cm 0cm 0cm},clip,width=\linewidth]{{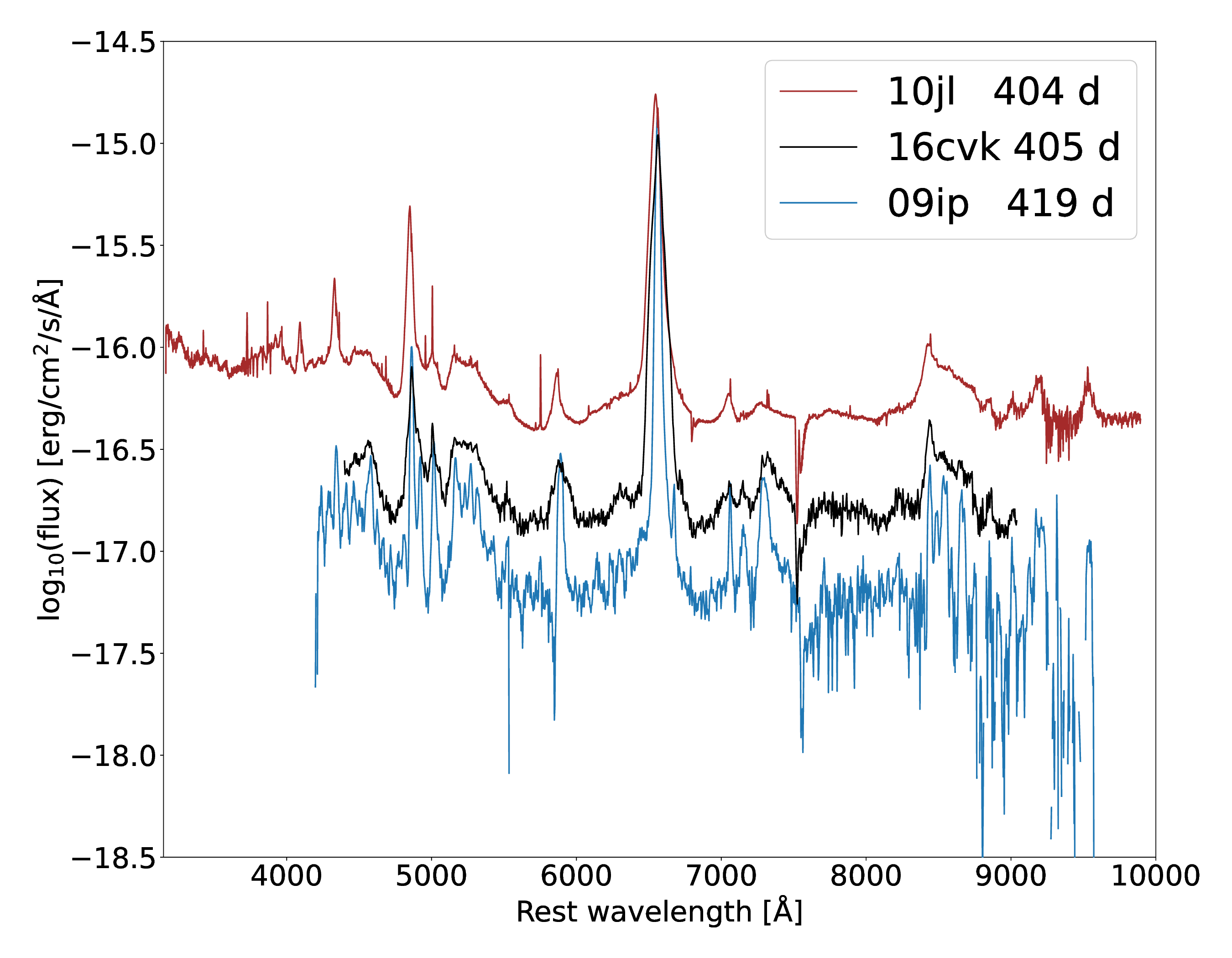}}
\caption{Late-time spectra of SN 2016cvk, SN 2009ip \citep{Fraser2015} and SN 2010jl \citep{Jencson2016}. Flux presented in logarithmic scale.}
\label{fig:latespec}
\end{figure}

\begin{figure}
    \centering
    \includegraphics[trim={0cm 0cm 0cm 0cm},clip,width=\linewidth]{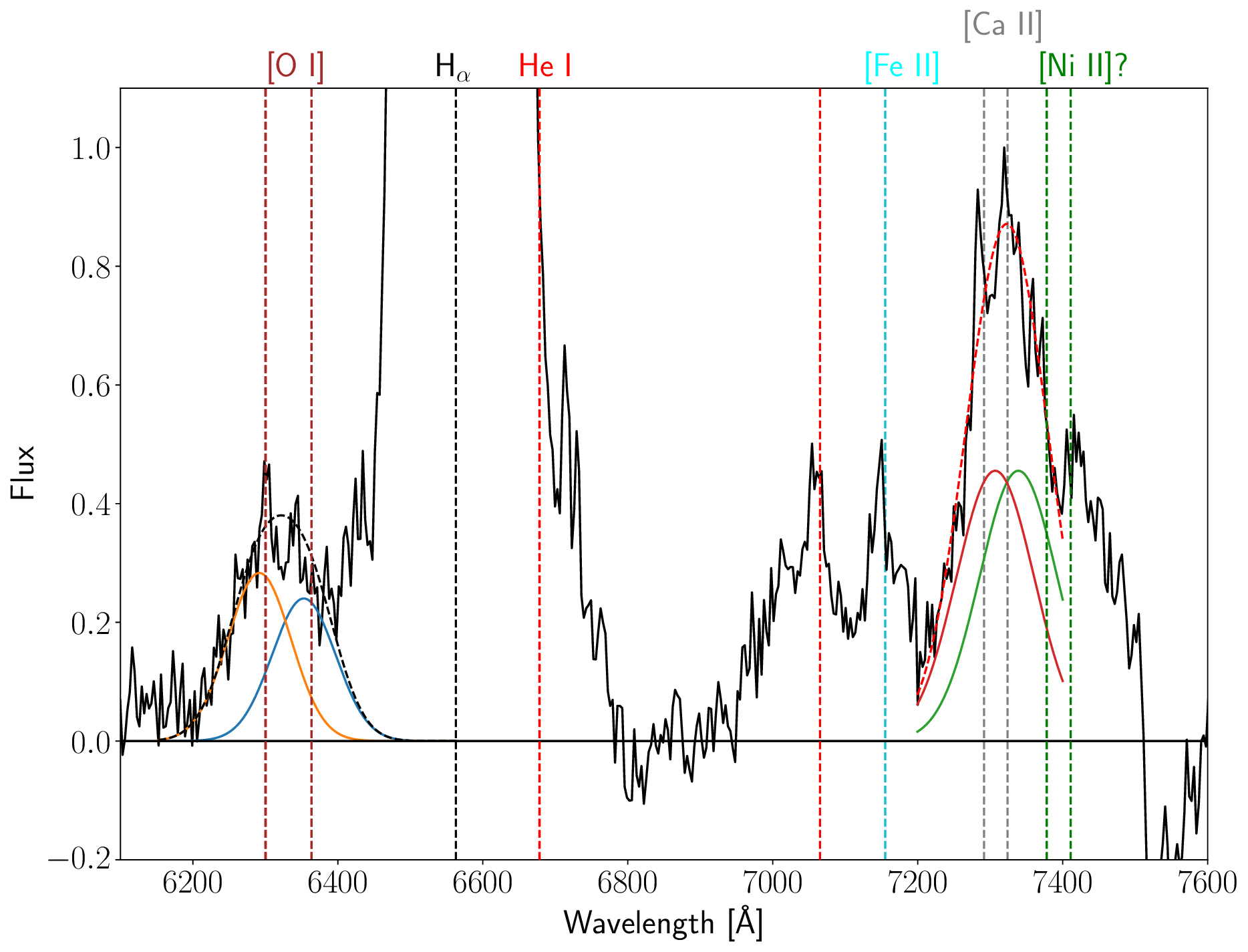}
    \caption{Partial spectrum of SN 2016cvk at +405~d along with Gaussian fits to forbidden lines of [O~{\sc i}] and Ca~{\sc ii}. The spectrum is continuum-subtracted and normalised to 1 at the peak value of the [Ca~{\sc ii}] doublet at $\sim$7300~Å. 
    }
    \label{fig:405dspec}
\end{figure}

The feature associated with the Ca~{\sc ii} NIR triplet has evolved to be increasingly asymmetric at the +223 and +405~d spectra of SN 2016cvk. We suggest that this indicates a prominent appearance of the O~{\sc i} line at 8446 Å, while no clear O~{\sc i} line at 7774 Å can be identified. This is similar to that reported for SN 2009ip, SN 2015bh, SN 2016bdu, and SN 2016jbu \citep[e.g. ][]{Fraser2015, EliasRosa2016, Pastorello2018, Brennan2022a}, though the likely O~{\sc i} 8446 Å line appears much less prominently in these events. It has been discussed in the context of SN 2009ip-like events, that such a strong flux difference between O~{\sc i} lines at 7774 and 8446 Å would not be produced by recombination or collisional excitation, but rather fluorescence by $\text{Ly}_\beta$ pumping \citep{Fraser2015, Graham2017, Brennan2022a}. Similar detection of O~{\sc i} features is also not unusual in other Type IIn SNe \citep[e.g. SN 2010jl; ][]{Fransson2014}. In fact, \cite{Graham2017} noted that the late-time spectra of SN 2009ip and SN 2010jl are quite identical, with one of the key differences of a much stronger O~{\sc i} line in the latter. However, the blended O~{\sc i} and Ca~{\sc ii} profile of SN 2016cvk at +405~d is actually quite similar to that of SN 2010jl at +404~d, and overall the late-time spectrum of SN 2016cvk strongly resembles that of the terminal Type IIn SN 2010jl, even more so than that of SN 2009ip, see Fig. \ref{fig:latespec}. 

We also detected a likely feature of [O~{\sc i}] in the +405~d spectrum of SN 2016cvk at $\sim$6300 Å, which could be evidence of nucleosynthesised material created in a terminal SN explosion. To our knowledge, this is perhaps the most prominent detection of the [O~{\sc i}] doublet at 6300 and 6364~Å in the late-time spectra of SN 2009ip-like events. We attempted a "boxy" line profile fit for the two [O~{\sc i}] doublet lines at 6300 and 6364~Å, similar to the fit created for the Type IIn SN 1995N by \cite{Fransson2002}; however, this approach did not yield meaningful results. A double Gaussian profile resulted in a notably better fit for our data with FWHM = 4700~km~s$^{-1}$ for the two components. The FWHM is similar to the intermediate emission components of SN 1995N, which was associated with the SN ejecta \citep{Fransson2002}. We created a similar double Gaussian fit for the [Ca~{\sc ii}] doublet at 7291 and 7324~Å, and measured a FWHM $= 5200$~km~s$^{-1}$ for the two components. The [Ca~{\sc ii}] has a red shoulder, which could have contribution for example from the [Ni~{\sc ii}] doublet at 7378 and 7412~Å. The integrated intensity ratio between the forbidden oxygen and calcium features is approximately $I$([Ca~{\sc ii}])/$I$([O~{\sc i}]) = 1.8. However, the integrated ratio of the two [O~{\sc i}] components is $I$(6300\AA)/$I$(6364\AA)$= 1.2$, which differs notably from the canonical 3:1 ratio. It is likely that the [O~{\sc i}] lines are collisionally suppressed; therefore, even if associated with nebular emission from a SN, the [O~{\sc i}] features cannot be used to reliably estimate the mass of the precursor star.

\section{Discussion}\label{sec:discussion}

The SN 2009ip-like targets in our sample have clearly distinguishable light curve phases: event A, event B, plateau, and tail phase. The event A light curves show notable differences in evolution and the associated $r$-band absolute peak magnitudes vary from $-13.9$ to $< -15.6$~mag with SN 2016cvk being the brightest, whereas the event B magnitudes are very similar and range from $-17.9$ to $-18.3$~mag. 
Therefore, if terminal events, it seems perhaps unlikely that the event A (rather than event B) with a diverse set of light curves would arise from a SN explosion and subsequently produce such similar event B phases via CSM interaction. 
Furthermore, the short plateau phase seems to be a quite common characteristic of SN 2009ip-like events; however, see also the recent study of SN 2023vbg \citep{Goto2025}. 
The absolute magnitude of the plateau is between $-15.8$ and $-16.2$~mag for the majority of the sample, SN 2016cvk being a brighter exception at $-16.9$~mag. 
The tail phase decline rate of SN 2016cvk is quite slow (e.g. 0.7~mag per 100~d in $r$-band) and interaction is likely ongoing during this phase between the expanding ejecta and CSM. Therefore, we could estimate only a conservative $^{56}$Ni mass upper limit of $<0.07 M_\odot$ for the event.

The precursor of SN 2016cvk is detected in archival data around $-1220$, $-700$, and $-440$~d with an absolute magnitude of roughly $-13$~mag. The relatively high luminosities of these events are consistent with eruptions rather than a quiescent star and are similar to those detected for SN 2009ip and SN 2016bdu \citep{Pastorello2013,Pastorello2018} suggesting a common characteristic among these transients. From the photometric data of the precursor around $-700$~d, we estimate $7600 \pm 1300$~K and $1.2 \pm 0.3 \times 10^{14}$~cm, indicating a major outburst. 

The H$_\alpha$ and H$_\beta$ emission lines of SN 2016cvk undergo a complex, multi-component evolution over time. They are best described using three main components: a strong, narrow Lorentzian emission component combined with a broader Gaussian emission, and only one P Cygni absorption component. 
During event A, the largest velocity suggested by the P Cygni absorption minimum of the H$_\alpha$ line profile is $-6500$~km~s$^{-1}$ and a maximum blue velocity at zero intensity is estimated to be $-10000$~km~s$^{-1}$ and $-12500$~km~s$^{-1}$ for H$_\alpha$ and H$_\beta$, respectively. 
During event B the largest H$_\alpha$ velocity at the absorption minimum is $-4100$~km~s$^{-1}$ and the maximum blue velocity at zero intensity is $-7500$~km~s$^{-1}$. This shows that material with relatively high velocity is present during both events A and B.

Spectra of SN 2016cvk taken during the rise to the event B peak ($-11$, $-6$, and $-1$ d) are dominated by narrow emission lines of hydrogen and helium. The Balmer decrement H$_\alpha$/H$_\beta$ at this time is around 1.8 to 1.2; such a low value is consistent with collisional de-excitation processes in high electron densities, which points to a concentrated, dense CSM. This also supports the assumption that the broad emission component in the line profile fit at these epochs is produced by electron scattering.
We also detect a blended flash ionisation feature of He~{\sc ii}, N~{\sc iii}, and C~{\sc iii} at around $\sim$4700 Å in these early spectra of SN 2016cvk. 
A similar flash feature has been reported in SN 2010mc \citep{Khazov2016} and SN 2015bh \citep{Thone2017}. We also identify the feature in the early archival spectra of SN 2016jbu; however, not in those of SN 2009ip. It is perhaps surprising that SN 2009ip with a high-quality spectroscopic data set does not show this flash feature. This could be related to the density or other characteristics of the CSM and the geometry of the system. SN 2009ip is also the event in our sample with the shortest constrained plateau length, and it is intriguing to speculate if these characteristics are connected. Further early observations of a larger sample of SN 2009ip-like events is clearly required to fully understand the nature of the complex ambient medium and the origin of SN 2009ip-like events.

We do not detect a double-peaked Balmer emission profile for SN 2016cvk at any epoch. This is in contrast to some SN 2009ip-like transients with an evolution to increasingly double-peaked  H$_\alpha$ line \citep[e.g. SN 2016jbu; ][]{Brennan2022b}. The only signature of double-peaked Balmer lines of SN 2016cvk is at the beginning of its tail phase at +82 d, where a second Gaussian emission component briefly appears. These differences in the strength and location of different Balmer emission components in SN 2009ip-like transients could be due to differences in the CSM structure or the viewing angle. \cite{Brennan2022b} presented a possible geometric model of the SN 2016jbu event as an explosion advancing in a disc-like CSM structure, which is viewed from an intermediately inclined angle. The redshifted emission component would originate from the side of the disc that is further away from the viewer, with the explosion advancing away from us, while the blueshifted component would be emitted from the side of the disc facing the viewer, with material expanding towards us. If analogous to SN 2016cvk, since this double-peaked emission or absorption structure is not seen, it is possible that such a CSM disc would be viewed in our case from a fairly edge-on viewing angle \citep[see e.g. ][]{Kurfurst2020}. Such a scenario is also suggested for SN 2009ip \citep[e.g. ][]{Reilly2017} with no clearly double-peaked H$_\alpha$ line.

\section{Conclusions}\label{sec:conclusions}

We have presented a spectrophotometric data set of the SN 2009ip-like transient SN 2016cvk and investigated the nature of the event. Comparisons to other SN 2009ip-like events suggest several shared characteristics among these transients in addition to the double-peaked light curve with fainter event A and the subsequent main event B. Historical outbursts appear to be common; SN 2016cvk was detected at an absolute magnitude of around $-13$~mag roughly $-1220$, $-700$, and $-440$ d before the main event peak. In particular, we estimated $7600 \pm 1300$~K and $1.2 \pm 0.3 \times 10^{14}$~cm ($\sim$1900~$R_\odot$) as the blackbody temperature and radius, respectively, for the outburst at $-700$ d. Material expanding with relatively high velocities appears to be present during both event A and B; specifically, the spectroscopic observations of SN 2016cvk suggest velocities up to $-12500$ and $-7500$~km~s$^{-1}$ during event A and B, respectively, with bulk velocities slower by roughly a factor of two. A short plateau phase appears to be common among SN 2009ip-like events, following the decline from the event B light curve maximum. The spectroscopic evolution of SN 2016cvk in the near-infrared region is dominated by hydrogen and helium features and is similar among the SN 2009ip-like events with such reported observations. The late-time spectra of SN 2016cvk and other SN 2009ip-like events appear interaction-dominated and resemble those of slowly-evolving Type IIn SNe with a high H$_\alpha$/H$_\beta$ Balmer decrement likely powered by collisional excitations. 

It has been suggested in the literature that the CSM of the SN 2009ip-like transients could be dense and distinctively disc-like with many events showing double-peaked emission profiles. Our observations of SN 2016cvk do not show such double-peaked spectral lines; however, the Balmer decrement H$_\alpha$/H$_\beta$ of SN 2016cvk near event B maximum is close to unity, consistent with collisional de-excitation processes and concentrated, dense CSM. Furthermore, it is possible that such a CSM disc would be viewed in the case of SN 2016cvk from a somewhat edge-on viewing angle.

We detect a blended ionisation feature of He~{\sc ii}, N~{\sc iii}, and C~{\sc iii} in the spectra of SN 2016cvk around $\sim$ 4700 Å at $-11$, $-6$, and $-1$~d from the event B maximum. We determine the onset of event B as $-11$ d from the light curve maximum, and measure that the flash features are visible for $16 \pm 5$~d. Such a long duration suggests that this feature is powered by radiation originating from ejecta-CSM interaction. Flash features have previously been reported in SN 2009ip-like events only in SN 2010mc \citep{Khazov2016} and SN 2015bh \citep{Thone2017}. However, we note that a similar, previously unreported, faint flash feature is also present in the early spectra of SN 2016jbu. Comparisons of the early $-11$~d spectrum of SN 2016cvk to the early SN models with flash features \citep{Boian2019} suggest either a low-mass RSG with a solar metallicity, or a high-mass RSG, YSG, or BSG with CNO-processed abundances, rather than an LBV progenitor.

The late-time +405~d spectrum of SN 2016cvk includes a feature that we identify as the [O~{\sc i}] doublet at $\sim 6300$ and 6364~Å, which could be evidence of nucleosynthesised material created in a terminal SN explosion. A Gaussian component fit suggests a FWHM velocity of 4700~km~s$^{-1}$ for the two components with a flux ratio close to unity. Therefore, the feature could be associated with the ejecta; however, the doublet is likely collisionally suppressed and cannot be used to estimate the mass of the progenitor star. While weak detections of this feature have been reported before in other SN 2009ip-like events, this is to our knowledge the most prominent detection of [O~{\sc i}] in these events.

\begin{acknowledgements}

We thank the anonymous referee for useful comments.

We thank Jesper Sollerman for helpful comments.

KKM and EK acknowledge financial support from the Emil Aaltonen foundation.

SM acknowledges support from the Research Council of Finland project 350458.

AR acknowledges financial support from the GRAWITA Large Program Grant (PI P. D’Avanzo). 

AR, AP and IS acknowledge financial support from the PRIN-INAF 2022 "Shedding light on the nature of gap transients: from the observations to the models".

SJB acknowledges their support by the European Research Council (ERC) under the European Union’s Horizon Europe research and innovation programme (grant agreement No. 10104229 - TransPIre)

This work was funded by ANID, Millennium Science Initiative, ICN12\_009.

BAI acknowledges support from National Agency for Research and Development (ANID) grants ANID-PFCHA/Doctorado Nacional/21221964.

PC acknowledges support via Research Council of Finland (grant 340613). 

TWC acknowledges the Yushan Fellow Program by the Ministry of Education, Taiwan for the financial support (MOE-111-YSFMS-0008-001-P1). 

MN is supported by the European Research Council (ERC) under the European Union’s Horizon 2020 research and innovation programme (grant agreement No.~948381).

CPG acknowledges financial support from the Secretary of Universities and Research (Government of Catalonia) and by the Horizon 2020 Research and Innovation Programme of the European Union under the Marie Sk\l{}odowska-Curie and the Beatriu de Pin\'os 2021 BP 00168 programme, from the Spanish Ministerio de Ciencia e Innovaci\'on (MCIN) and the Agencia Estatal de Investigaci\'on (AEI) 10.13039/501100011033 under the PID2023-151307NB-I00 SNNEXT project, from Centro Superior de Investigaciones Cient\'{i}ficas (CSIC) under the PIE project 20215AT016 and the program Unidad de Excelencia Mar\'{i}a de Maeztu CEX2020-001058-M, and from the Departament de Recerca i Universitats de la Generalitat de Catalunya through the 2021-SGR-01270 grant.

TEMB is funded by Horizon Europe ERC grant no. 101125877.

JLP acknowledges support from ANID, Millennium Science Initiative, AIM23-0001.

TMR is part of the Cosmic Dawn Center (DAWN), which is funded by the Danish National Research Foundation under grant DNRF140. TMR acknowledges support from the Research Council of Finland project 350458. 

Based on observations collected at the European Organisation for Astronomical Research in the Southern Hemisphere, Chile, as part of ePESSTO+ (the advanced Public ESO Spectroscopic Survey for Transient Objects Survey – PI: Inserra) and ePESSTO and PESSTO (PI: Smartt). ePESSTO+ observations were obtained under ESO program ID 1103.D-0328, while ePESSTO and PESSTO observations under ESO program IDs 097.D-0891, 191.D-0935 and 199.D-0143.

This work makes use of observations from the Las Cumbres Observatory global telescope network.

This research has made use of data obtained through the High Energy Astrophysics Science Archive Research Center online service, provided by the NASA/Goddard Space Flight Center.

This project used public archival data from the Dark Energy Survey \href{https://www.darkenergysurvey.org/}{(DES)}. Funding for the DES Projects has been provided by the U.S. Department of Energy, the U.S. National Science Foundation, the Ministry of Science and Education of Spain, the Science and Technology Facilities Council of the United Kingdom, the Higher Education Funding Council for England, the National Center for Supercomputing Applications at the University of Illinois at Urbana-Champaign, the Kavli Institute of Cosmological Physics at the University of Chicago, the Center for Cosmology and Astro-Particle Physics at the Ohio State University, the Mitchell Institute for Fundamental Physics and Astronomy at Texas A\&M University, Financiadora de Estudos e Projetos, Funda{\c c}{\~a}o Carlos Chagas Filho de Amparo {\`a} Pesquisa do Estado do Rio de Janeiro, Conselho Nacional de Desenvolvimento Cient{\'i}fico e Tecnol{\'o}gico and the Minist{\'e}rio da Ci{\^e}ncia, Tecnologia e Inova{\c c}{\~a}o, the Deutsche Forschungsgemeinschaft, and the Collaborating Institutions in the Dark Energy Survey. 

The Collaborating Institutions are Argonne National Laboratory, the University of California at Santa Cruz, the University of Cambridge, Centro de Investigaciones Energ{\'e}ticas, Medioambientales y Tecnol{\'o}gicas-Madrid, the University of Chicago, University College London, the DES-Brazil Consortium, the University of Edinburgh, the Eidgen{\"o}ssische Technische Hochschule (ETH) Z{\"u}rich, Fermi National Accelerator Laboratory, the University of Illinois at Urbana-Champaign, the Institut de Ci{\`e}ncies de l'Espai (IEEC/CSIC), the Institut de F{\'i}sica d'Altes Energies, Lawrence Berkeley National Laboratory, the Ludwig-Maximilians Universit{\"a}t M{\"u}nchen and the associated Excellence Cluster Universe, the University of Michigan, the National Optical Astronomy Observatory, the University of Nottingham, the Ohio State University, the OzDES Membership Consortium, the University of Pennsylvania, the University of Portsmouth, SLAC National Accelerator Laboratory, Stanford University, the University of Sussex, and Texas A\&M University.
 
Based in part on observations at Cerro Tololo Inter-American Observatory, National Optical Astronomy Observatory, which is operated by the Association of Universities for Research in Astronomy (AURA) under a cooperative agreement with the National Science Foundation. 
Database access and other data services are provided by the Astro Data Lab.

This publication makes use of data products from the Two Micron All Sky Survey, which is a joint project of the University of Massachusetts and the Infrared Processing and Analysis Center/California Institute of Technology, funded by the National Aeronautics and Space Administration and the National Science Foundation.

Funding for the Sloan Digital Sky Survey V has been provided by the Alfred P. Sloan Foundation, the Heising-Simons Foundation, the National Science Foundation, and the Participating Institutions. SDSS acknowledges support and resources from the Center for High-Performance Computing at the University of Utah. The SDSS web site is \url{www.sdss.org}.

SDSS is managed by the Astrophysical Research Consortium for the Participating Institutions of the SDSS Collaboration, including the Carnegie Institution for Science, Chilean National Time Allocation Committee (CNTAC) ratified researchers, the Gotham Participation Group, Harvard University, Heidelberg University, The Johns Hopkins University, L’Ecole polytechnique f\'{e}d\'{e}rale de Lausanne (EPFL), Leibniz-Institut für Astrophysik Potsdam (AIP), Max-Planck-Institut für Astronomie (MPIA Heidelberg), Max-Planck-Institut für Extraterrestrische Physik (MPE), Nanjing University, National Astronomical Observatories of China (NAOC), New Mexico State University, The Ohio State University, Pennsylvania State University, Smithsonian Astrophysical Observatory, Space Telescope Science Institute (STScI), the Stellar Astrophysics Participation Group, Universidad Nacional Aut\'{o}noma de M\'{e}xico, University of Arizona, University of Colorado Boulder, University of Illinois at Urbana-Champaign, University of Toronto, University of Utah, University of Virginia, Yale University, and Yunnan University.

This research used ASTROPY, a community-developed core Python package for Astronomy \citep{astropy}.
\end{acknowledgements}

\bibliographystyle{aa}
\bibliography{aa55740-25}

\begin{appendix}\label{sec:appendix}
\section{Additional notes, figures, and tables}
\subsection{Photometry}\label{app:photom}
The ESO facilities were used for the spectrophotometric observations of SN 2016cvk. Optical imaging was carried out using the 3.58~m NTT in La Silla, Chile, with EFOSC2 via the ePESSTO programme. 
The images were processed with the PESSTO pipeline \footnote{\url{https://github.com/svalenti/pessto/releases}} \citep{Smartt2015}, which applies the bias subtraction and flat-field correction to the images. 

Data was also used from the Las Cumbres Observatory network of automated 1~m telescopes \citep{Brown2013} located in the South African Astronomical Observatory (SAAO, South Africa), the Cerro Tololo Interamerican Observatory (CTIO, Chile), and the Siding Spring Observatory (SSO, Australia); site identifiers cpt, lsc, and coj, respectively. 
The data (programme IDs CON2016A-007 and CLN2016A-005) was automatically reduced by the \textsc{BANZAI} pipeline\footnote{\url{https://lco.global/documentation/data/BANZAIpipeline/}}, which performs bad-pixel masking, bias and dark subtraction, and flat field correction.  

Early observations of the field of SN 2016cvk by the CHASE survey were reported in the compilation of precursor activity of Type IIn SNe by \cite{Reguitti2024}.  
For completeness, we included their observations in our analysis, which provided an early $r$-band detection and limits of the precursor of SN 2016cvk from the 0.41~m Panchromatic Robotic Optical Monitoring and Polarimetry Telescopes (PROMPT) located in CTIO, Chile. 

Archival imaging was also obtained for the analysis, observed with the 4m Victor M. Blanco telescope (at CTIO) using the Dark Energy Camera (DECam). The images were automatically reduced using the Dark Energy Survey Image Processing Pipeline \citep{Morganson2018}. 
The processing of the data includes steps for bias subtraction, bad pixel masking, correcting nonlinear pixel response, and flat fielding. 
Independent measurements of the SN 2016cvk pre-outburst detections from the DECam data were previously reported by \cite{Reguitti2024}. 

Images were also obtained from the Savannah Skies Observatory in Australia (J. Brimacombe). The images were reduced using the MaxIm DL software, following the usual steps of bias subtraction as well as flat-field and dark corrections.

Some $V$-band observations of the SN 2016cvk location were publicly reported \citep{Brimacombe2016} by the ASAS-SN programme, which consists of a network of robotic 14~cm telescopes around the globe, equipped with identical CCD cameras \citep{Shappee2014, Jayasinghe2019}. These reported magnitudes were used in our analysis; however, we do not have access to the images. 

Similar to EFOSC2, the near-infrared data from SOFI was reduced using the PESSTO pipeline. In this case, the pipeline applies flat field, crosstalk, and illumination corrections, and performs sky subtractions and merging of the dithered images. 

We made use of archival data of SN 2016cvk from the Neil Gehrels Swift Observatory \citep{Gehrels2004} via programme IDs 34708 and 34717. The Level 2 pre-processed images from the Ultra-violet Optical Telescope \citep[UVOT;][]{Roming2005} instrument were downloaded for the analysis via the High Energy Astrophysics Science Archive Research Center (HEASARC) database. The data was processed with the HEASARC High Energy Astrophysics software (HEAsoft) package v6.34. The uvotimsum task in HEAsoft was used to sum the individual UVOT exposures of an observational epoch for each filter. The HEAsoft uvotsource task was used to carry out the aperture photometry of the transient. 
Some contamination from the host galaxy could not be excluded and the magnitudes are reported up to JD 2457732.94; after this epoch the host contamination was found to erratically affect the photometry.

\subsection{Spectroscopy}
The EFOSC2 spectra were reduced using the PESSTO pipeline. In this process, the two-dimensional spectra were trimmed, overscan and bias subtracted, and cosmic rays were removed from the data. The spectra were also flat-field corrected using halogen lamp flats, wavelength calibrated using arc lamp spectra, and flux calibrated using standard star observations. The SOFI spectra were reduced using the PESSTO pipeline, including the steps of flat fielding, sky subtraction, arc lamp based wavelength calibration, merging of the individual exposures, telluric correction, and flux calibration. The VLT spectrum was reduced using standard IRAF tasks, following the usual steps of master bias subtraction, normalised flat fielding, trace extraction, wavelength calibration using arc lamps, and relative flux calibration using observations of a spectroscopic standard star. All the spectra were absolute flux calibrated based on the photometry of SN 2016cvk.

\subsection{X-ray observations}
Swift's X-Ray Telescope \citep[XRT;][]{Burrows2005} archival images did not reveal a source associated with SN 2016cvk within the default energy range of 0.2~$-$~10~keV. The HEAsoft task sosta was used to calculate the count rates at the coordinates of the transient in the pre-processed Level 2 images from XRT, which did not result in any detection at S/N~$>$~2. The resulting luminosity upper limits are reported in Table \ref{tab:xray} based on the luminosity distance and the estimated unabsorbed flux limits converted from the sosta measurements using the HEASARC Portable, Interactive Multi-Mission Simulator (PIMMS) tool\footnote{\url{https://heasarc.gsfc.nasa.gov/docs/tools.html}} via the WebPIMMS v4.15 interface. The assumed parameters were a power law model with a photon index $\Gamma = 2$, and a weighted average Galactic H~{\sc i} column density of $N_\text{H} = 9.66 \times 10^{19}$~cm$^{-2}$ provided by the HEASARC nH tool for the coordinates. \cite{Ofek2013} reported X-ray detections of SN 2009ip at $\sim$10$^{39}$~erg~s$^{-1}$; our upper limits of $<$4~$\times$~10$^{40}$~erg~s$^{-1}$ and above for SN 2016cvk are not particularly constraining, and consistent with the X-ray luminosity of SN 2009ip.

\subsection{Trends between the A, B, and plateau phases}
\begin{figure*}
\centering
\begin{minipage}{0.5\linewidth}
\includegraphics[trim={0cm 0cm 0cm 0cm},clip,width=0.95\linewidth]{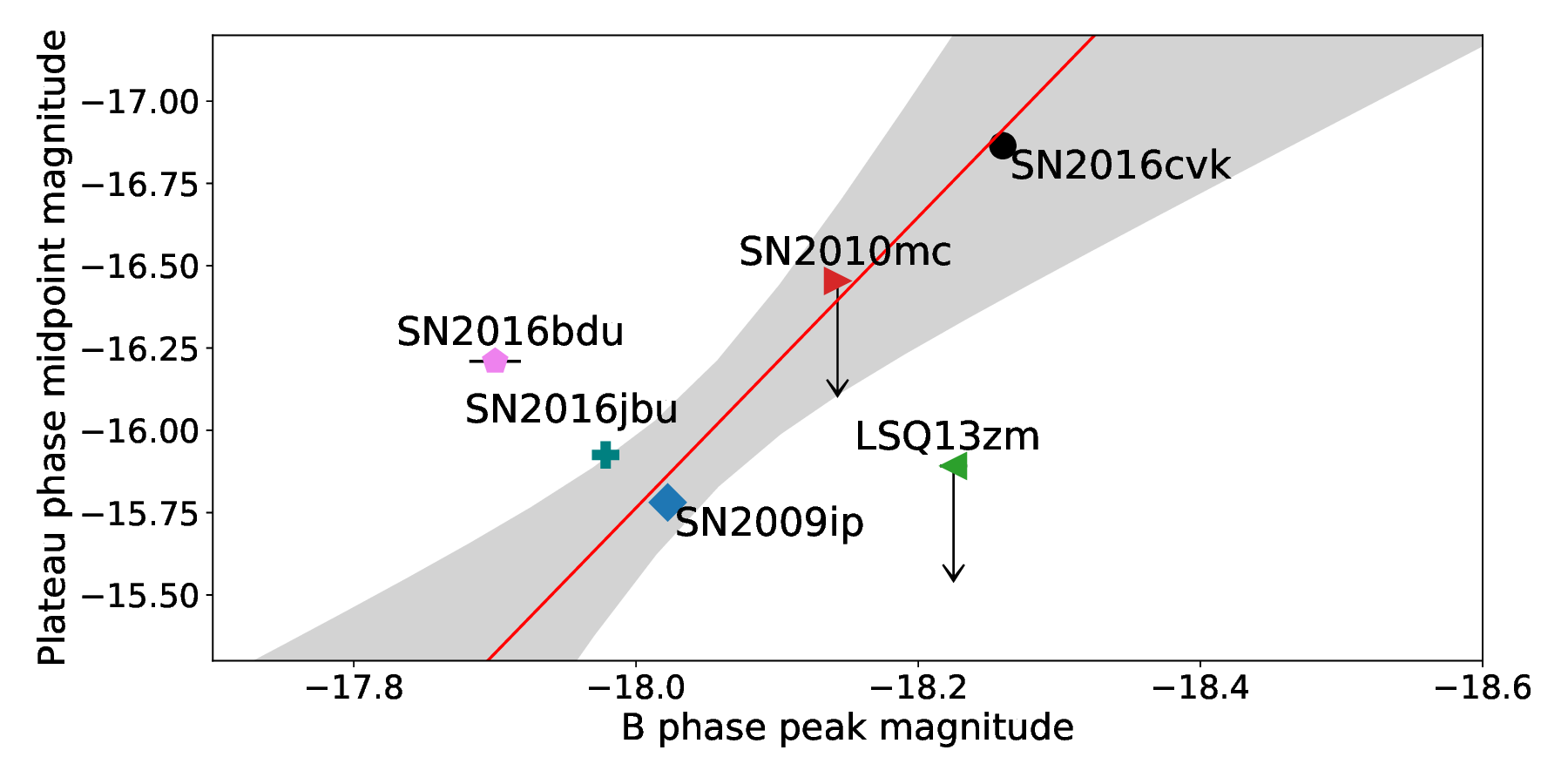}

\includegraphics[trim={-0.6cm 0cm -1.1cm 0cm},clip,width=0.95\linewidth]{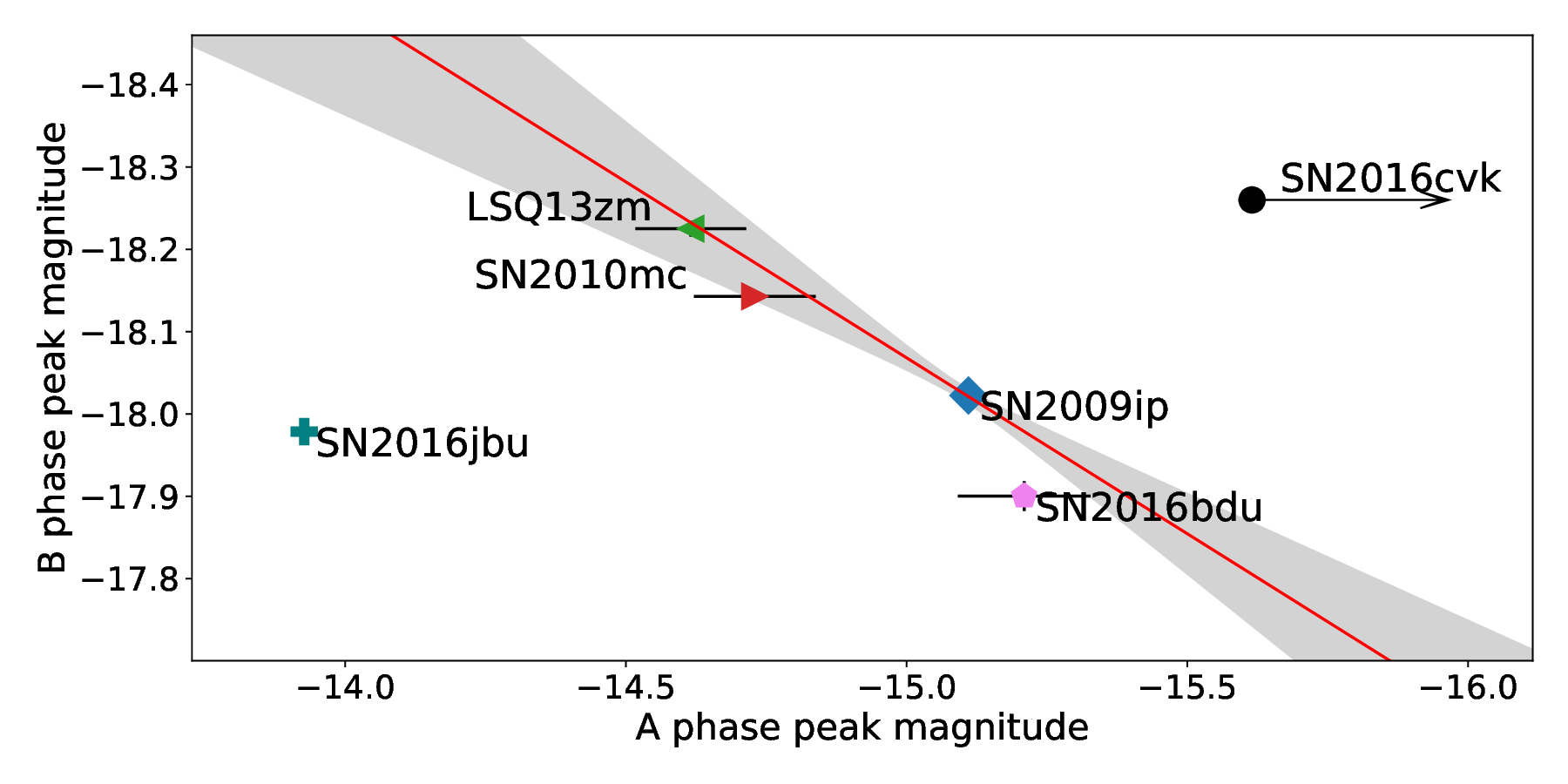}
\end{minipage}\begin{minipage}{0.5\linewidth}
\includegraphics[trim={-0.2cm 0cm -1.15cm 0cm},clip,width=0.95\linewidth]{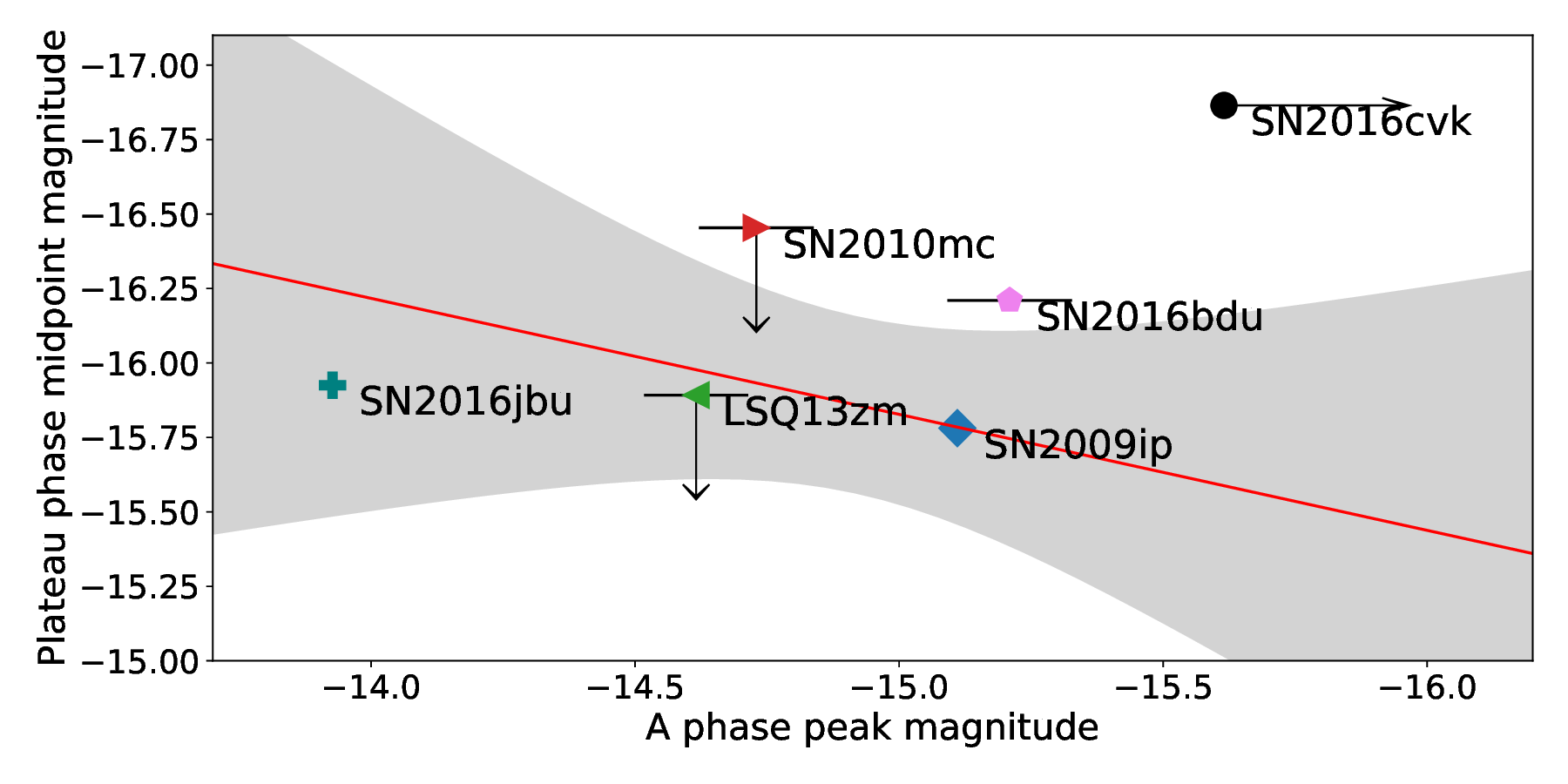}

\includegraphics[trim={-2cm 0cm -0.5cm 0cm},clip,width=0.97\linewidth]{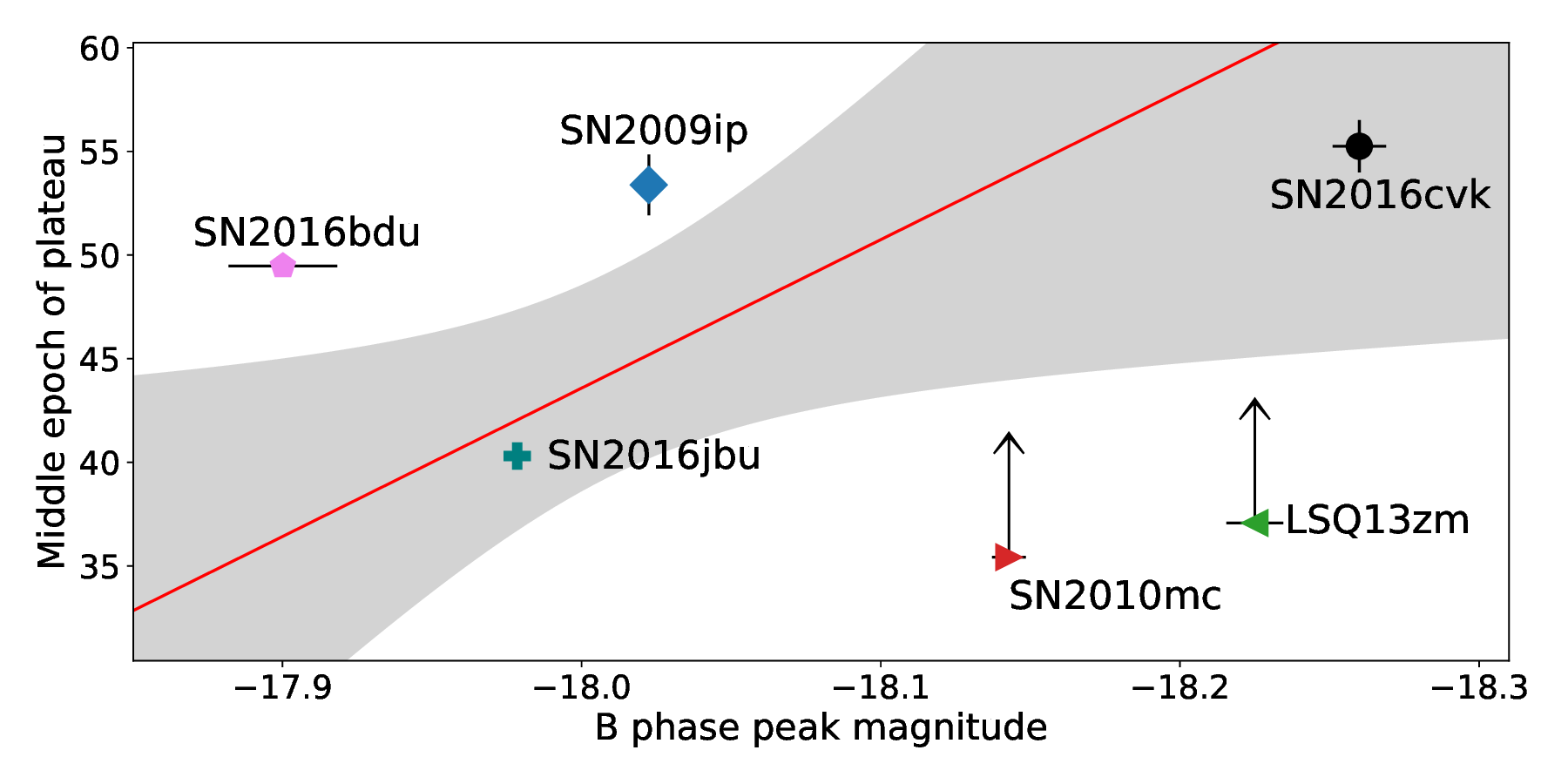}\end{minipage}
\caption{Comparisons and possible trends between the event A, event B, and plateau phases of SN 2009ip-like transients. Confidence band of $1\sigma$ is marked in the images with a gray band.}
\label{fig:phasecomp}
\end{figure*}

To check for any trends between the epochs or magnitudes of the different light curve phases, comparisons were carried out and a simple first degree polynomial was fit to the data with the help of the kmpfit fitting routine found in the Kapteyn package for Python 3. This routine takes in data that has errors in both x and y directions, and performs a fit using a least-squares method. The final results are presented in Fig. \ref{fig:phasecomp} with error limits marked for the original data points in x and y directions, and as a $1\sigma$ confidence band for the final fit as a gray coloured band. The data points marked with arrows are lower or upper limits, and were not included in the fits.

The data is quite scattered, the error margins are high, and the fit results do not support any robust trends between the magnitudes or epochs of the different light curve phases. As the number of observed SN 2009ip-like events increases, recreating these efforts with a larger sample size could yield more reliable results.

\onecolumn

\begin{small}
\setlength\tabcolsep{2.5pt}

\begin{longtable}{cccccccccccccc}
\caption{Ground-based photometry table of SN 2016cvk for optical bands with errors given in brackets.} 
\label{tab:photom}\\
\hline 
\hline
JD & $t$ & $m_\text{u}$ & $m_{B}$  & $m_\text{g}$ & $m_\text{V}$  & $m_\text{r}$  & $m_\text{i}$ & $m_\text{z}$ & $m_\text{Y}$ & Telescope \\
-2400000&(d)&(mag)&(mag)&(mag)&(mag)&(mag)&(mag)&(mag)&(mag)& + Instrument\\
\hline
\endfirsthead

\multicolumn{11}{c}{{\bfseries \tablename\ \thetable{} -- continued from previous page}} \\
\hline 
JD & $t$ & $m_{u}$  & $m_{B}$ & $m_{g}$ & $m_{V}$  & $m_{r}$  & $m_{i}$ & $m_{z}$ & $m_{Y}$ & Telescope \\
-2400000&(d)&(mag)&(mag)&(mag)&(mag)&(mag)&(mag)&(mag)&(mag)& + Instrument \\ \hline\\
\endhead

\hline \multicolumn{11}{|r|}{{Continued on next page}} \\ \hline
\endfoot

\hline \hline
\endlastfoot

55491.7 & -2151.7 & $\cdots$ & $\cdots$ & $\cdots$ & $\cdots$ & >18.8 & $\cdots$ & $\cdots$ & $\cdots$ & PROMPT+Apogee \\
55511.7 & -2131.7 & $\cdots$ & $\cdots$& $\cdots$  & $\cdots$ & >19.7 & $\cdots$ & $\cdots$ & $\cdots$ & PROMPT+Apogee \\
55696.8 & -1946.6 & $\cdots$ & $\cdots$& $\cdots$  & $\cdots$ & >18.6 & $\cdots$ & $\cdots$ & $\cdots$ & PROMPT+Apogee \\
55706.9 & -1936.5 & $\cdots$ & $\cdots$& $\cdots$  & $\cdots$ & >19.2 & $\cdots$ & $\cdots$ & $\cdots$ & PROMPT+Apogee \\
55713.9 & -1929.5 & $\cdots$ & $\cdots$& $\cdots$  & $\cdots$ & >18.4 & $\cdots$ & $\cdots$ & $\cdots$ & PROMPT+Apogee \\
55786.9 & -1856.5 & $\cdots$ & $\cdots$& $\cdots$  & $\cdots$ & >18.0 & $\cdots$ & $\cdots$ & $\cdots$ & PROMPT+Apogee \\
55797.7 & -1845.7 & $\cdots$ & $\cdots$ & $\cdots$ & $\cdots$ & >18.4 & $\cdots$ & $\cdots$ & $\cdots$ & PROMPT+Apogee \\
55813.7 & -1829.7 & $\cdots$ & $\cdots$& $\cdots$  & $\cdots$ & >18.4 & $\cdots$ & $\cdots$ & $\cdots$ & PROMPT+Apogee \\
55876.6 & -1766.8 & $\cdots$ & $\cdots$ & $\cdots$ & $\cdots$ & >18.3 & $\cdots$ & $\cdots$ & $\cdots$ & PROMPT+Apogee \\
56035.9 & -1607.5 & $\cdots$ & $\cdots$& $\cdots$  & $\cdots$ & >18.3 & $\cdots$ & $\cdots$ & $\cdots$ & PROMPT+Apogee \\
56052.8 & -1590.6 & $\cdots$ & $\cdots$& $\cdots$  & $\cdots$ & >17.4 & $\cdots$ & $\cdots$ & $\cdots$ & PROMPT+Apogee \\
56066.9 & -1576.5 & $\cdots$ & $\cdots$& $\cdots$  & $\cdots$ & >19.1 & $\cdots$ & $\cdots$ & $\cdots$ & PROMPT+Apogee \\
56081.7 & -1561.7 & $\cdots$ & $\cdots$& $\cdots$  & $\cdots$ & >18.2 & $\cdots$ & $\cdots$ & $\cdots$ & PROMPT+Apogee \\
56113.8 & -1529.6 & $\cdots$ & $\cdots$& $\cdots$  & $\cdots$ & >18.0 & $\cdots$ & $\cdots$ & $\cdots$ & PROMPT+Apogee \\
56126.6 & -1516.8 & $\cdots$ & $\cdots$& $\cdots$  & $\cdots$ & >18.9 & $\cdots$ & $\cdots$ & $\cdots$ & PROMPT+Apogee \\
56151.7 & -1491.7 & $\cdots$ & $\cdots$& $\cdots$  & $\cdots$ & >19.2 & $\cdots$ & $\cdots$ & $\cdots$ & PROMPT+Apogee \\
56166.7 & -1476.7 & $\cdots$ & $\cdots$& $\cdots$  & $\cdots$ & >19.2 & $\cdots$ & $\cdots$ & $\cdots$ & PROMPT+Apogee \\
56172.7 & -1470.7 & $\cdots$ & $\cdots$& $\cdots$  & $\cdots$ & >19.2 & $\cdots$ & $\cdots$ & $\cdots$ & PROMPT+Apogee \\
56191.8 & -1451.6 & $\cdots$ & $\cdots$& $\cdots$  & $\cdots$ & >19.5 & $\cdots$ & $\cdots$ & $\cdots$ & PROMPT+Apogee \\
56218.7 & -1424.7 & $\cdots$ & $\cdots$& $\cdots$  & $\cdots$ & >19.4 & $\cdots$ & $\cdots$ & $\cdots$ & PROMPT+Apogee \\
56412.9 & -1230.5 & $\cdots$ & $\cdots$& $\cdots$  & $\cdots$ & >18.4 & $\cdots$ & $\cdots$ & $\cdots$ & PROMPT+Apogee \\
56424.9 & -1218.5 & $\cdots$ & $\cdots$& $\cdots$  & $\cdots$ & 19.28(0.10) & $\cdots$ & $\cdots$ & $\cdots$ & PROMPT+Apogee \\
56468.9 & -1174.5 & $\cdots$ & $\cdots$& $\cdots$  & $\cdots$ & >19.4 & $\cdots$ & $\cdots$ & $\cdots$ & PROMPT+Apogee \\
56486.9 & -1156.5 & $\cdots$ & $\cdots$& $\cdots$  & $\cdots$ & >19.6 & $\cdots$ & $\cdots$ & $\cdots$ & PROMPT+Apogee \\
56497.9 & -1145.5 & $\cdots$ & $\cdots$& $\cdots$  & $\cdots$ & >19.1 & $\cdots$ & $\cdots$ & $\cdots$ & PROMPT+Apogee \\
56642.6 & -1000.8 & $\cdots$ & $\cdots$& $\cdots$  & $\cdots$ & >17.4 & $\cdots$ & $\cdots$ & $\cdots$ & PROMPT+Apogee \\
56643.6 & -999.8 & $\cdots$ & $\cdots$& $\cdots$  & $\cdots$ & >19.4 & $\cdots$ & $\cdots$ & $\cdots$ & PROMPT+Apogee \\
56647.6 & -995.8 & $\cdots$ & $\cdots$& $\cdots$  & $\cdots$ & >19.5 & $\cdots$ & $\cdots$ & $\cdots$ & PROMPT+Apogee \\
56916.7 & -726.7 & $\cdots$ & $\cdots$& 21.30(0.22) & $\cdots$ & $\cdots$ & $\cdots$ & $\cdots$ & $\cdots$ & CTIO4m+DECam \\
56926.6 & -716.8 & $\cdots$ & $\cdots$& $\cdots$  & $\cdots$ & $\cdots$ & 20.79(0.07) & $\cdots$ & $\cdots$ & CTIO4m+DECam \\
56927.6 & -715.8 & $\cdots$ & $\cdots$& 21.15(0.06)  & $\cdots$ & $\cdots$ & $\cdots$ & $\cdots$ & $\cdots$ & CTIO4m+DECam \\
56953.6 & -689.8 & $\cdots$ & $\cdots$& $\cdots$  & $\cdots$ & 20.33(0.05) & $\cdots$ & $\cdots$ & $\cdots$ & CTIO4m+DECam \\
56957.7 & -685.7 & $\cdots$ & $\cdots$& $\cdots$  & $\cdots$ & 20.43(0.33) & $\cdots$ & $\cdots$ & $\cdots$ & CTIO4m+DECam \\
56958.6 & -684.8 & $\cdots$ & $\cdots$& $\cdots$  & $\cdots$ & $\cdots$ & 21.05(0.05) & $\cdots$ & $\cdots$ & CTIO4m+DECam \\
56960.6 & -682.8 & $\cdots$ & $\cdots$& $\cdots$  & $\cdots$ & $\cdots$ & $\cdots$ & 20.89(0.12) & $\cdots$ & CTIO4m+DECam \\
56961.6 & -681.8 & $\cdots$ & $\cdots$& $\cdots$  & $\cdots$ & $\cdots$ & $\cdots$ & 21.02(0.27) & $\cdots$ & CTIO4m+DECam \\
57180.9 & -462.5 & $\cdots$ & $\cdots$& $\cdots$  & $\cdots$ & 20.18(0.25) & $\cdots$ & $\cdots$ & $\cdots$ & CTIO4m+DECam \\
57222.8 & -420.6 & $\cdots$& $\cdots$ & 20.72(0.46) & $\cdots$ & $\cdots$ & $\cdots$ & $\cdots$ & $\cdots$ & CTIO4m+DECam \\
57536.9 & -106.5 & $\cdots$ & $\cdots$& $\cdots$  & >17.0 & $\cdots$ & $\cdots$ & $\cdots$ & $\cdots$ & ASAS-SN \\
57545.8 & -97.6 & $\cdots$ & $\cdots$& $\cdots$  & >17.5 & $\cdots$ & $\cdots$ & $\cdots$ & $\cdots$ & ASAS-SN \\
57547.8 & -95.6 & $\cdots$ & $\cdots$& $\cdots$  & >17.5 & $\cdots$ & $\cdots$ & $\cdots$ & $\cdots$ & ASAS-SN \\
57555.7 & -87.7 & $\cdots$ & $\cdots$& $\cdots$  & >17.0 & $\cdots$ & $\cdots$ & $\cdots$ & $\cdots$ & ASAS-SN \\
57557.5 & -85.6 &  $\cdots$	&	18.39(0.29)	&  $\cdots$		&	17.90(0.38)	&	17.73(0.20)	&	17.94(0.34)	&  $\cdots$		&  $\cdots$	 &		cpt1m0-10+kb70 \\
57558.7 & -84.7 & $\cdots$ & $\cdots$& $\cdots$  & >16.6 & $\cdots$ & $\cdots$ & $\cdots$ & $\cdots$ & ASAS-SN \\
57559.8 & -83.6 & $\cdots$ & $\cdots$& $\cdots$  & 18.35(0.05) & $\cdots$ & $\cdots$ & $\cdots$ & $\cdots$ & NTT+EFOSC2 \\
57560.8 & -82.6 & $\cdots$ & $\cdots$& $\cdots$  & >16.0 & $\cdots$ & $\cdots$ & $\cdots$ & $\cdots$ & ASAS-SN \\
57565.2 & -78.2 & $\cdots$ & $\cdots$& $\cdots$  & 17.60(0.44) & $\cdots$ & $\cdots$ & $\cdots$ & $\cdots$ & CDK+STL-6303 \\
57569.3 & -73.8 &  $\cdots$ & 	18.12(0.14) & 	$\cdots$ 	& 	17.76(0.19) & 17.49(0.13) & 17.40(0.19)	&  $\cdots$	&  $\cdots$	 & 	coj1m0-03+kb77 \\
57576.8 & -66.6 & $\cdots$ & $\cdots$& $\cdots$  & >17.2 & $\cdots$ & $\cdots$ & $\cdots$ & $\cdots$ & ASAS-SN \\
57595.8 & -47.6 & $\cdots$ & $\cdots$& $\cdots$  & >17.4 & $\cdots$ & $\cdots$ & $\cdots$ & $\cdots$ & ASAS-SN \\
57607.8 & -35.6 & $\cdots$ & $\cdots$& $\cdots$  & >17.5 & $\cdots$ & $\cdots$ & $\cdots$ & $\cdots$ & ASAS-SN \\
57626.7 & -16.7 & $\cdots$ & $\cdots$& $\cdots$  & >17.8 & $\cdots$ & $\cdots$ & $\cdots$ & $\cdots$ & ASAS-SN \\
57631.6 & -11.8 & $\cdots$ & $\cdots$& $\cdots$  & 16.2(0.1) & $\cdots$ & $\cdots$ & $\cdots$ & $\cdots$ & ASAS-SN \\
57632.9 & -10.5 & $\cdots$ & $\cdots$& $\cdots$  & 15.47(0.17) & $\cdots$ & $\cdots$ & $\cdots$ & $\cdots$ & RCOS+STL-6303 \\
57633.0 & -10.4 & $\cdots$ & $\cdots$& $\cdots$  & 15.50(0.14) & $\cdots$ & $\cdots$ & $\cdots$ & $\cdots$ & RCOS+STL-6303 \\
57633.0 & -10.4 & $\cdots$ & $\cdots$& $\cdots$  & 15.48(0.14) & $\cdots$ & $\cdots$ & $\cdots$ & $\cdots$ & RCOS+STL-6303 \\
57634.1 & -9.3 & $\cdots$ & $\cdots$& $\cdots$  & 15.41(0.10) & $\cdots$ & $\cdots$ & $\cdots$ & $\cdots$ & RCOS+STL-6303 \\
57634.2 & -9.2 & $\cdots$ & $\cdots$& $\cdots$  & 15.41(0.12) & $\cdots$ & $\cdots$ & $\cdots$ & $\cdots$ & RCOS+STL-6303 \\
57634.8 & -8.6 & $\cdots$ & $\cdots$& 15.05(0.02)  & $\cdots$ & 15.14(0.02) & 15.38(0.03) & $\cdots$ & $\cdots$ & lsc1m0-05+kb69 \\
57635.1 & -8.3 & $\cdots$ & $\cdots$& $\cdots$  & 15.22(0.07) & $\cdots$ & $\cdots$ & $\cdots$ & $\cdots$ & RCOS+STL-6303 \\
57636.1 & -7.3 & $\cdots$ & $\cdots$& $\cdots$  & 15.09(0.10) & $\cdots$ & $\cdots$ & $\cdots$ & $\cdots$ & RCOS+STL-6303 \\
57637.0 & -6.4 & $\cdots$ & $\cdots$& $\cdots$  & 15.04(0.11) & $\cdots$ & $\cdots$ & $\cdots$ & $\cdots$ & RCOS+STL-6303 \\
57637.2 & -6.2 & $\cdots$ & $\cdots$ & 14.87(0.02) & $\cdots$ & 14.97(0.03) & 15.19(0.03) & $\cdots$ & $\cdots$ & coj1m0-03+kb77 \\
57637.5 & -5.9 & $\cdots$ & $\cdots$& $\cdots$  & 14.96(0.03) & $\cdots$ & $\cdots$ & $\cdots$ & $\cdots$ & NTT+EFOSC2 \\
57637.9 & -5.5 & $\cdots$ & $\cdots$& $\cdots$  & 15.00(0.10) & $\cdots$ & $\cdots$ & $\cdots$ & $\cdots$ & RCOS+STL-6303 \\
57638.2 & -5.2 & 15.24(0.03) & $\cdots$ & 14.83(0.04) & $\cdots$ & 14.93(0.03) & 15.12(0.07) & $\cdots$ & $\cdots$ & coj1m0-03+kb77 \\
57639.2 & -4.2 & $\cdots$ & $\cdots$& $\cdots$  & 14.97(0.09) & $\cdots$ & $\cdots$ & $\cdots$ & $\cdots$ & RCOS+STL-6303 \\
57640.2 & -3.2 & $\cdots$ & $\cdots$& $\cdots$  & 14.96(0.09) & $\cdots$ & $\cdots$ & $\cdots$ & $\cdots$ & RCOS+STL-6303 \\
57641.8 & -1.6 & 14.88(0.03) & $\cdots$ & 14.77(0.01) & $\cdots$ & 14.83(0.02) & 15.03(0.02) & $\cdots$ & $\cdots$ & lsc1m0-05+kb69 \\
57642.7 & -0.7 & $\cdots$ & $\cdots$& $\cdots$  & 14.86(0.01) & $\cdots$ & $\cdots$ & $\cdots$ & $\cdots$ & NTT+EFOSC2 \\
57644.6 & 1.2 & $\cdots$ & $\cdots$& $\cdots$  & $\cdots$ & $\cdots$ & $\cdots$ & $\cdots$ & 15.25(0.03) & CTIO4m+DECam \\
57644.8 & 1.4 & 14.91(0.07) & $\cdots$ & 14.82(0.02) & $\cdots$ & 14.85(0.03) & 15.02(0.03) & $\cdots$ & $\cdots$ & lsc1m0-05+kb69 \\
57645.5 & 2.1 & $\cdots$ & $\cdots$& $\cdots$  & $\cdots$ & $\cdots$ & $\cdots$ & $\cdots$ & 15.26(0.02) & CTIO4m+DECam \\
57647.8 & 4.4 & $\cdots$ & $\cdots$ & 14.95(0.07) & $\cdots$ & 14.90(0.07) & 15.02(0.08) & $\cdots$ & $\cdots$ & lsc1m0-05+kb69 \\
57650.5 & 7.1 & 15.01(0.15) & $\cdots$ & 15.06(0.07) & $\cdots$ & 15.02(0.08) & 15.18(0.09) & $\cdots$ & $\cdots$ & cpt1m0-13+kb76 \\
57650.8 & 7.4 & $\cdots$ & $\cdots$ & 15.13(0.03) & $\cdots$ & 14.98(0.03) & 15.06(0.04) & $\cdots$ & $\cdots$ & lsc1m0-05+kb69 \\
57652.5 & 9.1 & $\cdots$ & $\cdots$& $\cdots$  & 15.06(0.02) & $\cdots$ & $\cdots$ & $\cdots$ & $\cdots$ & NTT+EFOSC2 \\
57653.8 & 10.4 & $\cdots$& $\cdots$  & 15.22(0.03) & $\cdots$ & 15.11(0.03) & 15.20(0.02) & $\cdots$ & $\cdots$ & lsc1m0-05+kb69 \\
57656.5 & 13.1 & 15.53(0.10)& $\cdots$  & 15.43(0.02) & $\cdots$ & 15.18(0.01) & 15.25(0.02) & $\cdots$ & $\cdots$ & cpt1m0-13+kb76 \\
57659.8 & 16.4 & 16.32(0.09) & $\cdots$ & 15.60(0.04) & $\cdots$ & 15.31(0.03) & 15.34(0.03) & $\cdots$ & $\cdots$ & lsc1m0-05+kb69 \\
57664.7 & 21.3 & $\cdots$ & $\cdots$& $\cdots$  & 15.57(0.03) & $\cdots$ & $\cdots$ & $\cdots$ & $\cdots$ & NTT+EFOSC2 \\
57664.7 & 21.3 & 16.81(0.12) & $\cdots$ & 15.84(0.04) & $\cdots$ & 15.50(0.04) & 15.50(0.04) & $\cdots$ & $\cdots$ & lsc1m0-05+kb69 \\
57664.7 & 21.3 & 16.93(0.13) & $\cdots$ & 15.86(0.05) & $\cdots$ & 15.50(0.04) & 15.50(0.03) & $\cdots$ & $\cdots$ & lsc1m0-05+kb69 \\
57664.8 & 21.4 & $\cdots$ & $\cdots$ & 15.87(0.06) & $\cdots$ & 15.51(0.05) & 15.49(0.06) & $\cdots$ & $\cdots$ & lsc1m0-05+kb69 \\
57665.5 & 22.1 & $\cdots$ & $\cdots$ & 15.88(0.07) & $\cdots$ & 15.54(0.05) & 15.53(0.07) & $\cdots$ & $\cdots$ & cpt1m0-13+kb76 \\
57665.6 & 22.2 & $\cdots$ & $\cdots$ & 15.83(0.01) & $\cdots$ & 15.52(0.01) & 15.50(0.01) & $\cdots$ & $\cdots$ & CTIO4m+DECam \\
57667.7 & 24.3 & 16.98(0.08) & $\cdots$ & 15.98(0.02) & $\cdots$ & 15.59(0.02) & 15.60(0.02) & $\cdots$ & $\cdots$ & lsc1m0-05+kb69 \\
57671.3 & 27.9 & $\cdots$ & $\cdots$ & 16.23(0.07) & $\cdots$ & 15.77(0.03) & 15.73(0.06) & $\cdots$ & $\cdots$ & cpt1m0-10+kb70 \\
57671.7 & 28.3 & $\cdots$ & $\cdots$& $\cdots$  & 15.87(0.03) & $\cdots$ & $\cdots$ & $\cdots$ & $\cdots$ & NTT+EFOSC2 \\
57675.7 & 32.3 & 17.29(0.27)& $\cdots$  & 16.53(0.06) & $\cdots$ & 16.01(0.03) & 15.90(0.04) & $\cdots$ & $\cdots$ & lsc1m0-05+kb69 \\
57681.6 & 38.2 & $\cdots$ & $\cdots$ & 16.77(0.02) & $\cdots$ & $\cdots$ & $\cdots$ & $\cdots$ & $\cdots$ & CTIO4m+DECam \\
57681.7 & 38.3 & $\cdots$ & $\cdots$ & 16.82(0.09) & $\cdots$ & 16.20(0.09) & 16.15(0.10) & $\cdots$ & $\cdots$ & lsc1m0-05+kb69 \\
57683.5 & 40.1 & $\cdots$ & $\cdots$& $\cdots$  & $\cdots$ & 16.23(0.14) & 16.21(0.11) & $\cdots$ & $\cdots$ & CTIO4m+DECam \\
57684.6 & 41.2 & $\cdots$& $\cdots$ & $\cdots$ & $\cdots$ & $\cdots$ & $\cdots$ & $\cdots$ & 16.09(0.02) & CTIO4m+DECam \\
57686.5 & 43.1 & 18.12(0.10) & $\cdots$  & 16.97(0.07) & $\cdots$ & 16.27(0.09) & 16.22(0.11) & $\cdots$ & $\cdots$ & cpt1m0-13+fl14 \\
57686.5 & 43.1 & $\cdots$ & $\cdots$& $\cdots$  & $\cdots$ & 16.30(0.06) & $\cdots$ & $\cdots$ & $\cdots$ & CTIO4m+DECam \\
57687.5 & 44.1 & $\cdots$ & $\cdots$& $\cdots$  & $\cdots$ & 16.32(0.04) & $\cdots$ & $\cdots$ & $\cdots$ & CTIO4m+DECam \\
57688.5 & 45.1 & $\cdots$ & $\cdots$& $\cdots$  & $\cdots$ & $\cdots$ & 16.22(0.02) & 16.10(0.01) & $\cdots$ & CTIO4m+DECam \\
57689.7	&	46.3	&	18.53(0.13) &	$\cdots$	&	16.94(0.03) &	$\cdots$	&	16.28(0.01) &	16.27(0.02) &	$\cdots$	&	$\cdots$	& lsc1m0-09+fl03	\\
57690.3 & 46.9 	& 	$\cdots$ & 	17.31(0.08) & 	17.08(0.12) & 	16.72(0.15) & 	16.32(0.06) & 	16.29(0.07)	&  $\cdots$		&  $\cdots$	 & 	cpt1m0-12+fl06 \\
57694.6 & 51.2 & $\cdots$ & $\cdots$& $\cdots$  & 16.79(0.09) & $\cdots$ & $\cdots$ & $\cdots$ & $\cdots$ & NTT+EFOSC2 \\
57695.4 & 52.3 	&	$\cdots$ & 	17.71(0.14) & 	17.17(0.06) & 	16.79(0.07) & 	16.36(0.05) & 	16.37(0.06)	&  $\cdots$		&  $\cdots$	 & 	cpt1m0-12+fl06 \\
57696.4	&	53.0	&	18.71(0.11) &	$\cdots$	&	17.09(0.05) &	$\cdots$	&	16.31(0.03) &	16.23(0.10) &	$\cdots$	&	$\cdots$ &  cpt1m0-12+fl06	\\
57698.4 & 55.0 & 18.67(0.16) & $\cdots$  & 17.38(0.22) & $\cdots$ & 16.39(0.06) & 16.34(0.03) & $\cdots$ & $\cdots$ & cpt1m0-10+fl16 \\
57698.7 & 55.6 	&  $\cdots$ & 	17.90(0.19) & 	17.30(0.17) & 	16.86(0.08) & 	16.40(0.08) & 	16.42(0.06)	&  $\cdots$		&  $\cdots$	 & 	lsc1m0-04+fl04 \\
57702.0	&	58.6	& 18.78(0.13) &	$\cdots$	&	17.16(0.08) &	$\cdots$	&	16.32(0.04) &	16.28(0.04) &	$\cdots$	&	$\cdots$	& coj1m0-03+fl11	\\
57703.5 & 60.1 & $\cdots$ & $\cdots$& $\cdots$  & $\cdots$ & $\cdots$ & $\cdots$ & $\cdots$ & 16.26(0.01) & CTIO4m+DECam \\
57704.7 & 61.6 	&  $\cdots$ & 	$\cdots$ & 	$\cdots$		& 	$\cdots$ 	& 	16.41(0.04) & 	16.40(0.04)	&  $\cdots$		&  $\cdots$	 & 	lsc1m0-05+fl15 \\
57705.4 & 62.3 	&  $\cdots$ &  17.92(0.17) & 17.34(0.12) & 16.84(0.13) & 16.42(0.08) & 16.43(0.08)	&  $\cdots$	&  $\cdots$	 & 	cpt1m0-10+fl16 \\
57708.0 & 64.6 & $\cdots$ & $\cdots$ & 17.45(0.22) & $\cdots$ & 16.39(0.04) & 16.41(0.04) & $\cdots$ & $\cdots$ & coj1m0-03+fl11 \\
57710.5 & 67.1 & $\cdots$ & $\cdots$& $\cdots$  & 16.92(0.02) & $\cdots$ & $\cdots$ & $\cdots$ & $\cdots$ & NTT+EFOSC2 \\
57710.6 & 67.5 	&	$\cdots$ & 	18.24(0.23) & 	17.58(0.12) & 	17.22(0.10) & 	16.53(0.09) & 	16.59(0.08)	&  $\cdots$		&  $\cdots$	 & 	lsc1m0-09+fl03 \\
57713.0 & 69.6 & 18.95(0.22) & $\cdots$  & 17.83(0.16) & $\cdots$ & 16.72(0.06) & 16.80(0.07) & $\cdots$ & $\cdots$ & coj1m0-03+fl11 \\
57718.3 & 75.2 	&  $\cdots$ & 18.48(0.20) & 18.07(0.22) & 17.82(0.18) & 17.07(0.12) & 17.47(0.16) &  $\cdots$	&  $\cdots$	 & 	cpt1m0-12+fl06 \\
57722.3 & 78.9 & 19.76(0.40) & $\cdots$  & 18.56(0.15) & $\cdots$ & 17.35(0.08) & 17.87(0.04) & $\cdots$ & $\cdots$ & cpt1m0-12+fl06 \\
57725.3 & 81.9 & $\cdots$ & $\cdots$ & 18.54(0.21) & $\cdots$ & 17.45(0.13) & 17.99(0.16) & $\cdots$ & $\cdots$ & cpt1m0-13+fl14 \\
57725.5 & 82.1 & $\cdots$ & $\cdots$& $\cdots$  & 18.39(0.04) & $\cdots$ & $\cdots$ & $\cdots$ & $\cdots$ & NTT+EFOSC2 \\
57728.6 & 85.2 & 19.85(0.16) & $\cdots$ & 18.69(0.23) & $\cdots$ & 17.57(0.08) & 18.07(0.09) & $\cdots$ & $\cdots$ & lsc1m0-05+fl15 \\
57732.9 & 89.8 	&  $\cdots$ &	19.56(0.09) & 	18.99(0.07) & 	18.82(0.08) &  17.62(0.07) & 	18.27(0.09)	&  $\cdots$		&  $\cdots$	 & 	coj1m0-11+fl12 \\
57734.3 & 90.9 & $\cdots$ & $\cdots$ & 18.77(0.36) & $\cdots$ & 17.58(0.13) & 18.25(0.20) & $\cdots$ & $\cdots$ & cpt1m0-10+fl16 \\
57740.6 & 97.2 & $\cdots$  & $\cdots$& 19.07(0.21) & $\cdots$ & 17.73(0.06) & 18.30(0.07) & $\cdots$ & $\cdots$ & lsc1m0-04+fl04 \\
57742.6 & 99.2 & $\cdots$ & $\cdots$ & 18.95(0.19) & $\cdots$ & 17.77(0.10) & 18.42(0.07) & $\cdots$ & $\cdots$ & lsc1m0-05+fl15 \\
57745.9 & 102.5 & $\cdots$& $\cdots$  & 19.10(0.20) & $\cdots$ & 17.68(0.04) & 18.38(0.22) & $\cdots$ & $\cdots$ & coj1m0-03+fl11 \\
57749.9 & 106.5 & $\cdots$ & $\cdots$ & 19.05(0.13) & $\cdots$ & 17.78(0.04) & 18.53(0.07) & $\cdots$ & $\cdots$ & coj1m0-11+fl12 \\
57837.9 & 194.5 & $\cdots$ & $\cdots$& $\cdots$  & $\cdots$ & 18.44(0.21) & 19.29(0.31) & $\cdots$ & $\cdots$ & NTT+EFOSC2 \\
57865.8 & 222.4 & $\cdots$ & $\cdots$ & 20.35(0.10) & $\cdots$ & 18.54(0.02) & 19.84(0.10) & $\cdots$ & $\cdots$ & NTT+EFOSC2 \\
57873.8 & 230.4 & $\cdots$ & $\cdots$ & 20.46(0.23) & 20.10(0.09) & 18.72(0.04) & 19.89(0.11) & $\cdots$ & $\cdots$ & NTT+EFOSC2 \\
57905.8 & 262.4 & $\cdots$ & $\cdots$ & 20.84(0.29) & 20.49(0.25) & 18.93(0.14) & 20.27(0.27) & $\cdots$ & $\cdots$ & NTT+EFOSC2 \\
57982.7 & 339.3 & $\cdots$ & $\cdots$ & 20.99(0.40) & $\cdots$ & 19.49(0.15) & 20.84(0.39) & $\cdots$ & $\cdots$ & CTIO4m+DECam \\
57998.6 & 355.2 & $\cdots$ & $\cdots$& $\cdots$  & $\cdots$ & $\cdots$ & $\cdots$ & 20.46(0.16) & $\cdots$ & CTIO4m+DECam \\
58083.6 & 440.2 & $\cdots$ & $\cdots$& $\cdots$  & $\cdots$ & $\cdots$ & $\cdots$ & $\cdots$ & $\cdots$ & NTT+SOFI \\
58371.7 & 728.3 & $\cdots$ & $\cdots$& $\cdots$  & 21.84(0.24) & 20.42(0.07) & 22.24(0.27) & $\cdots$ & $\cdots$ & NTT+EFOSC2 \\
58468.6 & 825.2 & $\cdots$ & $\cdots$& $\cdots$  & 22.00(0.11) & 20.48(0.05) & 22.50(0.18) & $\cdots$ & $\cdots$ & NTT+EFOSC2 \\
59323.9 & 1680.5 & $\cdots$ & $\cdots$& $\cdots$  & $\cdots$ & 21.33(0.23) & $\cdots$ & $\cdots$ & $\cdots$ & NTT+EFOSC2 \\
\end{longtable}
\end{small}

\begin{table*}[hbt!]
\centering
\caption{Photometry table of SN 2016cvk for near-infrared bands with errors given in brackets.}
\label{tab:photomJHK}
\begin{tabular}{cccccc}
\hline
\hline
JD & $t$ & $m_{J}$ & $m_{H}$ & $m_{K}$ & Telescope \\
(-2400000.0)&(d)&(mag)&(mag)&(mag)& + Instrument \\
\hline
57641.6 & -1.7 & 14.65(0.04) & 14.54(0.05) & 14.33(0.04) & NTT+SOFI \\
57656.7 & 13.4 & 14.62(0.04) & 14.48(0.05) & 14.28(0.03) & NTT+SOFI \\
57666.7 & 23.4 & 14.82(0.04) & 14.60(0.05) & 14.39(0.02) & NTT+SOFI \\
57683.7 & 40.4 & 15.18(0.24) & 14.78(0.22) & 14.50(0.22) & NTT+SOFI \\
57684.6 & 41.3 & 15.32(0.05) & $\cdots$ & $\cdots$ & NTT+SOFI \\
57695.7 & 52.4 & 15.37(0.05) & 15.13(0.05) & 14.82(0.03) & NTT+SOFI \\
57724.5 & 81.2 & 16.82(0.04) & 16.70(0.07) & 16.10(0.03) & NTT+SOFI \\
57872.9 & 229.6 & 18.85(0.31) & 19.08(0.12) & 18.20(0.15) & NTT+SOFI \\
58083.6 & 440.3 & 20.03(0.20) & $\cdots$ & $\cdots$ & NTT+SOFI \\
\hline
\end{tabular}
\end{table*}

\begin{table*}[hbt!]
\centering
\caption{Photometry table of SN 2016cvk in Swift UV ($UVW2$, $UVM2$, $UVW1$, $U$, $B$, and $V$) and X-ray (0.2~$-$~10~keV) bands with errors given in brackets.}
\label{tab:xray}
\begin{tabular}{ccccccccc}
\hline
\hline
JD & $t$ & $m_{UVW2}$ & $m_{UVM2}$ & $m_{UVW1}$ & $m_U$ & $m_B$ & $m_V$ & $L_{0.2 - 10~\text{keV}}$ \\
(-2400000.0)&(d)&(mag)&(mag)&(mag)&(mag)&(mag)&(mag)&($10^{40}$erg s$^{-1}$)\\
\hline
57633.0 & -10.4 &	13.99(0.09) & 13.93(0.07) & 14.10(0.07) & 14.39(0.07) & 15.54(0.06) & 15.51(0.07) & $<$5.8  \\
57636.6 & -6.8 & 13.11(0.09) & 13.23(0.07) & 13.28(0.07) & 13.69(0.07) & 14.97(0.06) & 15.02(0.07) & $<$5.3  \\
57640.2 & -3.2 & 13.24(0.09) & 13.26(0.07) & 13.23(0.07) & 13.55(0.07) & 14.85(0.07) & 15.04(0.08) & $<$12.0  \\
57642.1 & -1.3 & 13.41(0.09) & 13.40(0.07) & 13.34(0.07) & 13.64(0.07) & 14.88(0.07) & 14.95(0.07) & $<$8.2  \\
57643.6 & 0.2 & 13.87(0.09) & 13.81(0.08) & 13.64(0.08) & 13.67(0.06) & 15.02(0.06) & 14.83(0.09) & $<$3.5  \\
57646.1 & 2.7 & 14.21(0.09) & 14.17(0.07) & 13.94(0.07) & 13.88(0.05) & 15.09(0.06) & 14.97(0.05) & $<$5.6  \\
57648.3 & 4.9 & 14.34(0.09) & 14.22(0.07) & 14.06(0.07) & 14.02(0.07) & 15.10(0.06) & 14.92(0.07) & $<$6.8  \\
57652.3 & 8.9 & 15.03(0.09) & 14.91(0.08) & 14.65(0.07) & 14.41(0.07) & 15.27(0.07) & 15.06(0.07) & $<$12.9  \\
57655.6 & 12.2 & 15.57(0.10) & 15.49(0.08) & 15.15(0.08) & 14.75(0.08) & 15.53(0.07) & 15.27(0.08) & $<$7.5  \\
57659.9 & 16.5 & 16.20(0.10) & 16.14(0.08) & 15.76(0.08) & 15.17(0.08) & 15.74(0.07) & 15.45(0.08) & $<$4.2  \\
57664.1 & 20.7 & 16.67(0.10) & 16.74(0.09) & 16.17(0.09) & 15.61(0.08) & 15.99(0.07) & 15.58(0.08) & $<$10.5  \\
57667.6 & 24.2 & 17.12(0.11) & 16.98(0.09) & 16.63(0.09) & 15.83(0.08) & 16.19(0.07) & 15.74(0.08) & $<$8.0  \\
57671.8 & 28.4 & 17.46(0.12) & 17.35(0.10) & 16.96(0.10) & 16.21(0.09) & 16.42(0.08) & 15.89(0.08) & $<$8.1  \\
57682.7 & 39.3 & 17.81(0.12) & 17.81(0.11) & 17.49(0.11) & 16.96(0.10) & 16.96(0.08) & 16.33(0.09) & $<$6.6  \\
57689.2 & 45.8 & 18.14(0.14) & 18.19(0.16) & 18.00(0.14) & 17.25(0.12) & 17.29(0.08) & 16.62(0.11) & $<$7.1  \\
\hline
\end{tabular}
\end{table*}

\begin{table*}
\caption{Spectroscopic log of observations for SN 2016cvk}
\centering
\begin{tabular}{cccccccc}
\hline
\hline
JD & Epoch & Grism & Slit & $\lambda / \Delta \lambda$ & $\lambda$ & $t_\text{exp}$ & Telescope + instrument \\
($-2400000$) & (d) & & (") & & (\AA) & (s) & \\
\hline
57557.8 & $-$86 & Blue & 1.65 & 800 & 3606 - 9163 & 900 & duPont+WFCCD \\
57632.8 & $-$11 & Blue & 1.65 & 800 & 3650 - 9200 & 900 & duPont+WFCCD \\
57637.5 & $-6$ & Gr\#11, Gr\#16 & 1.0 & 390, 595 & 3345 - 9995 & 1000, 1000 & NTT+EFOSC2 \\ 
57641.5 & $-2$ & Blue, Red & $1.0$ & 930, 980 & 9500 - 25200 & 960, 1500 & NTT+SOFI \\
57642.7 & $-1$ & Gr\#11, Gr\#16 & 1.0 & 390, 595 & 3345 - 9995 & 1500, 1500 & NTT+EFOSC2 \\ 
57651.5 & +9 & Gr\#11, Gr\#16 & 1.0 & 390, 595 & 3345 - 9995 & 1500, 1500 & NTT+EFOSC2 \\ 
57656.5 & $+13$ & Blue, Red & $1.0$ & 930, 980 & 9500 - 25200 & 960, 1500 & NTT+SOFI \\ 
57664.7 & +21 & Gr\#11, Gr\#16 & 1.0 & 390, 595 & 3345 - 9995 & 900, 900 & NTT+EFOSC2 \\ 
57671.7 & +28 & Gr\#11, Gr\#16 & 1.0 & 390, 595 & 3345 - 9995 & 900, 900 & NTT+EFOSC2 \\
57683.6 & $+40$ & Blue, Red & $1.0$ & 930, 980 & 9500 - 25200 & 3150, 2160 & NTT+SOFI \\
57694.6 & +51 & Gr\#11, Gr\#16 & 1.0 & 390, 595 & 3345 - 9995 & 1500, 1500 & NTT+EFOSC2 \\
57710.5 & +67 & Gr\#11, Gr\#16 & 1.0 & 390, 595 & 3345 - 9995 & 1500, 1500 & NTT+EFOSC2 \\
57725.5 & +82 & Gr\#11, Gr\#16 & 1.0 & 390, 595 & 3345 - 9995 & 1800, 1800 & NTT+EFOSC2 \\
57866.9 & +223 & Gr\#11, Gr\#16 & 1.0 & 390, 595 & 3345 - 9995 & 2700, 2700 & NTT+EFOSC2 \\
58048.6 & +405 & GRIS\_300V+10 & 1.0 & 440 & 4450 - 9050 & 2700, 2700 & VLT-U1+FORS2 \\
\hline
\end{tabular}

\label{tab:specinstr}
\end{table*}

 \begin{table}[]
 \caption{Peak blackbody temperatures and corresponding epochs measured from the event B $r$-band maximum for SN 2009ip-like targets.}
     \centering
     \begin{tabular}{cccc}
     \hline
     Transient & Epoch & $T_{\text{BB, peak}}$ & Reference \\
     & (d) & (K) & \\
     \hline
         SN 2016cvk & $-6$ & 14000 & Sect. \ref{sec:bb} \\
         SN 2016jbu & $-10$ & 15000 & \cite{Brennan2022b} \\
         SN 2009ip & $-10$ & 16500 & \cite{Fraser2013} \\
         SN 2010mc & $-12$ & $>$15800 & \cite{Ofek2013} \\
         LSQ13zm & $-7$ & 12500 & \cite{Tartaglia2016} \\
         SN 2015bh & $-14$ & 18500 & \cite{Thone2017} \\
     \hline
     \end{tabular}
     \label{tab:maxt}
 \end{table}

\newpage
\twocolumn

\begin{table*}
\caption{Results of the blackbody fits to the $UVW2, UVM2, UVW1, u, U, B, g, V, r, i, J, H, K$ band photometry of SN 2016cvk with errors given in brackets.} 
 \centering\begin{tabular}{lcccc}
\hline\hline
Epoch & $L_\text{BB}$ & $T_\text{BB}$ & $R_\text{BB}$ & Data \\ 
(d) & ($10^{41}$ erg s$^{-1}$ cm$^{-2}$) & (K) & ($10^{14}$ cm) &\\ 
\hline
$-8.6$ & $142.0(49.8)$  & $13200(900)$  & $8.1(0.8)$ & Swift + optical \\ 
$-6.2$ & $189.0(75.6)$  & $13800(1100)$  & $8.5(1.0)$ & Swift + optical \\ 
$-5.2$ & $193.0(73.1)$  & $13600(1000)$  & $8.9(1.0)$ & Swift + optical \\ 
$-1.6$ & $198.0(46.9)$  & $13300(600)$  & $9.5(0.7)$ & Swift + optical + near-infrared \\ 
$1.4$ & $148.0(24.9)$  & $11400(400)$  & $11.1(0.6)$ & Swift + optical + near-infrared \\ 
$4.4$ & $130.0(19.3)$  & $10800(300)$  & $11.6(0.6)$ & Swift + optical + near-infrared \\ 
$7.2$ & $108.0(15.0)$  & $10000(200)$  & $12.3(0.6)$ & Swift + optical + near-infrared \\ 
$7.4$ & $107.0(14.5)$  & $9900(200)$  & $12.6(0.6)$ & Swift + optical + near-infrared \\ 
$10.4$ & $86.1(11.3)$  & $9000(200)$  & $13.5(0.7)$ & Swift + optical + near-infrared \\ 
$13.1$ & $72.1(10.2)$  & $8400(200)$  & $14.4(0.8)$ & Swift + optical + near-infrared \\ 
$16.4$ & $58.6(9.5)$  & $7800(200)$  & $15.0(0.9)$ & Swift + optical + near-infrared \\ 
$21.3$ & $45.5(8.3)$  & $7200(200)$  & $15.6(1.1)$ & Swift + optical + near-infrared \\ 
$21.4$ & $44.8(9.1)$  & $7100(200)$  & $15.6(1.3)$ & Swift + optical + near-infrared \\ 
$21.4$ & $44.7(9.1)$  & $7100(200)$  & $15.6(1.3)$ & Swift + optical + near-infrared \\ 
$22.1$ & $43.6(8.6)$  & $7100(200)$  & $15.6(1.2)$ & Swift + optical + near-infrared \\ 
$22.2$ & $43.9(8.6)$  & $7100(200)$  & $15.7(1.2)$ & Swift + optical + near-infrared \\ 
$24.3$ & $40.4(7.7)$  & $6900(200)$  & $15.8(1.2)$ & Swift + optical + near-infrared \\ 
$27.9$ & $34.6(7.2)$  & $6700(200)$  & $15.6(1.3)$ & Swift + optical + near-infrared \\ 
$32.3$ & $28.9(7.4)$  & $6600(200)$  & $14.8(1.5)$ & Swift + optical + near-infrared \\ 
$38.3$ & $22.9(7.7)$  & $6300(300)$  & $14.2(1.9)$ & Swift + optical + near-infrared \\ 
$40.1$ & $21.6(7.6)$  & $6200(300)$  & $14.3(2.1)$ & Swift + optical + near-infrared \\ 
$43.1$ & $19.9(7.0)$  & $6000(300)$  & $14.5(2.1)$ & Swift + optical + near-infrared \\ 
$45.2$ & $19.1(7.0)$  & $5800(300)$  & $15.2(2.3)$ & Swift + optical + near-infrared \\ 
$46.3$ & $18.0(7.1)$  & $5800(300)$  & $14.9(2.4)$ & optical + near-infrared \\ 
$46.9$ & $17.6(6.9)$  & $5800(300)$  & $14.7(2.4)$ & optical + near-infrared \\ 
$52.0$ & $16.9(6.0)$  & $5600(300)$  & $15.4(2.2)$ & optical + near-infrared \\ 
$53.0$ & $17.4(6.2)$  & $5600(300)$  & $15.6(2.2)$ & optical + near-infrared \\ 
$55.0$ & $15.9(5.2)$  & $5600(300)$  & $14.9(1.9)$ & optical + near-infrared \\ 
$55.3$ & $15.6(4.9)$  & $5600(300)$  & $14.8(1.8)$ & optical + near-infrared \\ 
$55.3$ & $15.6(4.9)$  & $5600(300)$  & $14.8(1.8)$ & optical + near-infrared \\ 
$58.6$ & $15.0(4.6)$  & $5600(300)$  & $14.4(1.7)$ & optical + near-infrared \\ 
$61.3$ & $13.5(3.8)$  & $5700(300)$  & $13.6(1.5)$ & optical + near-infrared \\ 
$61.3$ & $13.5(3.8)$  & $5700(300)$  & $13.6(1.5)$ & optical + near-infrared \\ 
$62.0$ & $13.2(3.8)$  & $5700(300)$  & $13.4(1.4)$ & optical + near-infrared \\ 
$62.0$ & $13.2(3.7)$  & $5700(300)$  & $13.4(1.4)$ & optical + near-infrared \\ 
$64.6$ & $12.1(3.5)$  & $5700(300)$  & $12.6(1.4)$ & optical + near-infrared \\ 
$67.3$ & $10.7(2.8)$  & $5800(300)$  & $11.7(1.1)$ & optical + near-infrared \\ 
$67.3$ & $10.7(2.8)$  & $5800(300)$  & $11.7(1.1)$ & optical + near-infrared \\ 
$69.6$ & $9.3(2.3)$  & $5800(200)$  & $11.0(1.0)$ & optical + near-infrared \\ 
$74.9$ & $6.5(1.5)$  & $5500(200)$  & $9.8(0.8)$ & optical + near-infrared \\ 
$74.9$ & $6.5(1.5)$  & $5500(200)$  & $9.8(0.8)$ & optical + near-infrared \\ 
$78.9$ & $4.8(1.2)$  & $5400(200)$  & $9.1(0.8)$ & optical + near-infrared \\ 
$81.9$ & $4.4(1.1)$  & $5400(200)$  & $8.5(0.8)$ & optical + near-infrared \\ 
$85.2$ & $4.0(1.1)$  & $5300(200)$  & $8.3(0.8)$ & optical \\ 
$89.6$ & $3.4(1.1)$  & $5100(300)$  & $8.4(0.9)$ & optical \\ 
$89.6$ & $3.4(1.1)$  & $5100(300)$  & $8.4(0.9)$ & optical \\ 
$90.9$ & $3.5(1.1)$  & $5200(300)$  & $8.2(0.9)$ & optical \\ 
$97.2$ & $3.2(1.0)$  & $5100(300)$  & $8.1(0.9)$ & optical \\ 
$99.2$ & $3.0(1.0)$  & $5200(300)$  & $7.8(0.9)$ & optical \\ 
$102.5$ & $3.0(1.0)$  & $5100(300)$  & $7.8(0.9)$ & optical \\ 
$106.5$ & $2.8(1.0)$  & $5200(300)$  & $7.4(0.9)$ & optical \\ 
\hline\end{tabular}\label{tab:bb}\end{table*}

\begin{figure*}
\begin{minipage}{0.5\linewidth}
\includegraphics[width=\linewidth]{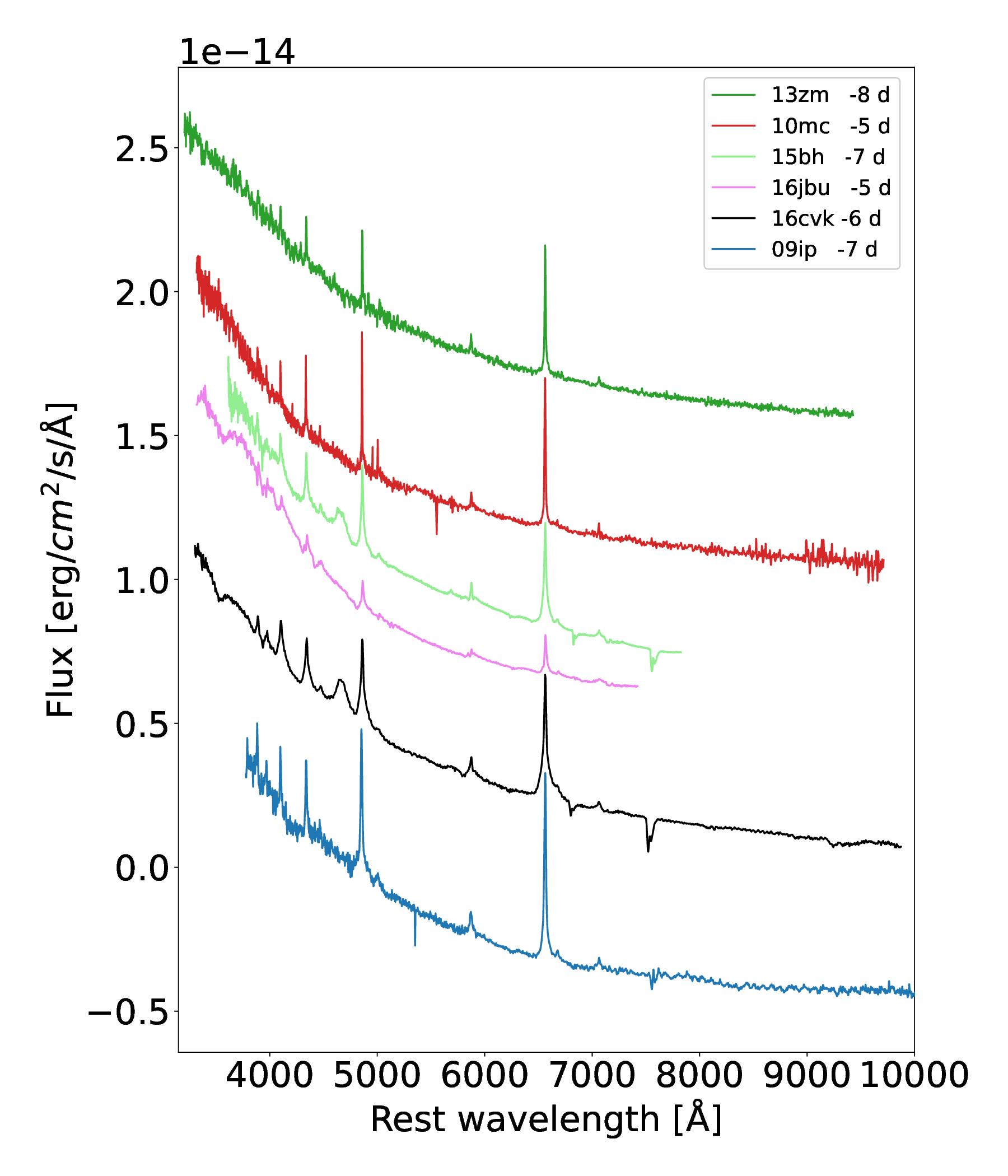}

\includegraphics[width=\linewidth]{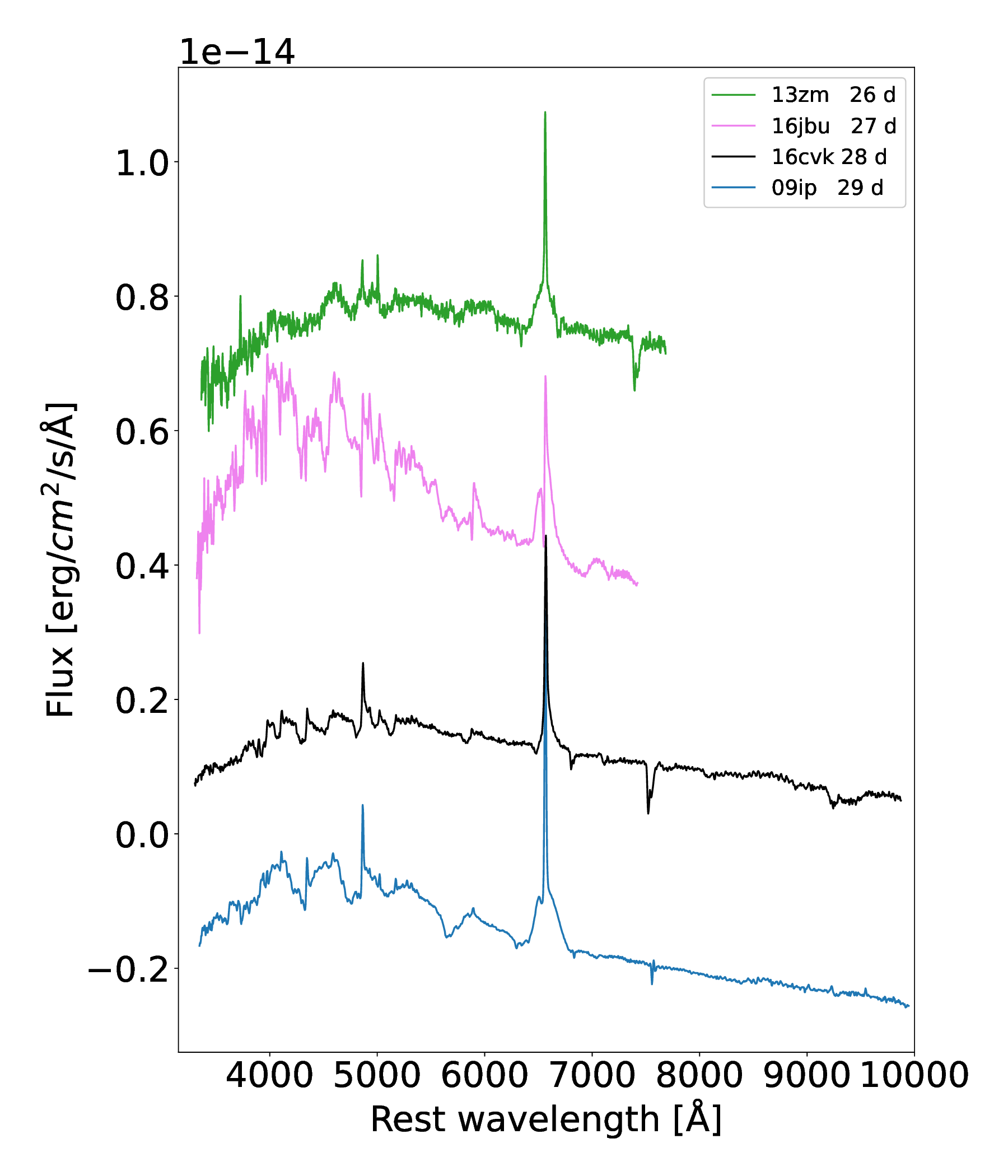}
\end{minipage}\begin{minipage}{0.5\linewidth}
\includegraphics[width=\linewidth]{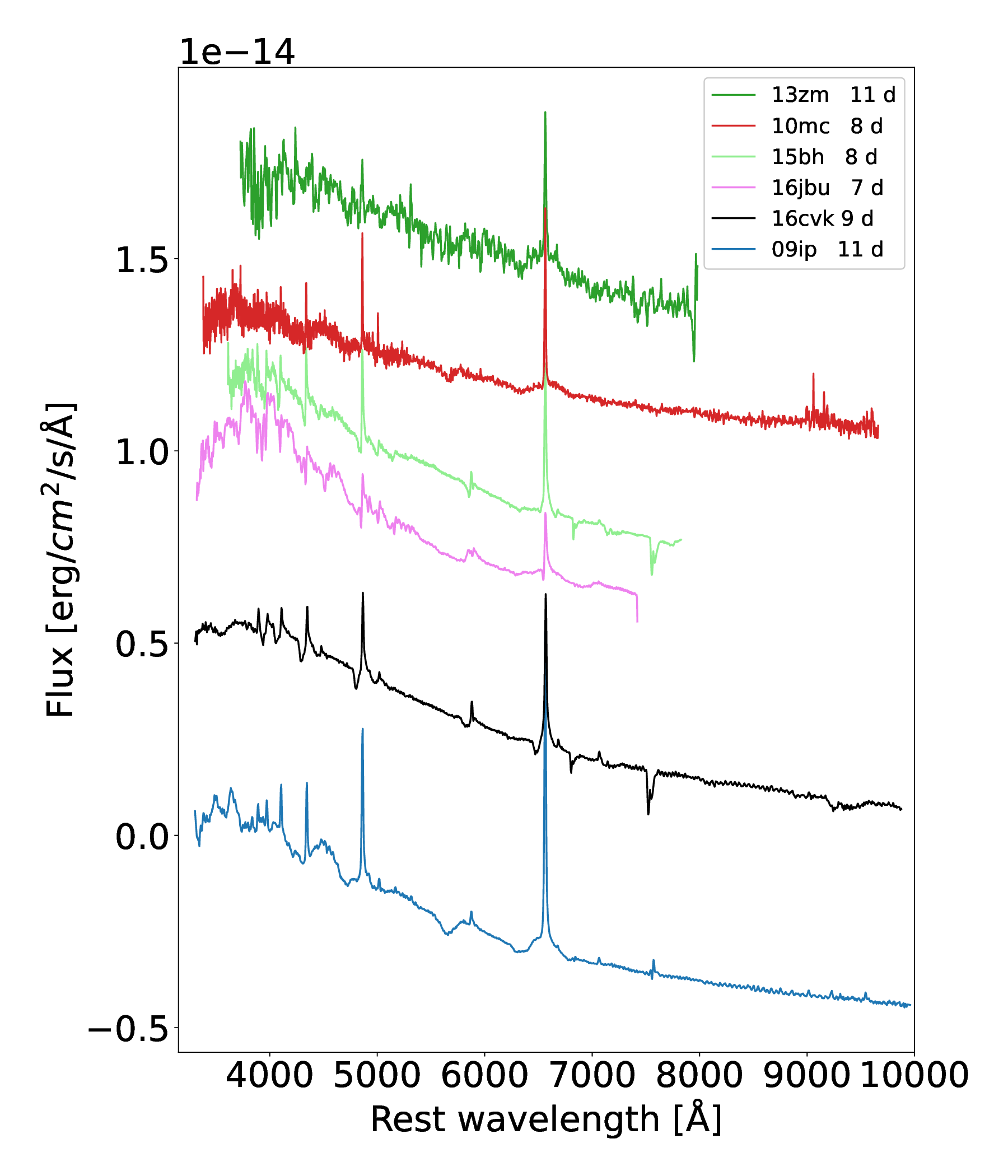}

\includegraphics[width=\linewidth]{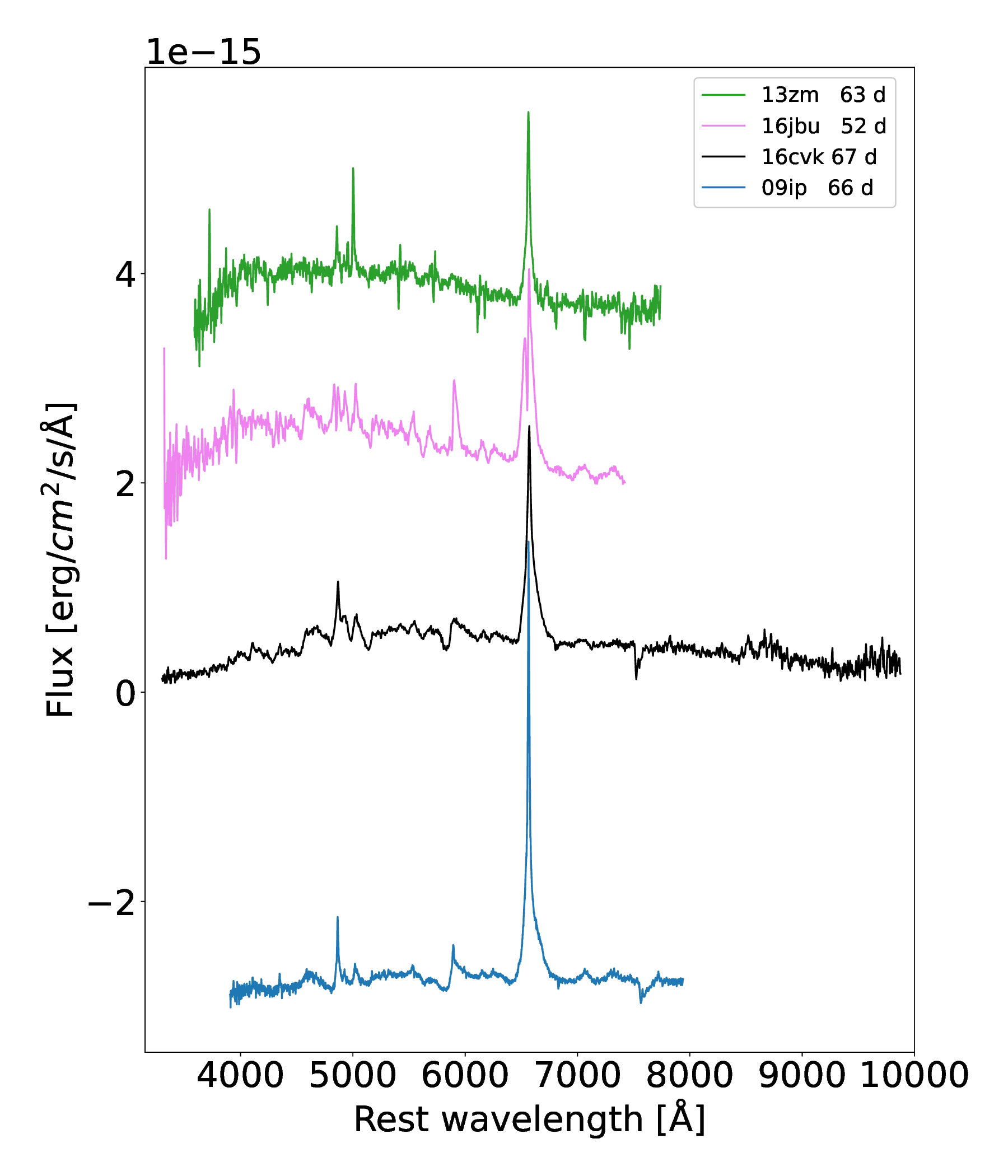}
\end{minipage}

\caption{Selected spectra of SN 2016cvk at $-6$, $+9$, $+28$, and $+67$~d from the $r$-band maximum compared to those of other SN 2009ip-like events. The spectra have been dereddened, corrected to the rest frame wavelengths, and rescaled and shifted vertically for clarity.}
\label{fig:all_spectra_allSN}
\end{figure*}

\begin{table*}
\centering
\caption{Parameters of multi-component fit for the H\protect$_\alpha$ line for different Gaussian $(G1-G3)$ and Lorentzian $(L)$ components with the errors given in brackets, along with the Balmer decrement $D = I(H_\alpha)/I(H_\beta)$. }

\begin{tabular}{ccccccccc}
\hline
Epoch & $\text{FWHM}_{\text{L1}}$ & $\text{FWHM}_{\text{G1}}$ & $v_{0, \text{G1}}$ & $\text{FWHM}_{\text{G2}}$ & $v_{0, \text{G2}}$ & $\text{FWHM}_{\text{G3}}$ & $v_{0,\text{G3}}$  & $I(H_\alpha)/I(H_\beta)$  \\
$(\text{d})$ & $(\text{km} \, \text{s}^{-1})$ & $(\text{km} \, \text{s}^{-1})$ & $(\text{km} \, \text{s}^{-1})$ & $(\text{km} \, \text{s}^{-1})$ & $(\text{km} \, \text{s}^{-1})$ & $(\text{km} \, \text{s}^{-1})$ & $(\text{km} \, \text{s}^{-1})$ & \\\hline
 $-86$  & $850(30)$  & $4700(200)$  & $250(100)$  & $-$ & $-$ & $3600(400)$ & $-6500(200)$ & $2.4 (0.1)$ \\
 $-11$  & $550(30)$  & $6300(200)$  & $300(100)$  & $-$ & $-$ & $-$ & $-$ & $1.7 (0.1)$ \\
 $-6$  & $720(30)$  & $8600(300)$  & $1400(100)$  & $-$ & $-$ & $-$ & $-$ & $1.5 (0.1)$ \\
$-1$  & $540(30)$  & $4400(200)$  & $300(100)$  & $-$  & $-$  & $-$  & $-$  & $1.2 (0.1)$  \\
 $9$  & $600(20)$ & $3200(200)$  & $1200(200)$  & $-$  & $-$ & $1800(100)$  & $-4000(100)$  & $1.8 (0.1)$  \\
 $21$  & $750(30)$  & $3200(500)$  & $1900(300)$& $-$ & $-$ & $4300(600)$  & $-2900(300)$ & $2.6 (0.1)$  \\
 $28$  & $830(30)$  & $4400(300)$  & $2200(200)$ & $-$ & $-$ & $1700(200)$  & $-4100(100)$  & $3.4 (0.1)$ \\
 $51$  & $1200(100)$  & $6300(100)$  & $1100(100)$ & $-$ & $-$ & $-$ & $-$  & $3.6 (0.1)$ \\
 $67$  & $1100(100)$  & $6500(200)$  & $900(100)$ & $-$ & $-$ & $-$ & $-$  & $3.8 (0.1)$ \\
 $82$  & $1200(100)$  & $4200(200)$  & $-2000(100)$ & $5200(200)$  & $2100(200)$  & $-$ & $-$ & $5.0 (0.1)$  \\
 $223$  & $1200(100)$ & $4500(100)$  & $-100(10)$ & $-$ & $-$ & $-$ & $-$  & $10.1 (0.1)$ \\
\hline
\end{tabular}
\label{tab:MCMCHalpha}
\end{table*}

\begin{table*}
\centering
\caption{Parameters of multi-component fit for the H\protect$_\beta$ line for different Gaussian $(G1-G3)$ and Lorentzian $(L)$ components with the errors given in brackets.}
\begin{tabular}{cccccccccc}
\hline
Epoch & $\text{FWHM}_{\text{L1}}$ & $\text{FWHM}_{\text{G1}}$ & $v_{0, \text{G1}}$ & $\text{FWHM}_{\text{G2}}$ & $v_{0, \text{G2}}$ & $\text{FWHM}_{\text{G3}}$ & $v_{0,\text{G3}}$ \\
$(\text{d})$ & $(\text{km} \, \text{s}^{-1})$ & $(\text{km} \, \text{s}^{-1})$ & $(\text{km} \, \text{s}^{-1})$ & $(\text{km} \, \text{s}^{-1})$ & $(\text{km} \, \text{s}^{-1})$ & $(\text{km} \, \text{s}^{-1})$ & $(\text{km} \, \text{s}^{-1})$ \\\hline
  $-86$  & $830(70)$ & $3200(200)$  & $300(100)$  & $-$ & $-$ & $5600(300)$  & $-6500(100)$  \\
  $-11$  & $500(100)$ & $4700(100)$  & $100(100)$  & $-$ & $-$ & $-$  & $-$  \\
  $-6$  & $<800$ & $4100(200)$  & $-200(100)$  & $-$ & $-$ & $-$  & $-$  \\
 $-1$  & $<800$ & $4900(600)$  & $-300(400)$  & $-$ & $-$ &$3000(500)$  & $-3500(300)$  \\
 $9$  & $<800$ & $2100(800)$  & $1400(400)$  & $-$ & $-$ & $2600(200)$  & $-3800(100)$  \\
 $21$  & $<800$  & $2000(300)$  & $2300(100)$  & $-$ & $-$ & $4000(300)$  & $-3000(100)$ \\
 $28$  & $800(100)$ & $2300(300)$  & $1800(200)$  & $-$ & $-$ & $3700(300)$  & $-3000(100)$ \\
 $51$  & $1200(100)$ & $3800(200)$  & $1500(200)$  & $-$ & $-$ & $1500(100)$  & $-4000(100)$  \\
 $67$  & $800(100)$ & $4600(200)$  & $700(100)$  & $-$ & $-$ & $2300(300)$  & $-3800(200)$ \\
 $82$  & $1500(100)$ & $2900(100)$  & $-2300(100)$  & $4300(200)$  & $1700(200)$ & $-$ & $-$ \\
 $223$  & $<800$ & $3700(200)$  & $-100(100)$  & $-$ & $-$ & $-$ & $-$ \\
\hline
\end{tabular}
\label{tab:MCMCHbeta}
\end{table*}

\begin{figure*}
\begin{minipage}{0.26\linewidth}
\includegraphics[trim={2cm 3.3cm 3.5cm 3cm},clip,width=\linewidth]
{{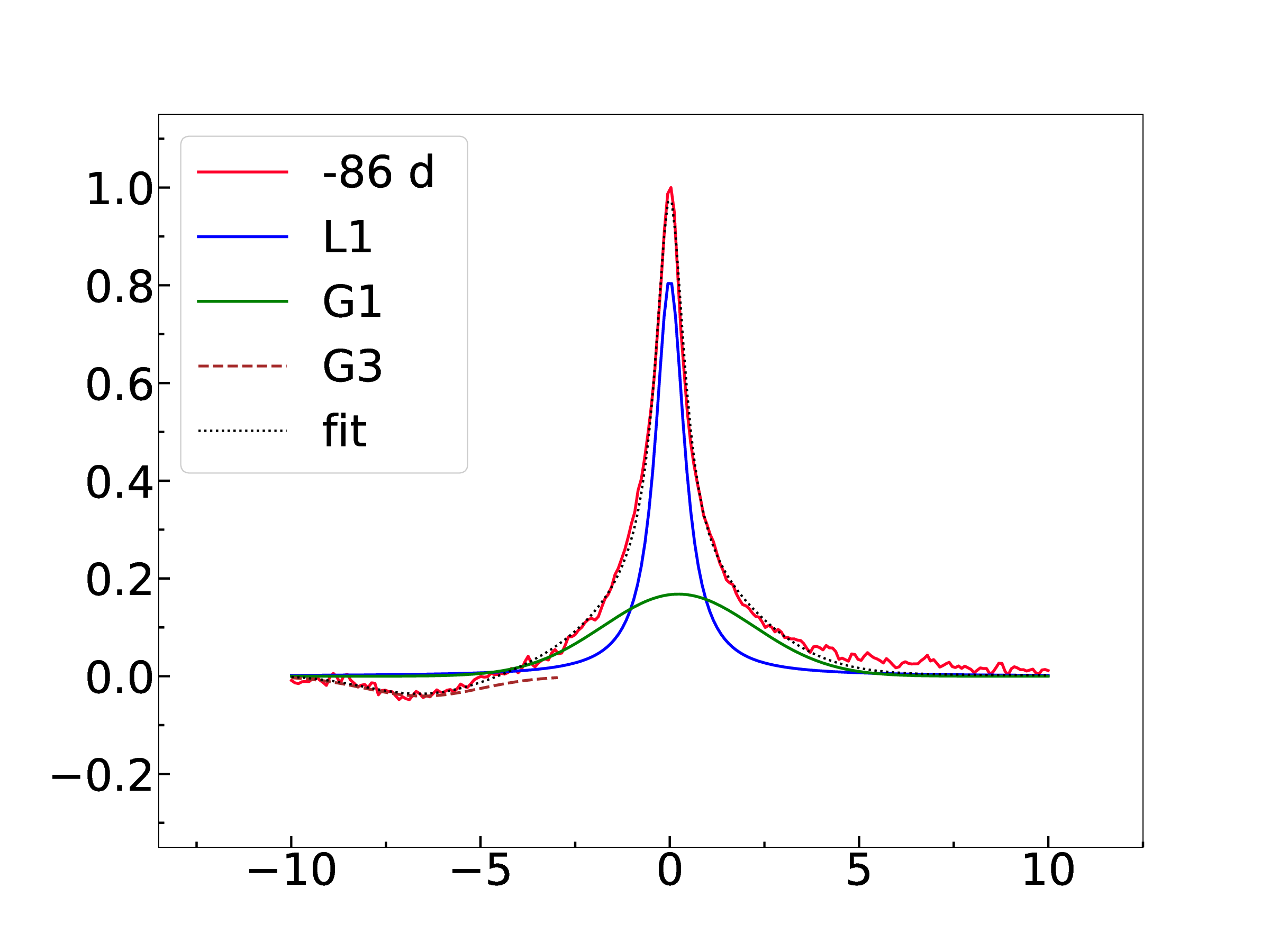}}

\includegraphics[trim={2cm 3.3cm 3.5cm 3cm},clip,width=\linewidth]
{{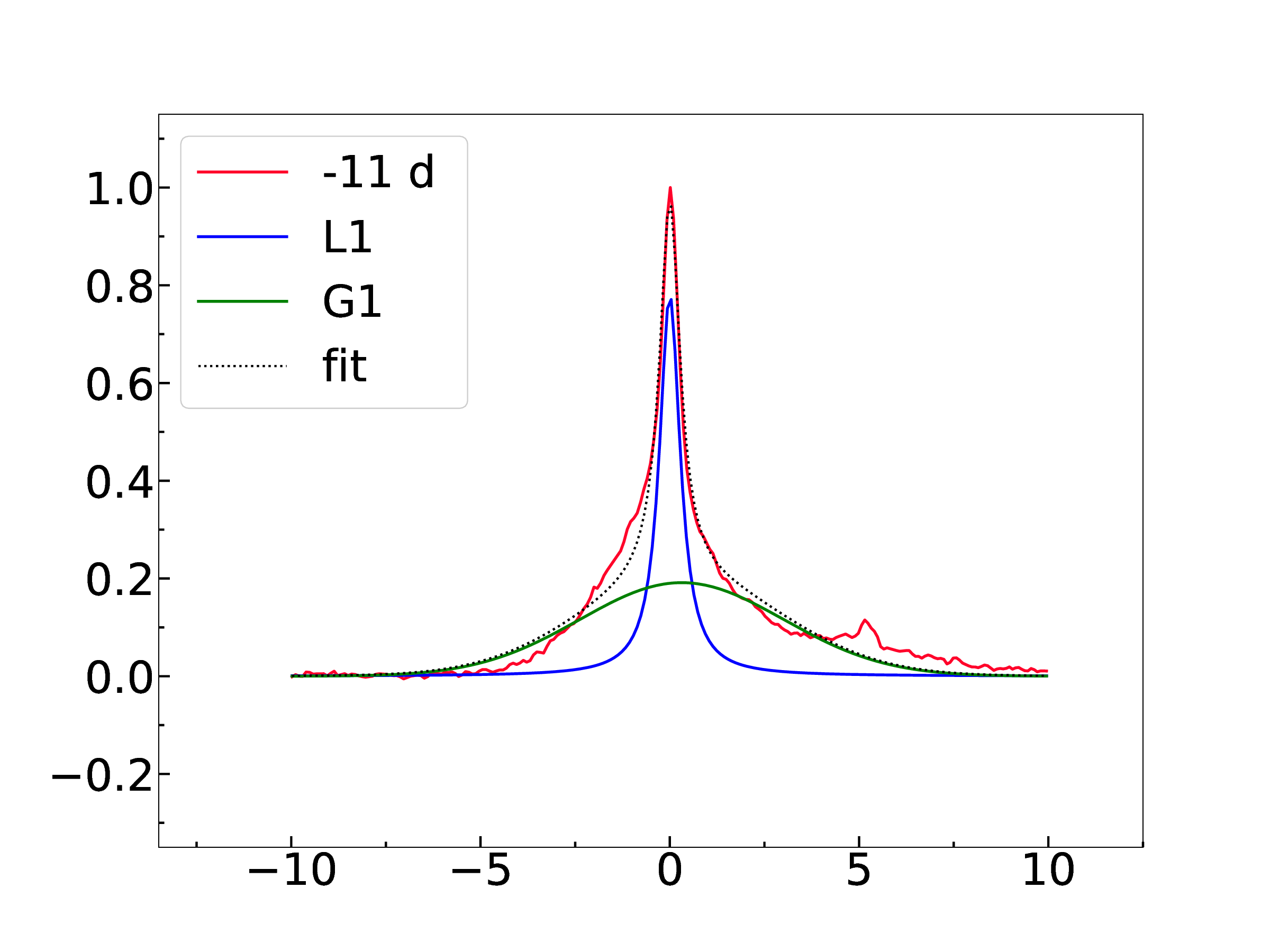}}

\includegraphics[trim={2cm 3.3cm 3.5cm 3cm},clip,width=\linewidth]
{{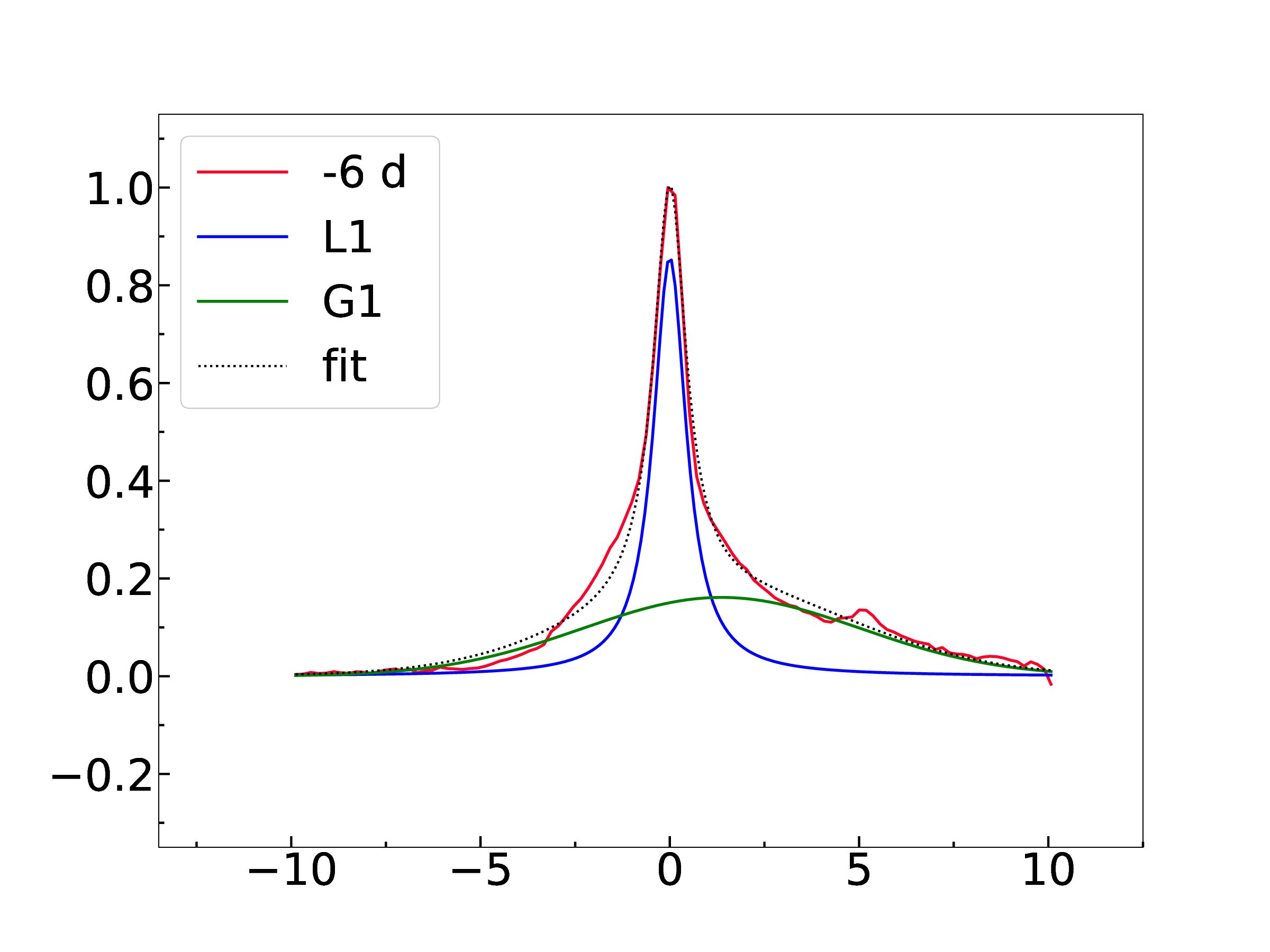}}

\includegraphics[trim={2cm 3.3cm 3.5cm 3cm},clip,width=\linewidth]
{{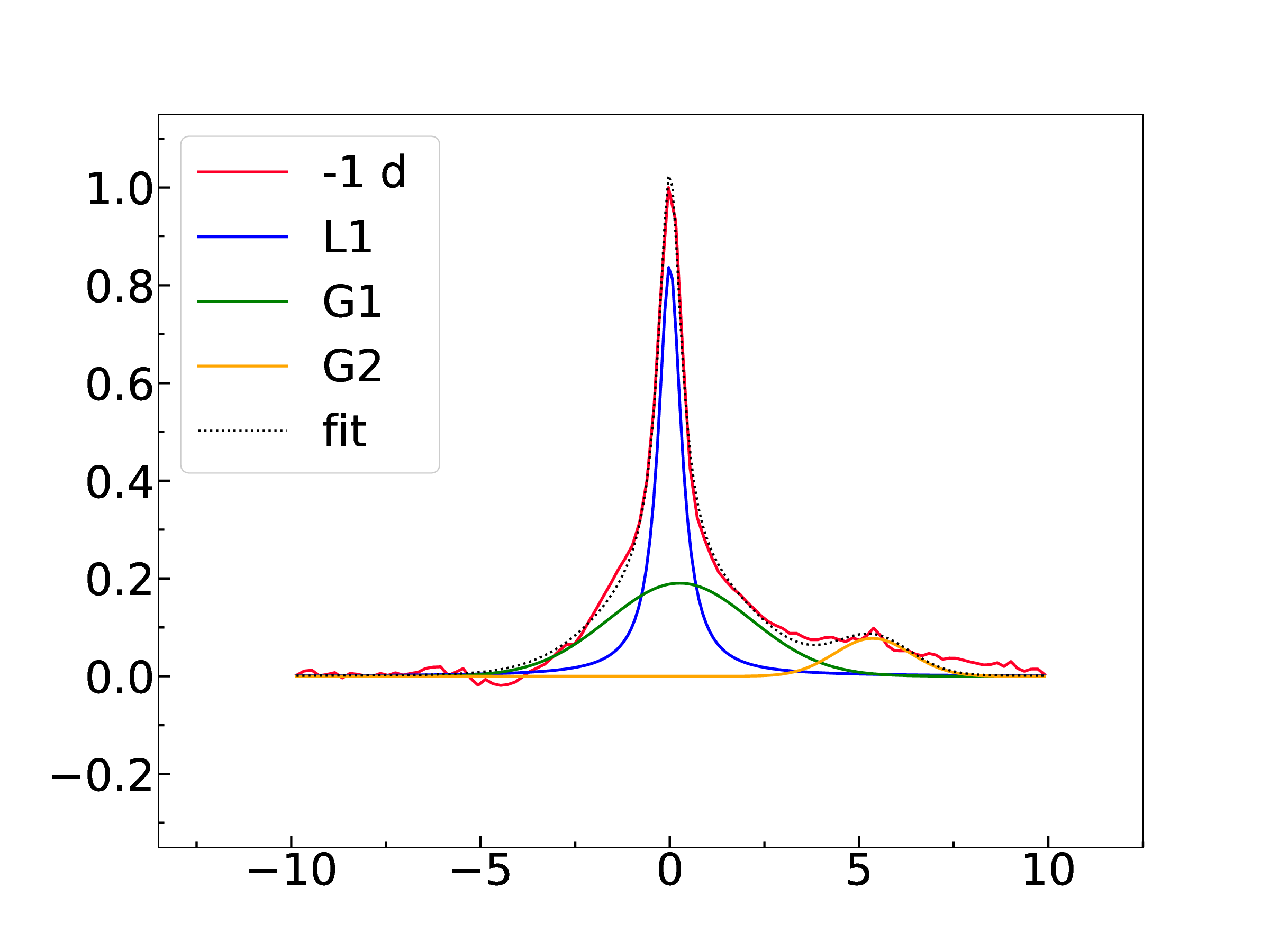}}1

\includegraphics[trim={2cm 3.3cm 3.5cm 3cm},clip,width=\linewidth]
{{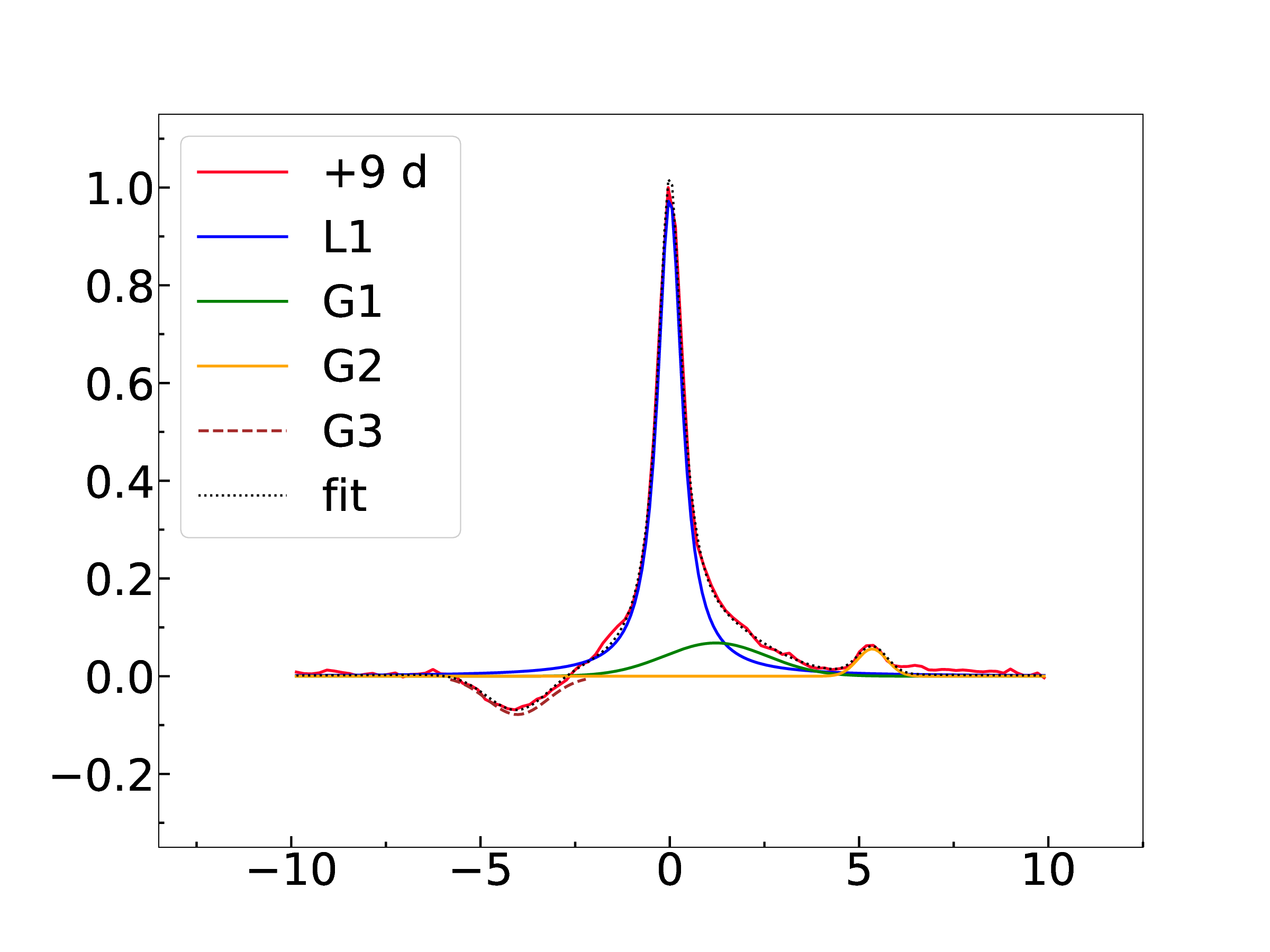}}

\includegraphics[trim={2cm 1.4cm 3.5cm 3cm},clip,width=\linewidth]
{{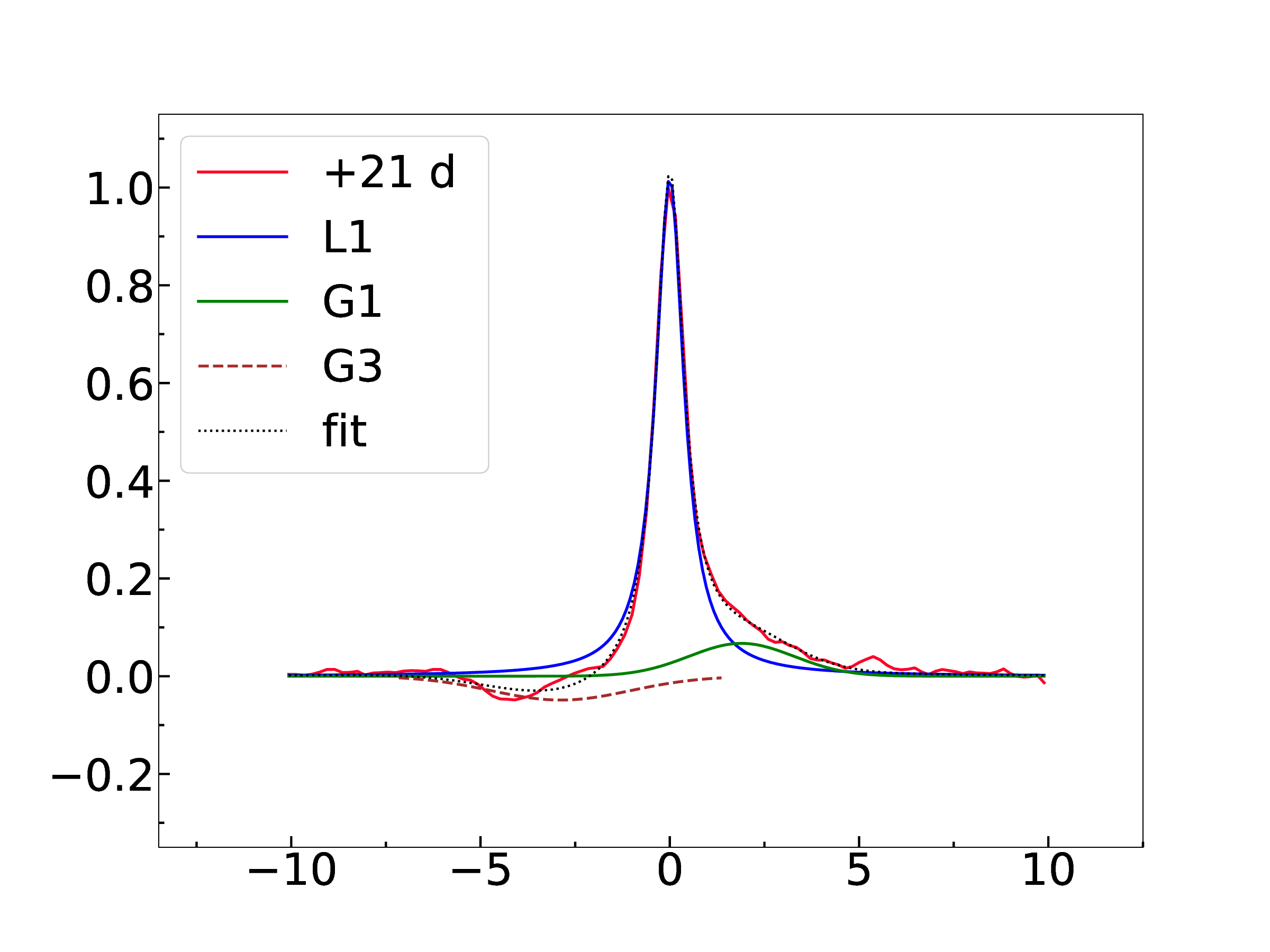}}
\end{minipage}\begin{minipage}{0.238\linewidth}
\includegraphics[trim={4.8cm 3.2cm 3.5cm 3cm},clip,width=\linewidth]
{{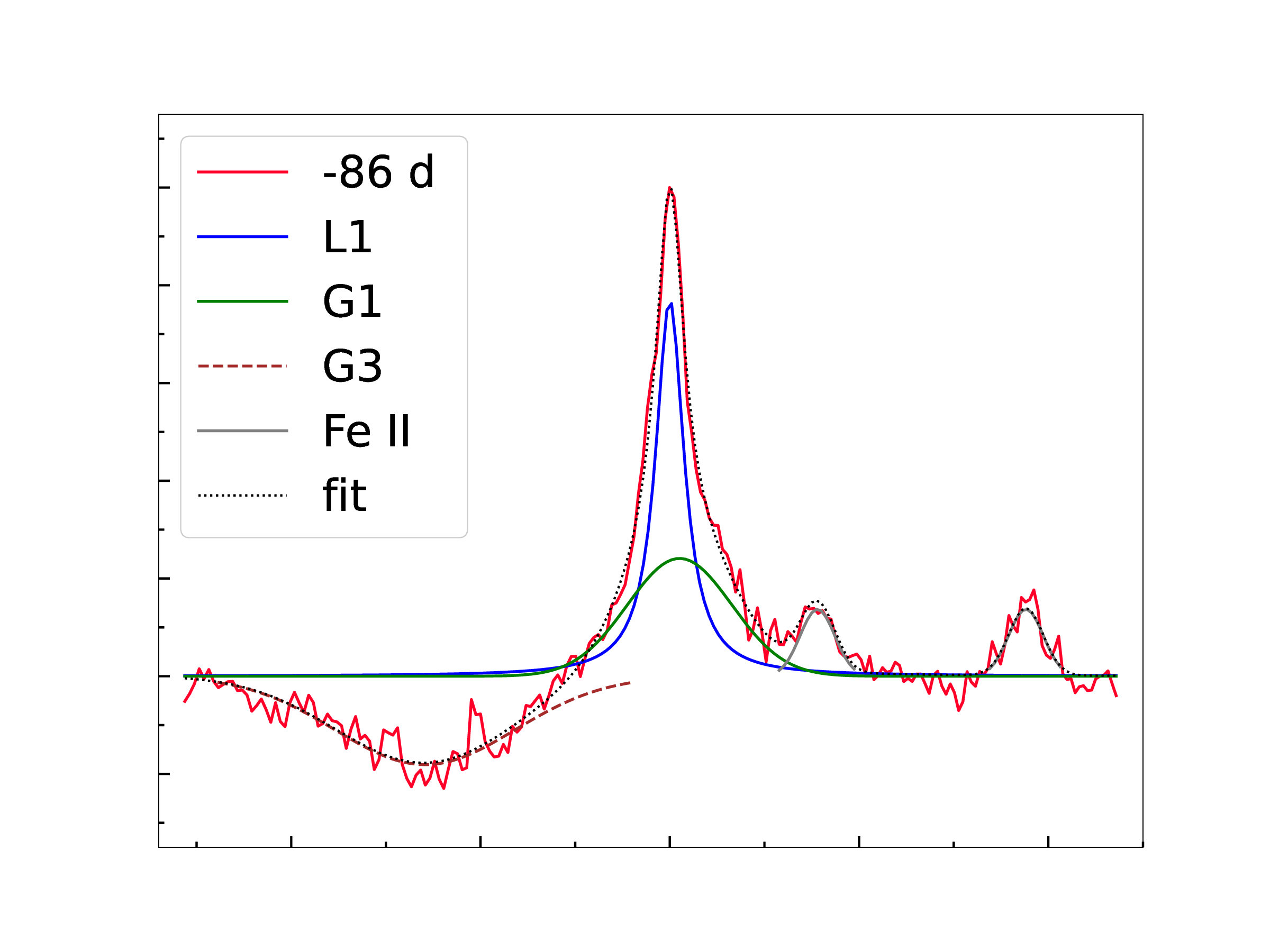}}

\includegraphics[trim={4.8cm 3.2cm 3.5cm 3cm},clip,width=\linewidth]
{{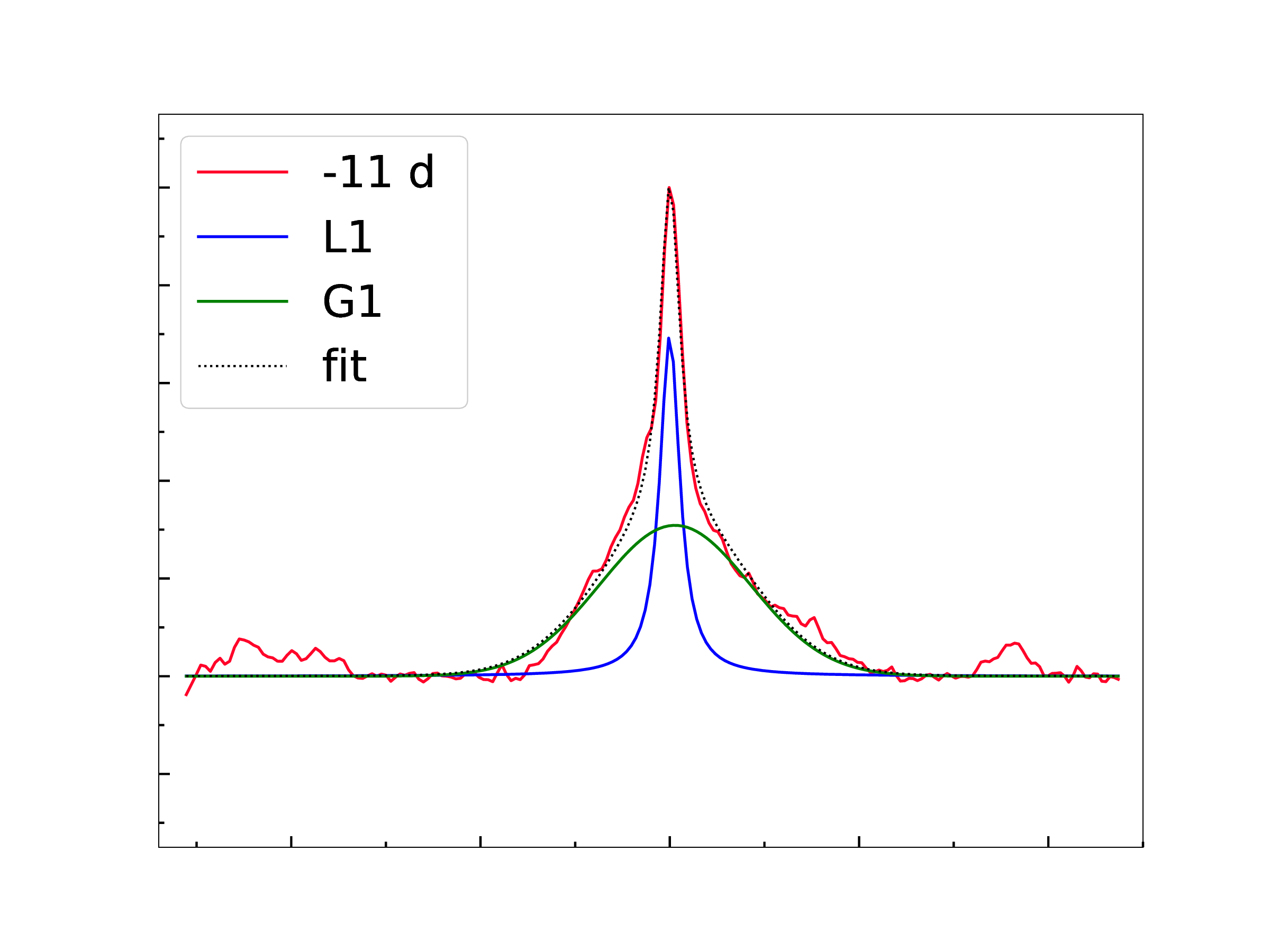}}

\includegraphics[trim={4.8cm 3.2cm 3.5cm 3cm},clip,width=\linewidth]
{{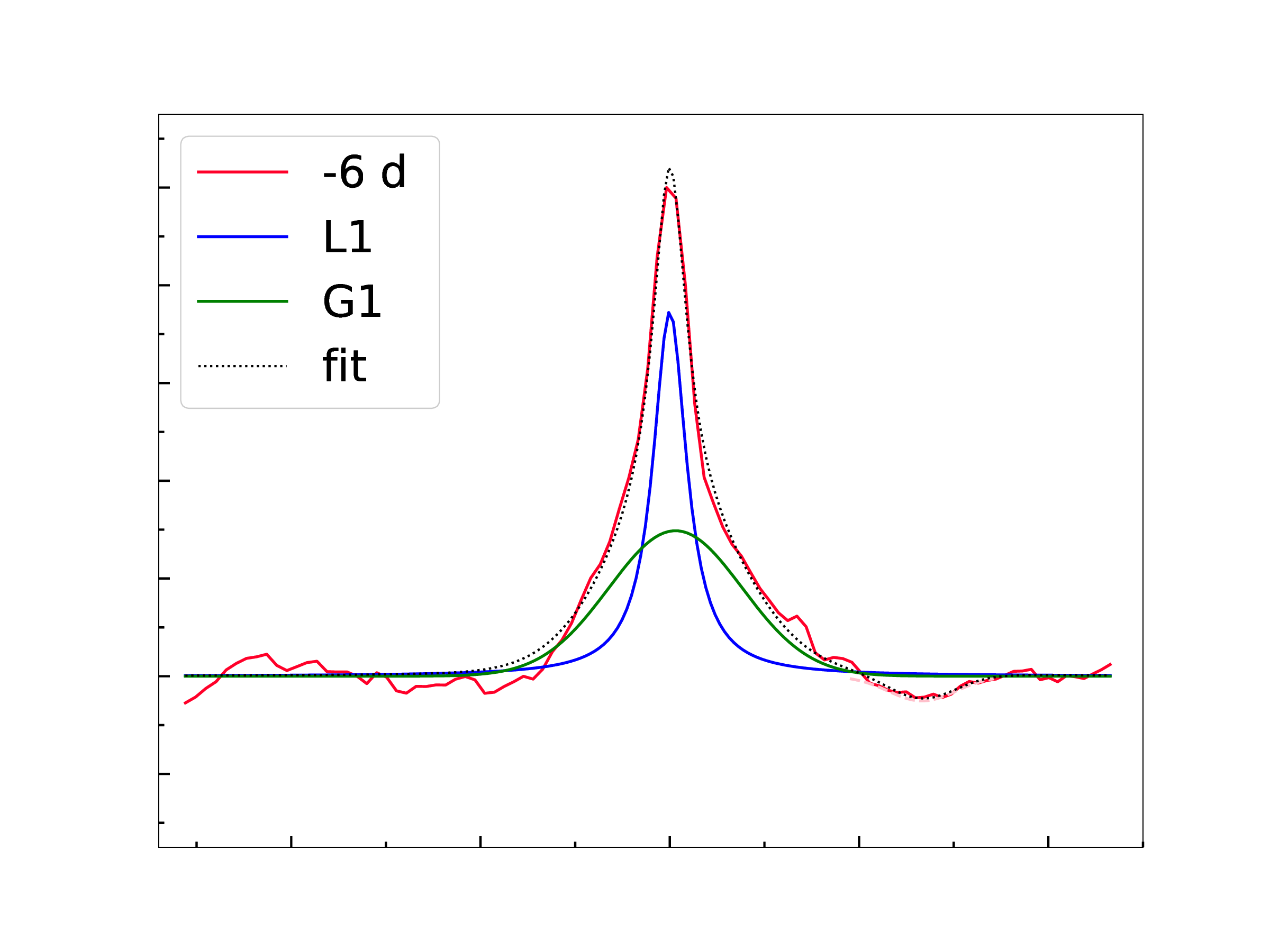}}

\includegraphics[trim={4.8cm 3.2cm 3.5cm 3cm},clip,width=\linewidth]
{{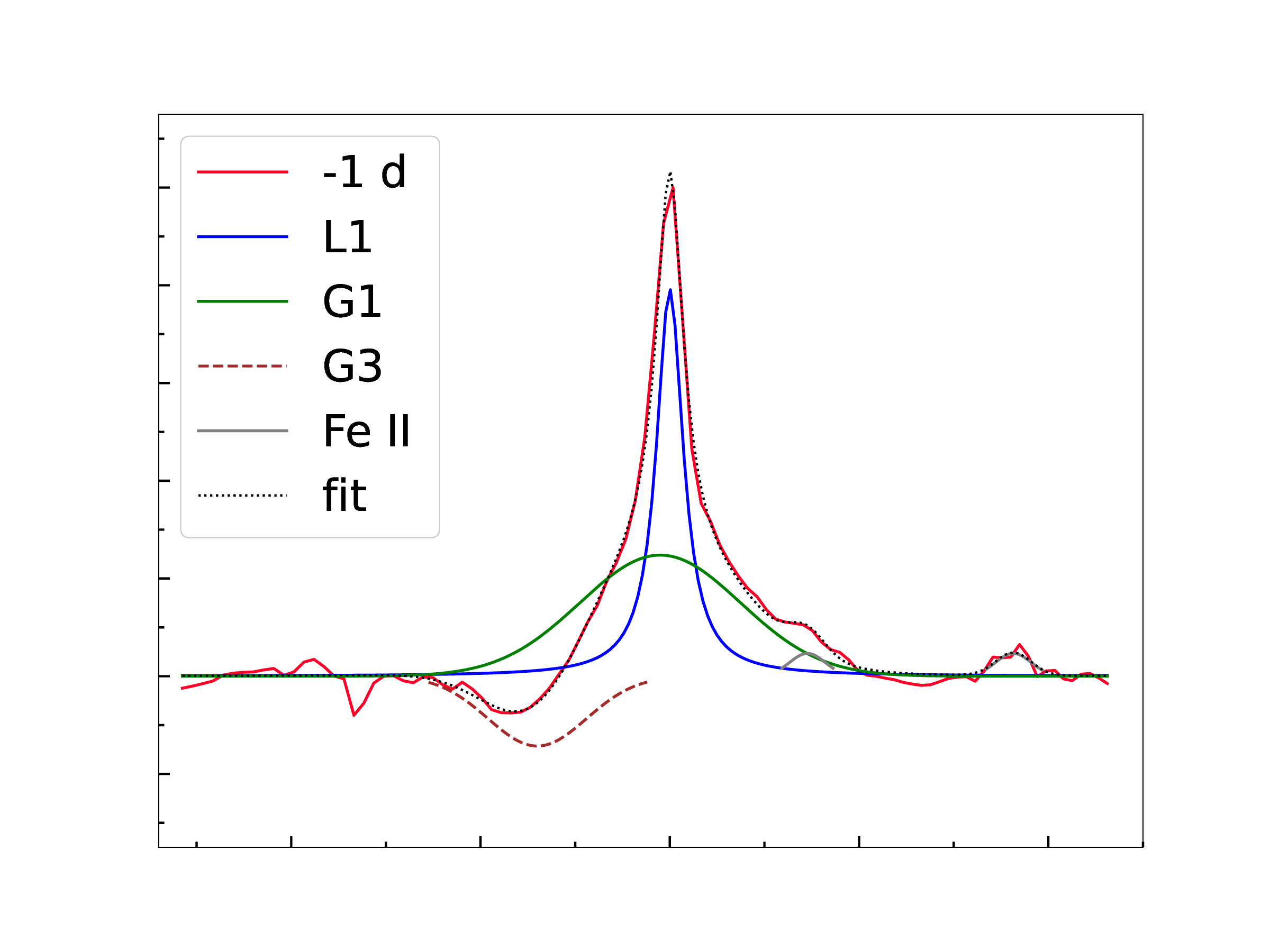}}

\includegraphics[trim={4.8cm 3.2cm 3.5cm 3cm},clip,width=\linewidth]
{{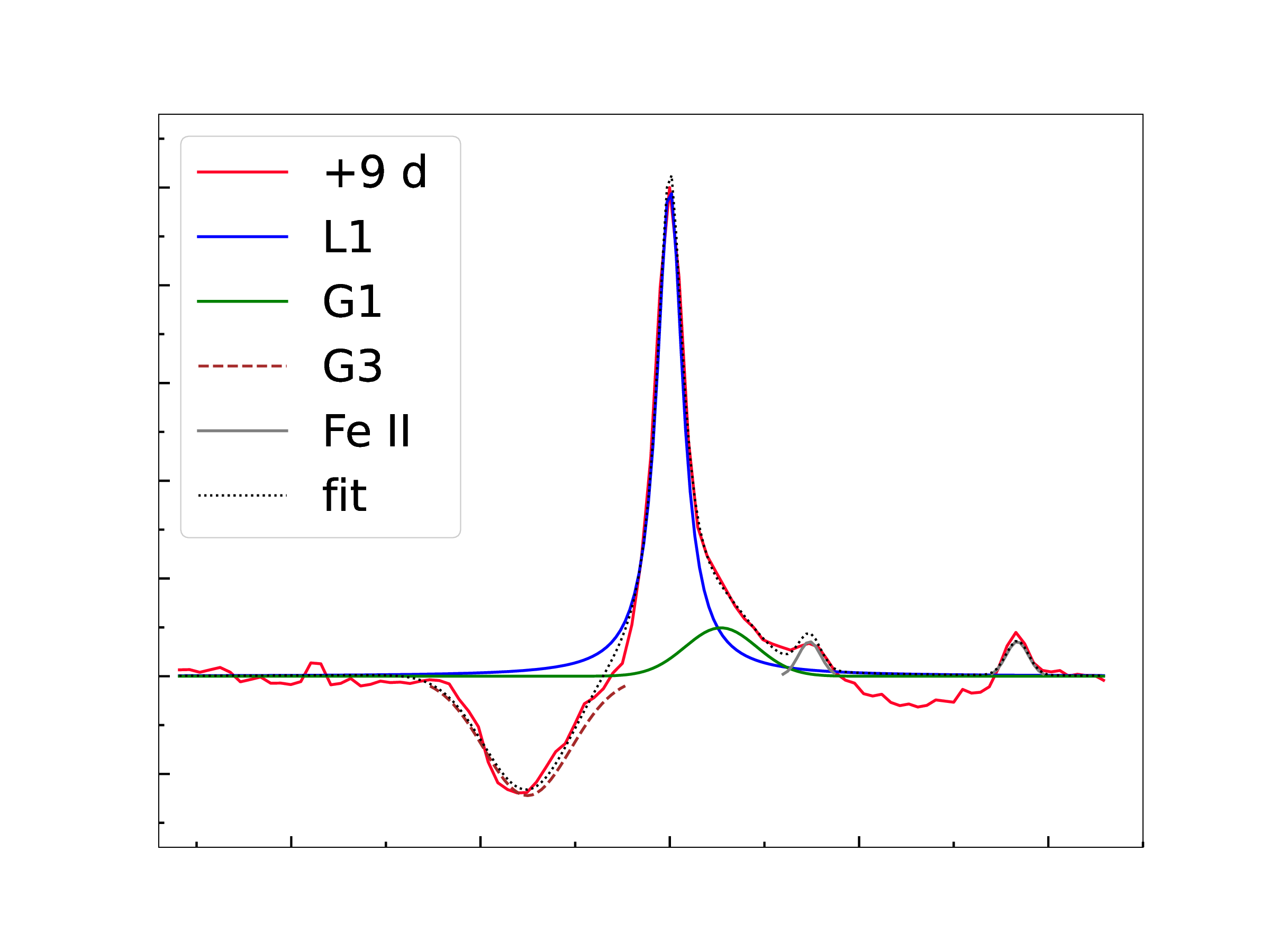}}

\includegraphics[trim={4.8cm 1.4cm 3.5cm 3cm},clip,width=\linewidth]
{{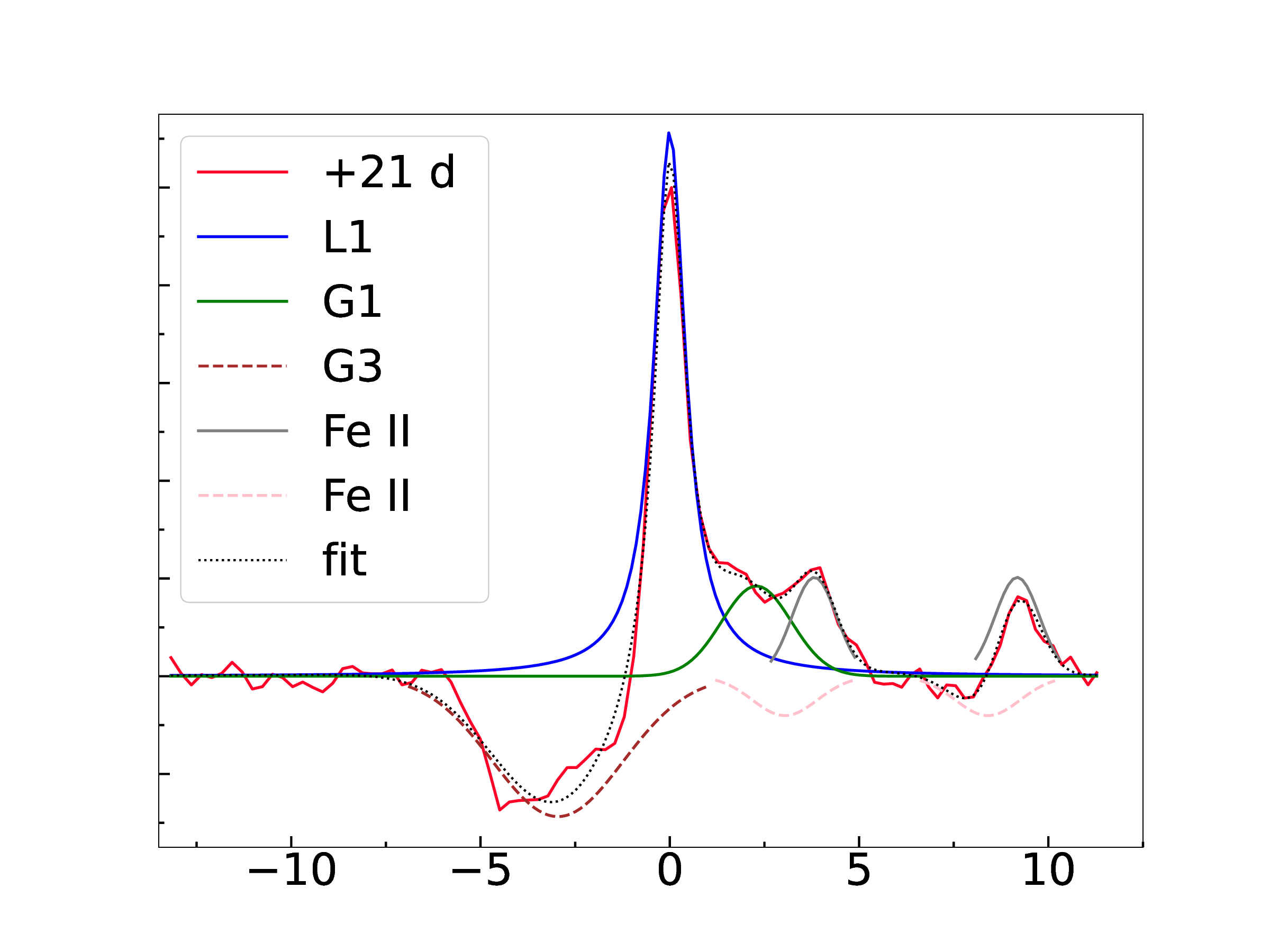}}
\end{minipage}
\begin{minipage}{0.239\linewidth}
\vspace*{-3.3cm}
\includegraphics[trim={4.8cm 3.3cm 3.5cm 3cm},clip,width=\linewidth]
{{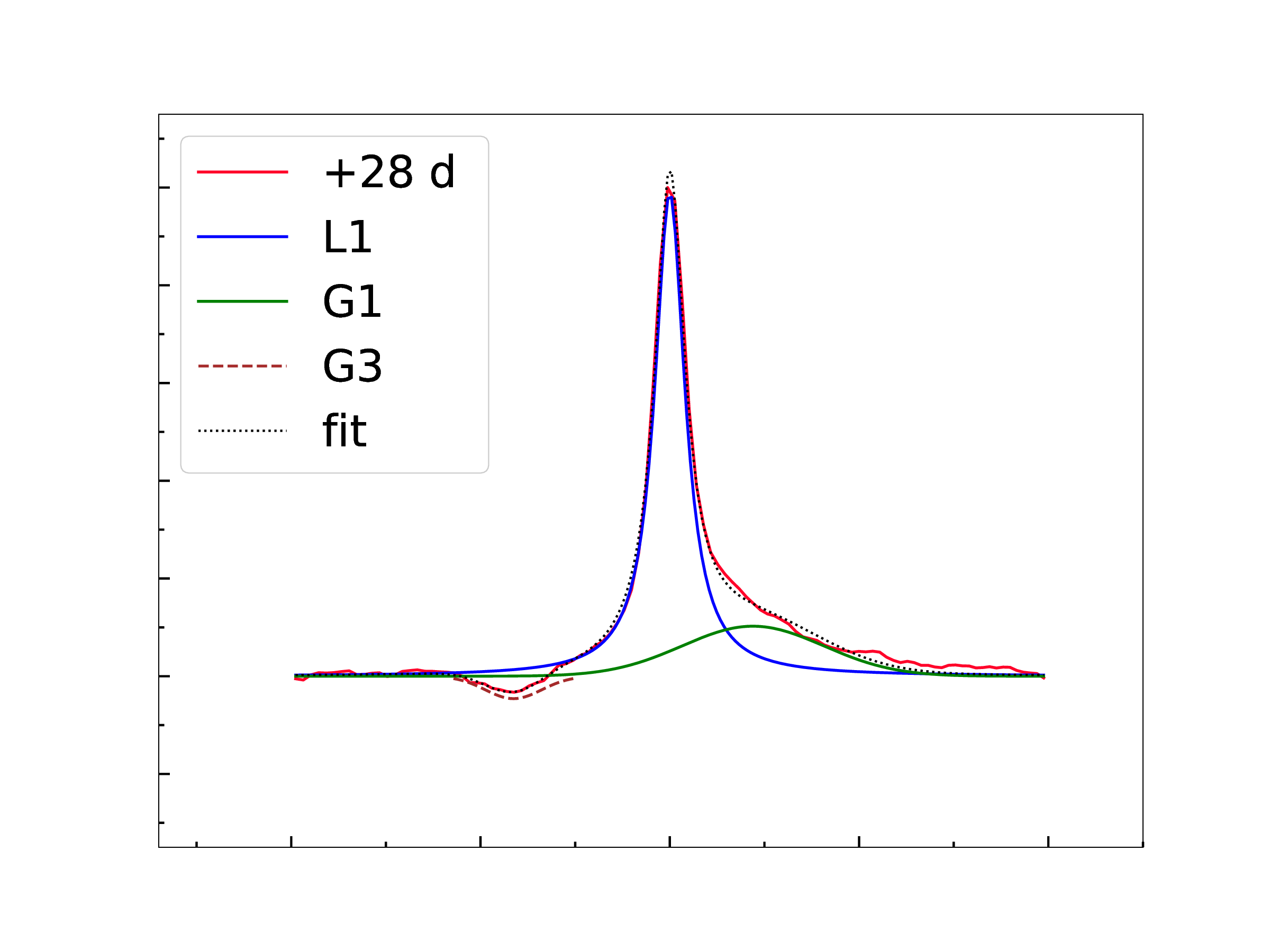}}1

\includegraphics[trim={4.8cm 3.3cm 3.5cm 3cm},clip,width=\linewidth]
{{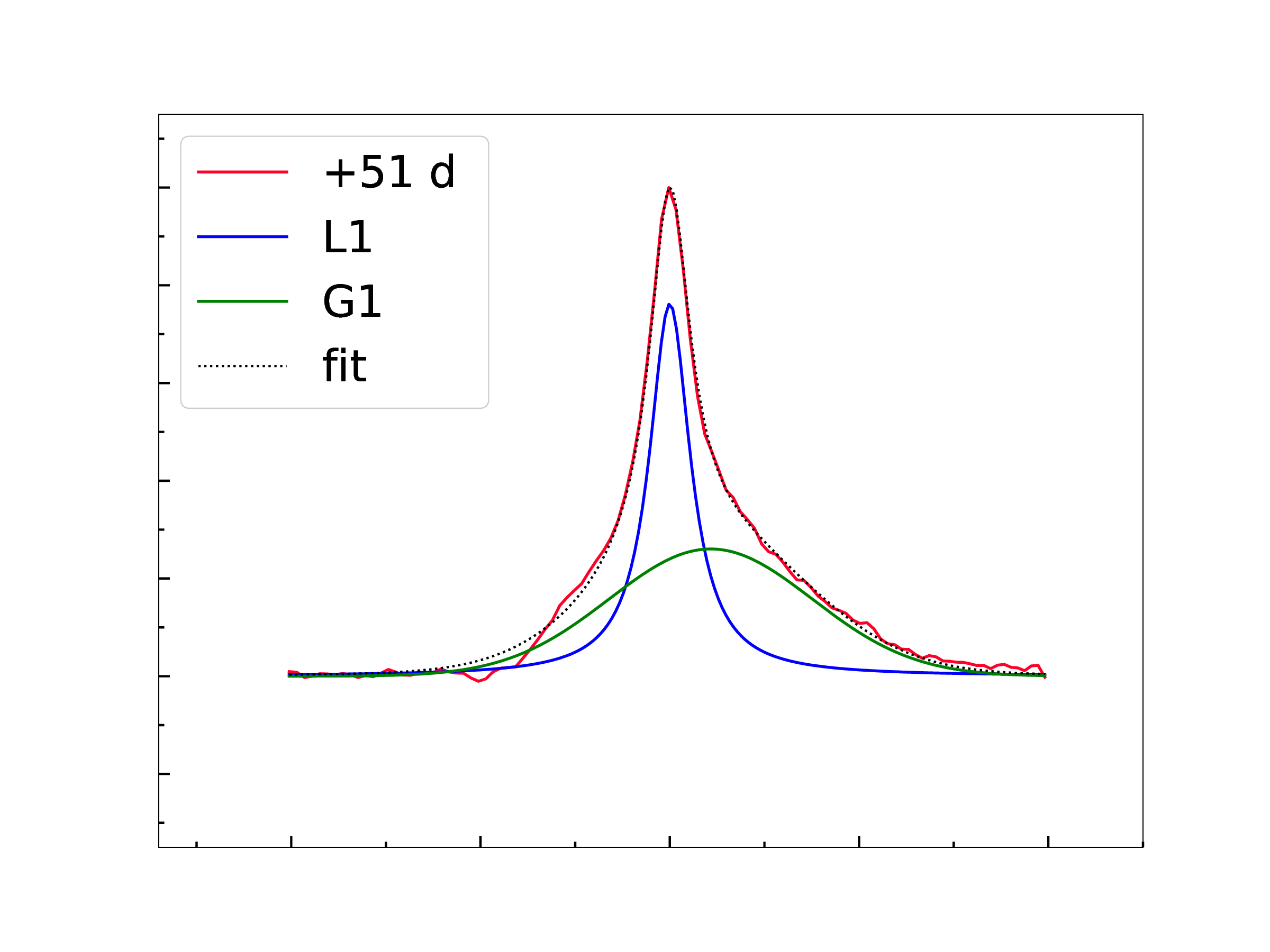}}

\includegraphics[trim={4.8cm 3.3cm 3.5cm 3cm},clip,width=\linewidth]
{{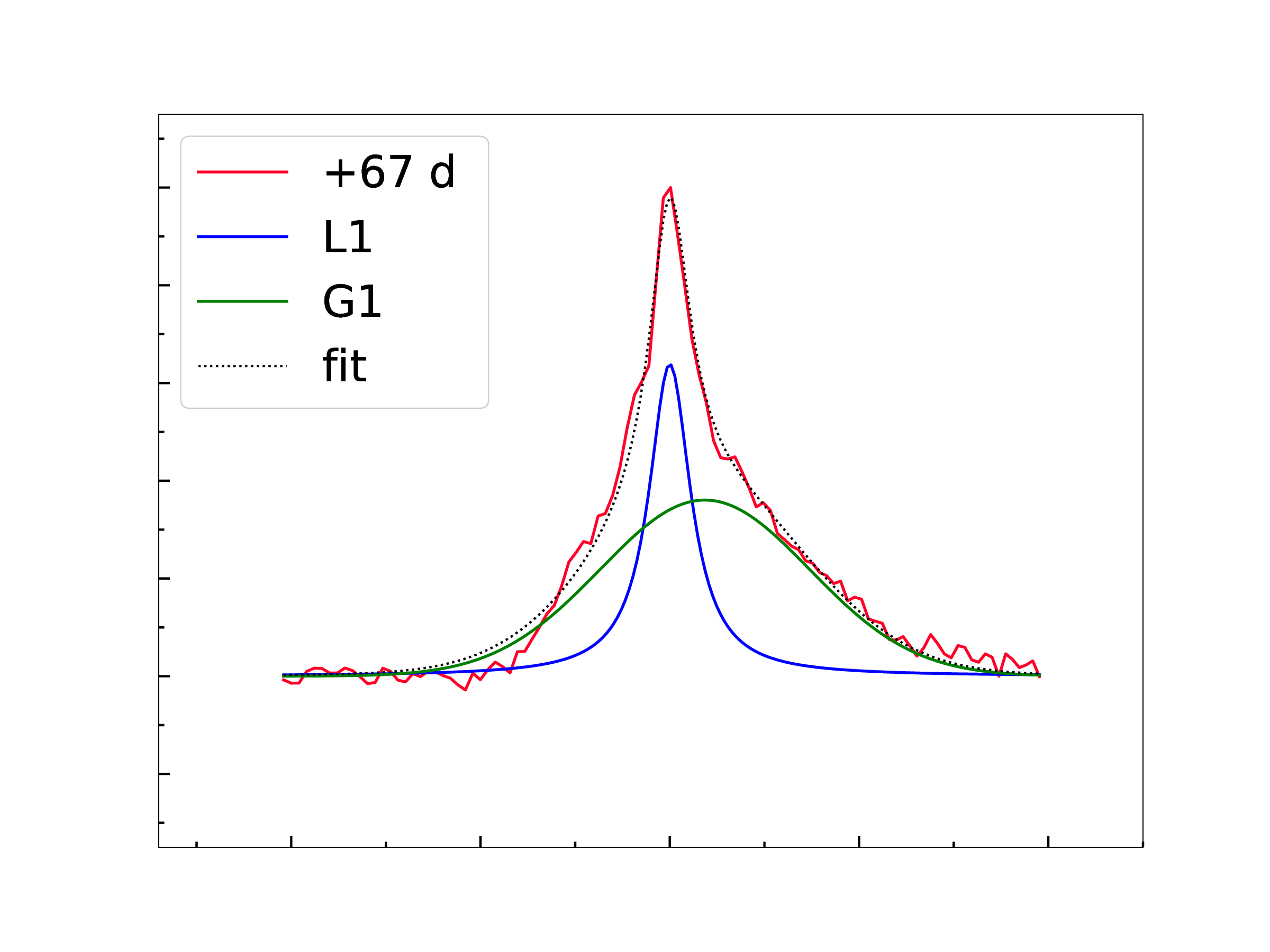}}

\includegraphics[trim={4.8cm 3.3cm 3.5cm 3cm},clip,width=\linewidth]
{{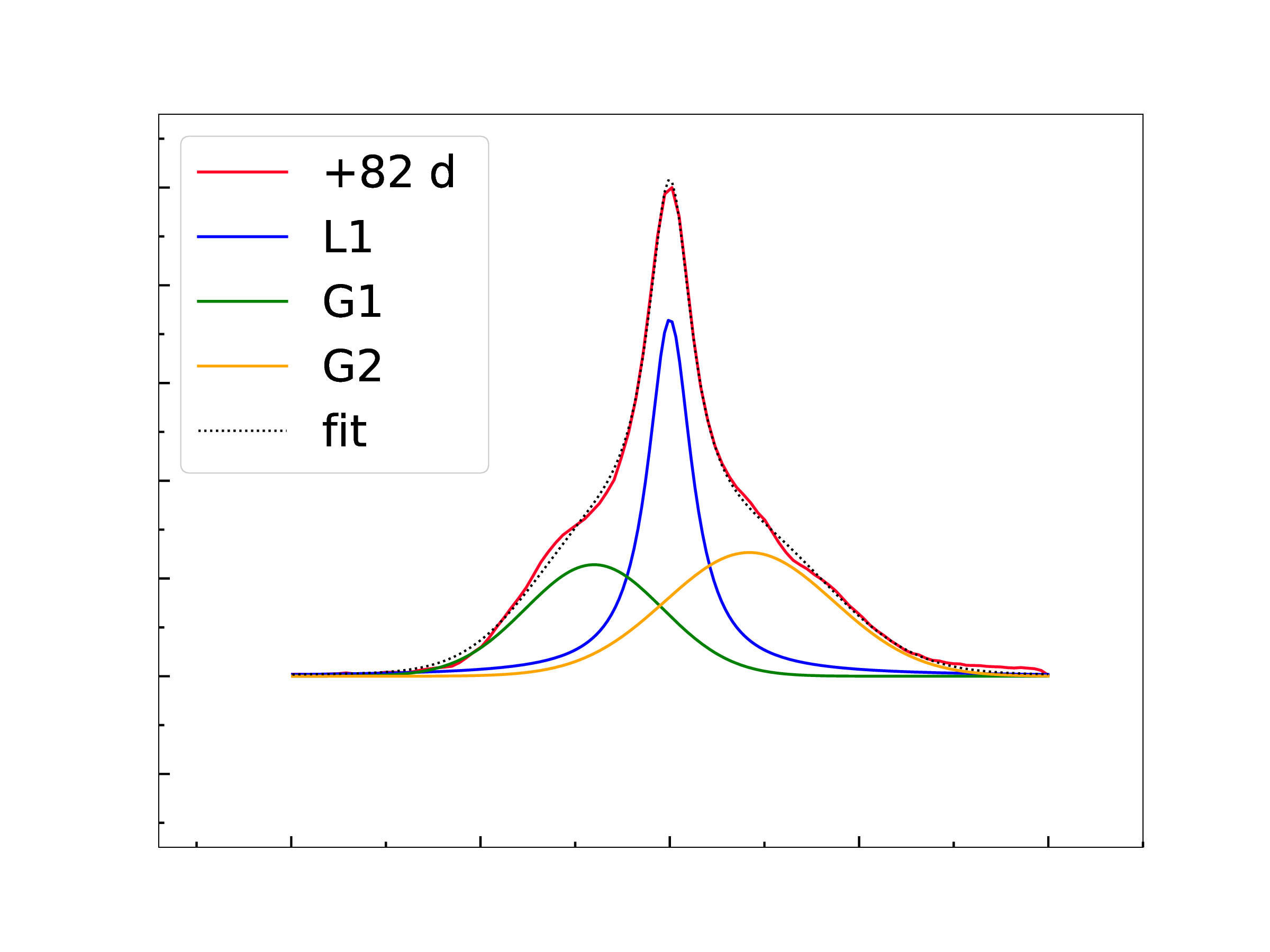}}

\includegraphics[trim={4.8cm 1.3cm 3.5cm 3cm},clip,width=\linewidth]
{{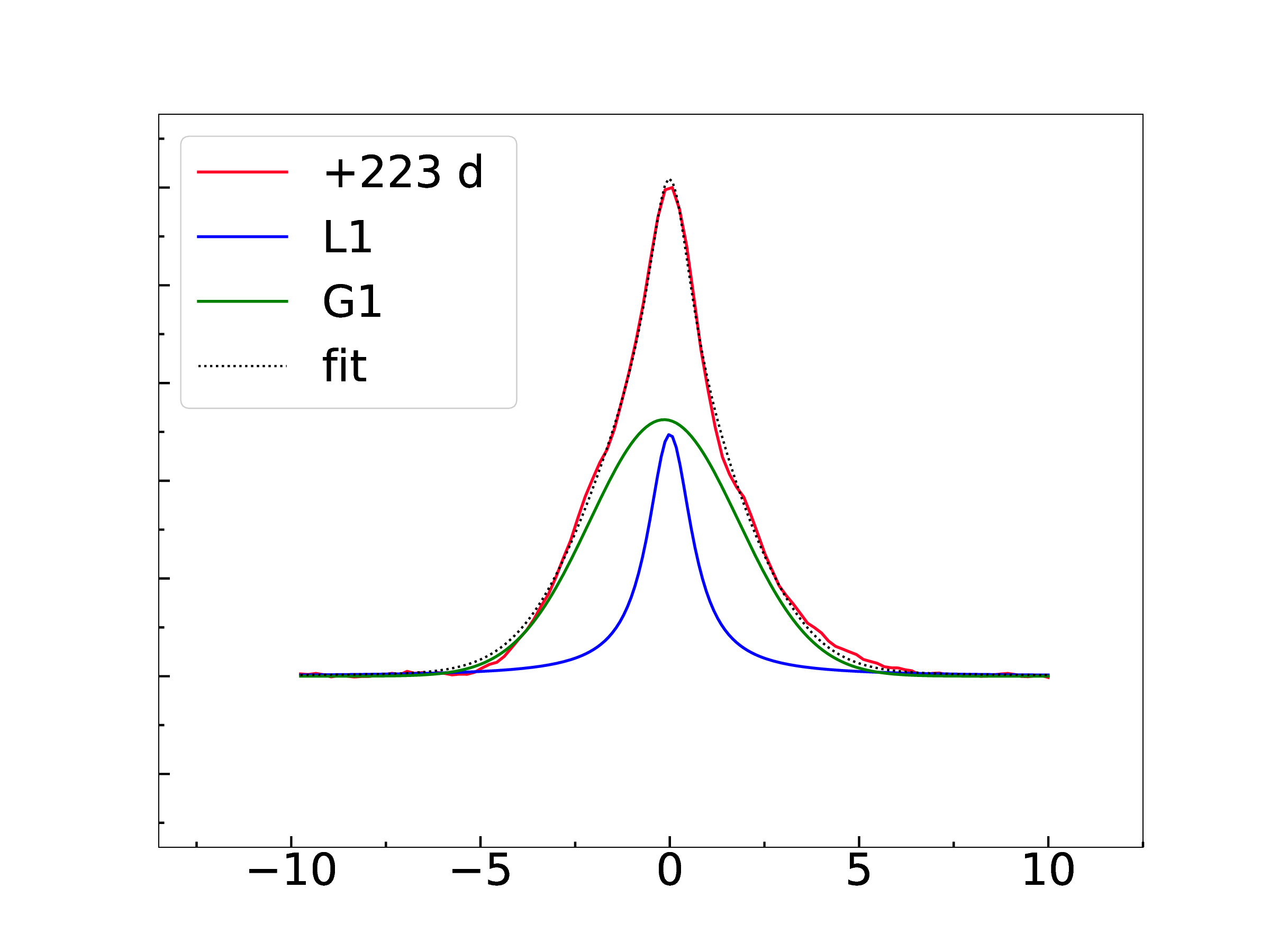}}
\end{minipage}\begin{minipage}{0.238\linewidth}
\vspace*{-3.3cm}
\includegraphics[trim={4.8cm 3.3cm 3.5cm 3cm},clip,width=\linewidth]
{{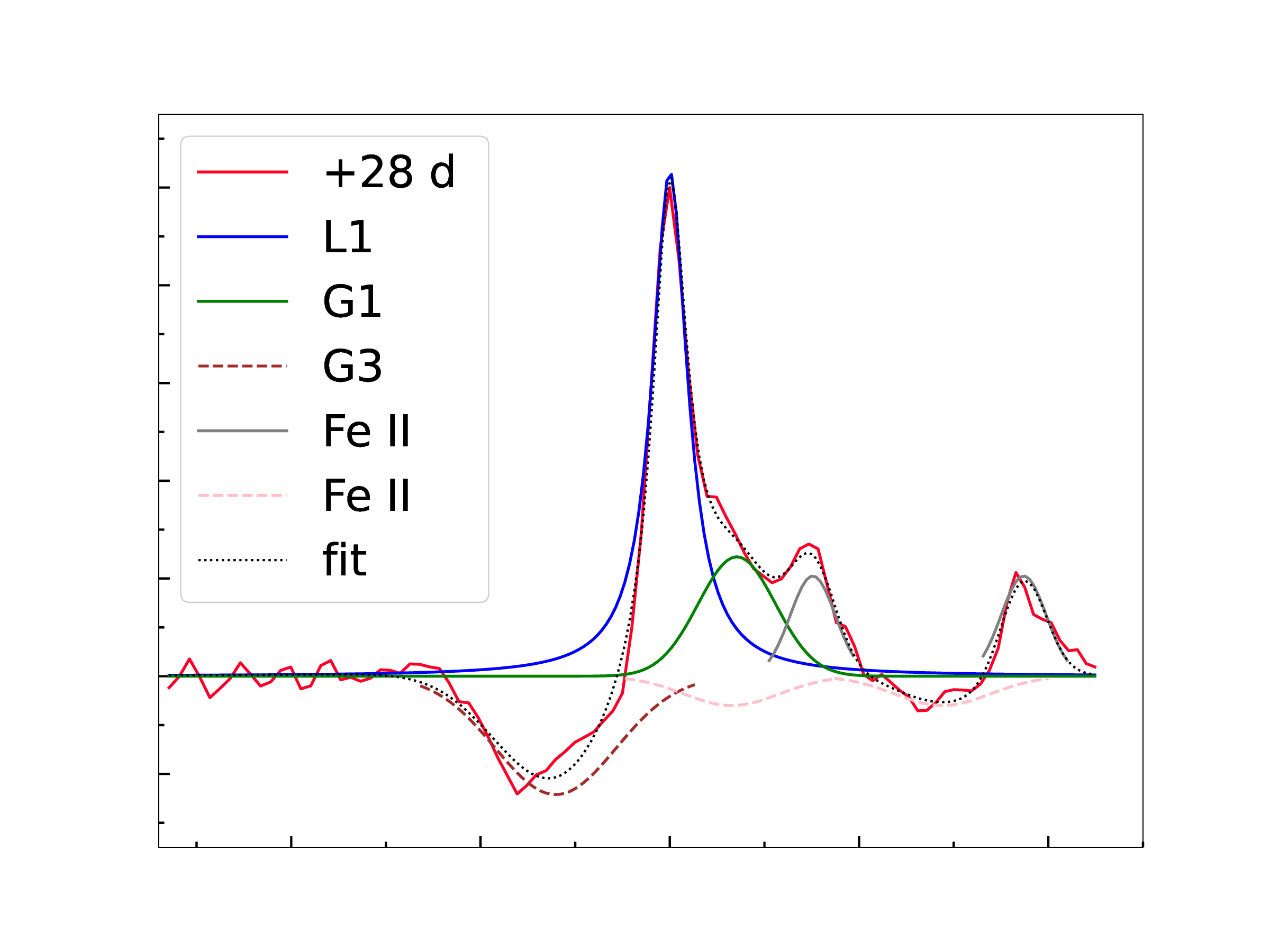}}

\includegraphics[trim={4.8cm 3.3cm 3.5cm 3cm},clip,width=\linewidth]
{{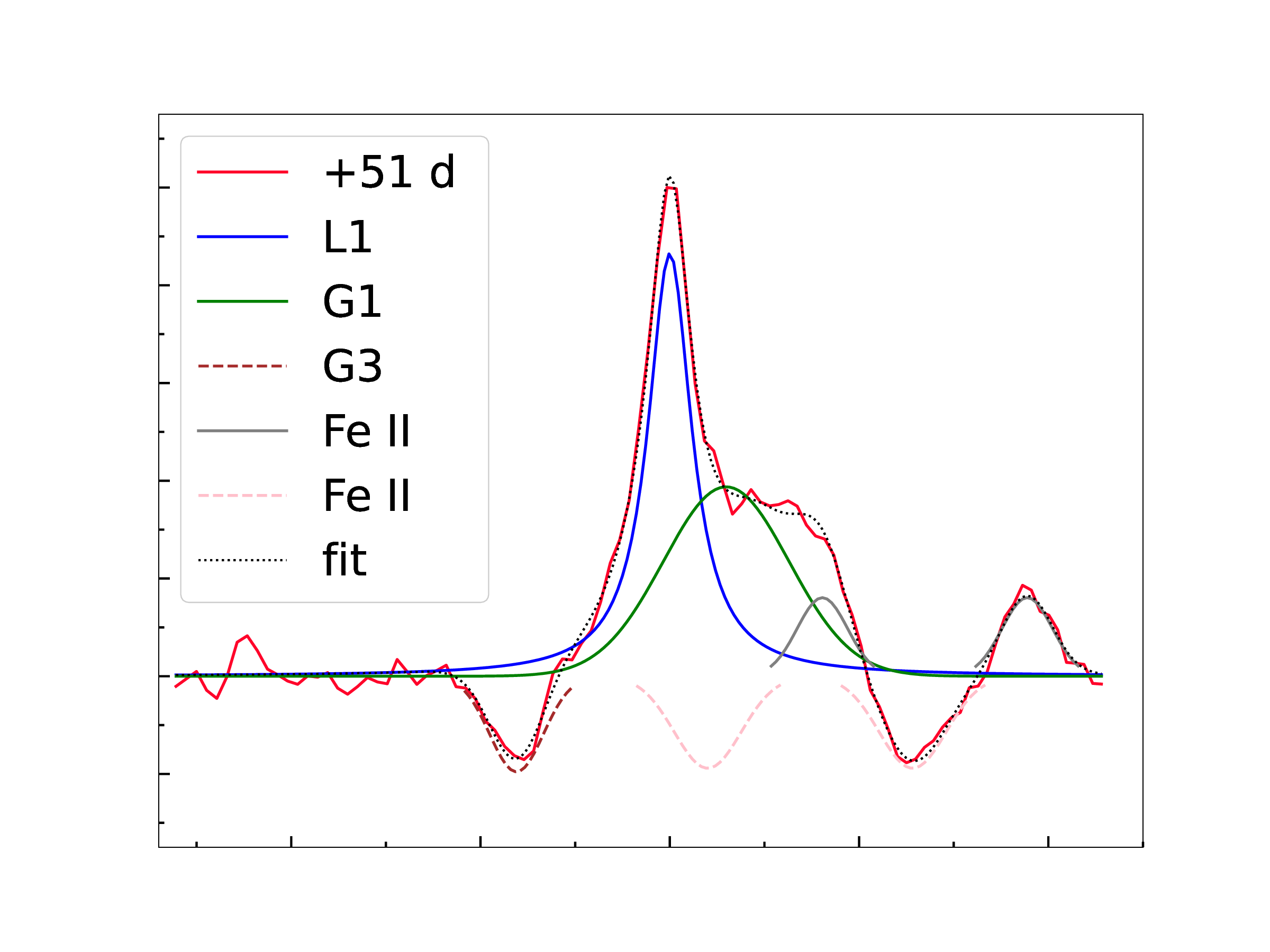}}

\includegraphics[trim={4.8cm 3.3cm 3.5cm 3cm},clip,width=\linewidth]
{{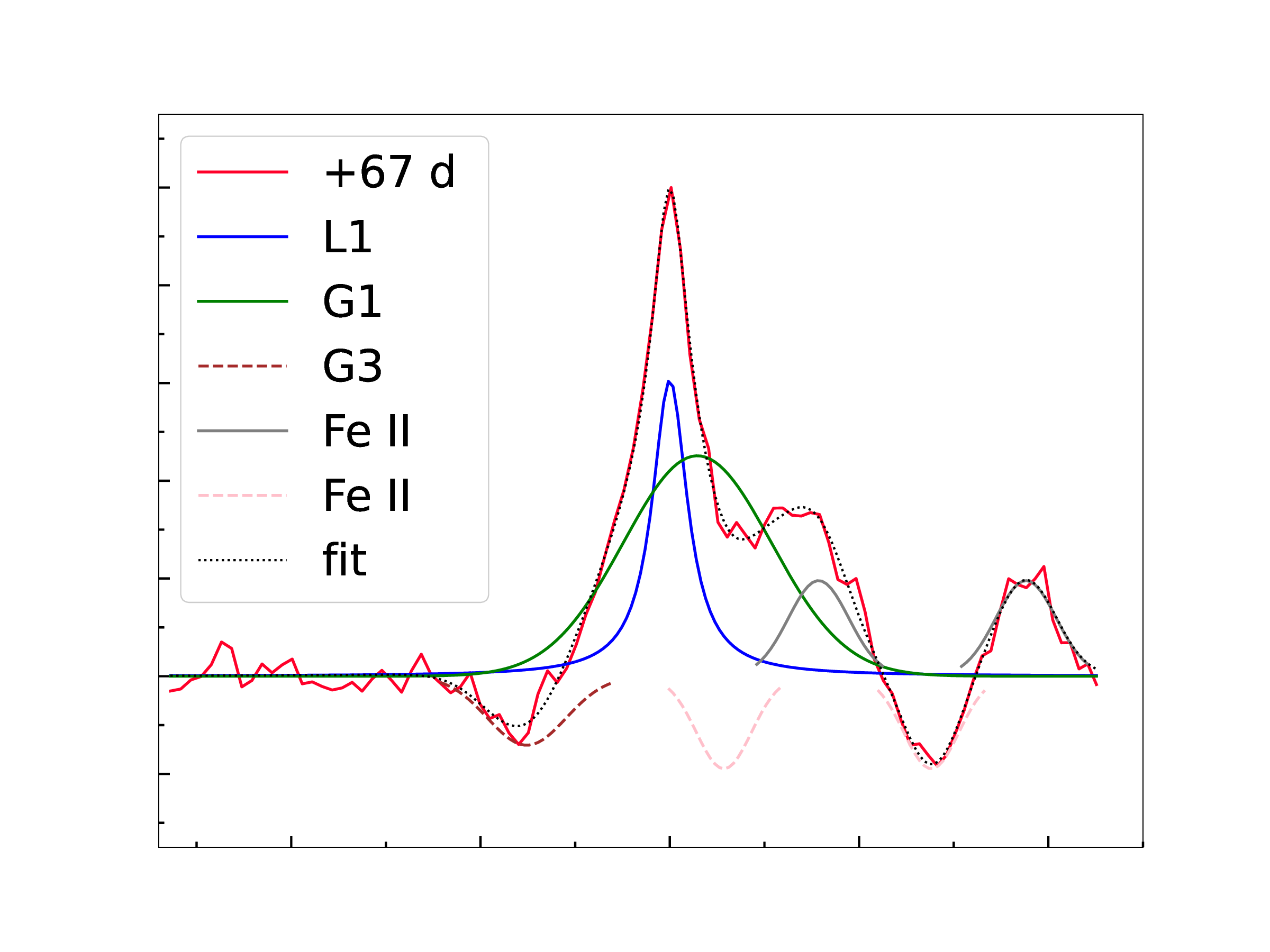}}

\includegraphics[trim={4.8cm 3.3cm 3.5cm 3cm},clip,width=\linewidth]
{{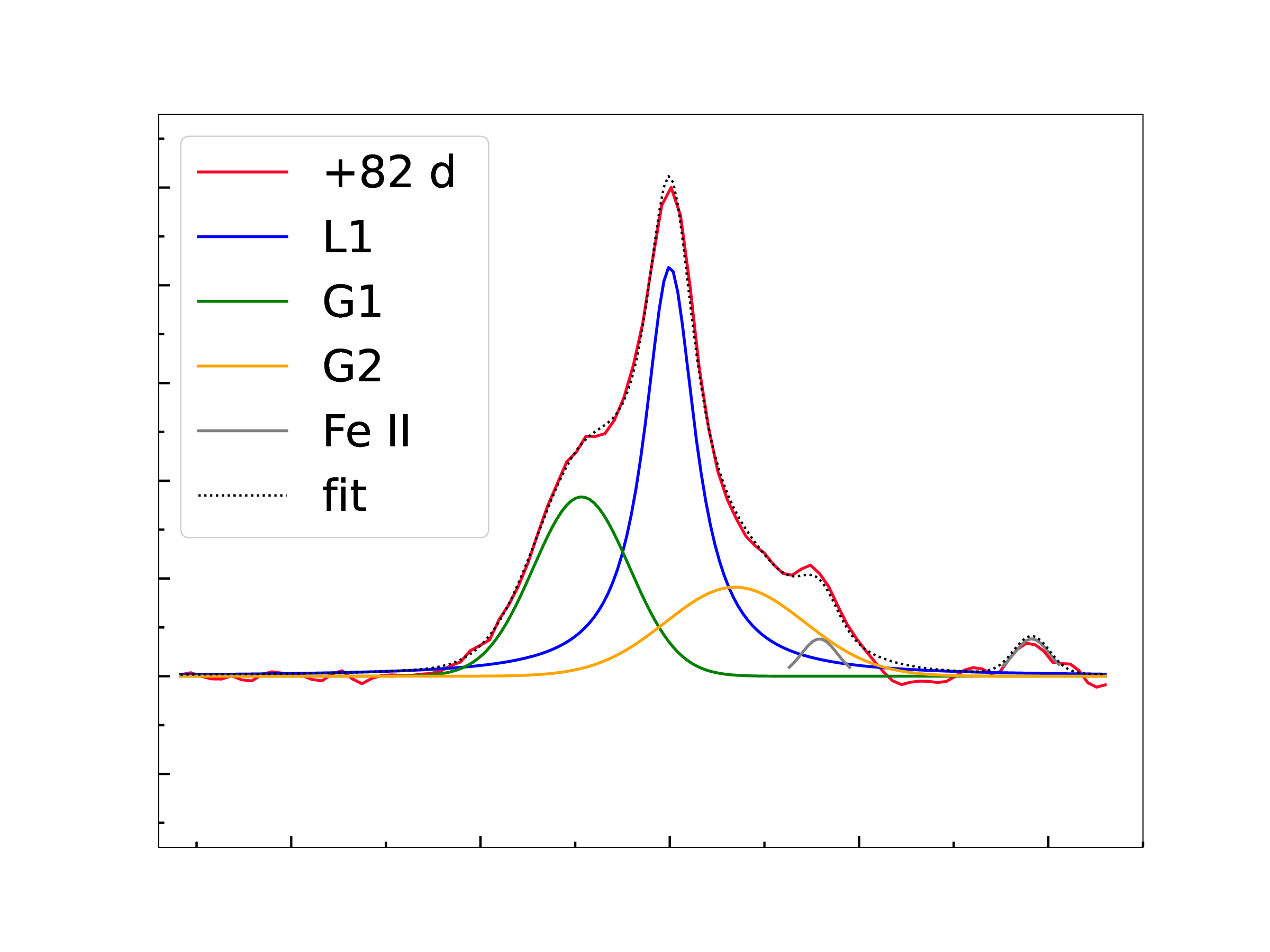}}

\includegraphics[trim={4.8cm 1.3cm 3.5cm 3cm},clip,width=\linewidth]
{{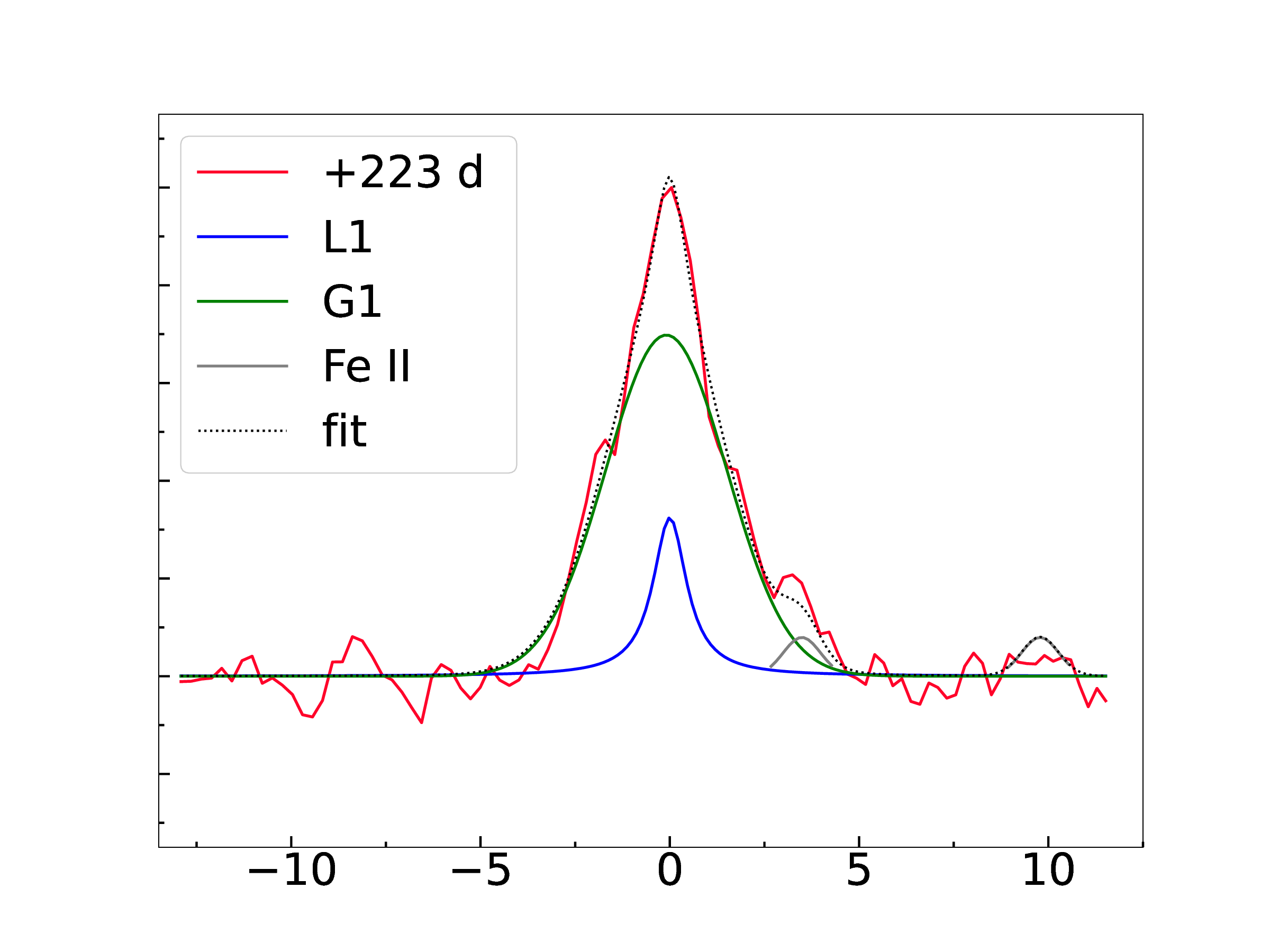}}
\end{minipage}
\vfill
\caption{Multi-component fits to H$_\alpha$ (first and third column) and H$_\beta$ (second and fourth column) lines in spectra of SN 2016cvk for different epochs. Unit of velocity (x axis) is $10^3$km s$^{-1}$, and the flux (y axis) is scaled to peak intensity of the spectral line.}\label{fig:MCMC}
\end{figure*}

\end{appendix}

\label{lastpage}

\end{document}